\documentclass[aps,twocolumn,nofootinbib,preprintnumbers]{revtex4}
\usepackage{bm} 
\usepackage{amsmath}
\usepackage{amssymb}
\usepackage{bbm}
\usepackage{slashed}
\usepackage{graphicx}
\def\qph{\mathrm{q}}
\def\bea{\begin{eqnarray}}
\def\eea{\end{eqnarray}}
\def\sfrac#1#2{{\textstyle \frac{#1}{#2}}}

\providecommand{\abs}[1]{\lvert#1\rvert}

\begin{document}

\title{Covariant spectator theory for the electromagnetic three-nucleon form factors: Complete impulse approximation}

\author{S\'ergio Alexandre Pinto}
\affiliation{Centro de F\'isica Nuclear da Universidade de Lisboa, 1649-003 Lisboa, Portugal \\and Departamento de F\'isica da Universidade de \'Evora, 7000-671 \'Evora, Portugal}

\author{Alfred Stadler}
\affiliation{Centro de F\'isica Nuclear da Universidade de Lisboa, 1649-003 Lisboa, Portugal \\and Departamento de F\'isica da Universidade de \'Evora, 7000-671 \'Evora, Portugal}

\author{Franz Gross}
\affiliation{Thomas Jefferson National Accelerator Facility, Newport News, VA 23606\\ and College of William and Mary, Williamsburg, VA 23187}

\date{\today}

\begin{abstract}
We present the first calculations of the electromagnetic form factors of $^3$He and $^3$H within the framework of the Covariant Spectator Theory (CST).  This first exploratory study concentrates on the sensitivity of the form factors to the strength of the scalar meson-nucleon off-shell coupling, known from previous studies to have a strong influence on the three-body binding energy.  Results presented here were obtained  using the complete impulse approximation (CIA), which includes contributions of relativistic origin that appear as two-body corrections in a non-relativistic framework, such as ``Z-graphs'', but omits other two and three-body currents.  We compare our results to non-relativistic calculations augmented by relativistic corrections of ${\cal O}(v/c)^2$.
\end{abstract}
\preprint{JLAB-THY-09-939}

\maketitle

\section{Introduction}

The Covariant Spectator Theory (CST) 
\cite{CST,Gro87,Sta97b} 
is a manifestly covariant formalism developed in particular for the description of few-nucleon systems. One of its most characteristic features is that four-momenta are conserved in intermediate states and all particles but one are confined to their mass shells. The remaining particle is off mass shell with its propagation described by the corresponding full Feynman propagator. One can derive a closed set of integral equations for the scattering amplitude, which effectively sum an infinite set of Feynman diagrams that consist of iterations of a kernel with an off-shell propagator after the on-shell constraints have been incoporated.  The kernel can be symmetrized (or antisymmetrized) as required for the treatment of identical particles.

For the nucleon-nucleon (NN) scattering amplitude, a number of CST potential models of One-Boson-Exchange (OBE) form have been developed that give a good description of the deuteron and the elastic NN scattering observables below 350 MeV \cite{Gro92,Gro07,Gro08}. 
The presence of off-mass-shell nucleons allows for a richer structure of the OBE kernels than what is possible in nonrelativistic frameworks. The general structure of the CST kernels contains terms for the coupling of bosons to nucleons that are exactly zero when the nucleon is on mass shell but contribute for off-mass-shell nucleons. For the scalar-isoscalar ($\sigma$ or $\sigma_0$) and scalar-isovector ($\delta$ or $\sigma_1$) mesons these couplings were investigated for the first time in \cite{Sta97}, and it was found that they increase the quality of the fit to the observables significantly.

When the CST equation for the three-nucleon (3N) bound state \cite{Sta97b} was solved for the first time, it was realized that the scalar off-shell coupling strongly influences the 3N binding energy \cite{Sta97}. When the scalar off-shell coupling strength was varied systematically, it turned out that the value that gave the best fit to the NN observables simultaneously also produced the best agreement with the experimental triton bound state energy of $E_t=-8.48$ MeV. What is also remarkable about this result is that nonrelativistic calculations with so-called ``realistic'' potentials, \textit{i.e.}, potentials that fit the NN scattering data with a $\chi^2/N_\mathrm{data} \approx 1$, do not reproduce $E_t$ without the addition of irreducible 3N forces specially adjusted for that purpose.

This somewhat unexpected success of the CST NN potential models in explaining $E_t$ appeared at first to be to some extent accidental. However, recently we found two new high-precision OBE models that fit the world data on $np$ scattering below 350 MeV with an essentially perfect $\chi^2/N_\mathrm{data} \approx 1$ \cite{Gro07,Gro08}, and in both cases we again found the triton binding energy to be very close to the measured value. Thus we are lead to conjecture that our relativistic OBE kernels capture the essential part of the physics responsible for the binding of the 3N system.

Of course one expects more from a realistic description of a bound state than merely to reproduce the correct binding energy. It should also reproduce its internal structure, which can be accessed through the corresponding electromagnetic form factors. While the interaction of nuclei with electromagnetic probes introduces some new problems in the form of ambiguities in the definition of the nuclear currents, the calculation of the electromagnetic 3N form factors certainly presents an interesting and necessary test of the three-body CST.

The exact form of the electromagnetic currents of a 3N system in CST was derived in a previous paper \cite{Gro04}. In the present work we use this current in order to calculate the electromagnetic form factors of $^3$H and $^3$He, as well as their isoscalar and isovector combinations, in the so-called complete impulse approximation (CIA), \textit{i.e.}, in an approximation that includes all one-body currents but leaves out two- and three-body currents.  As will be discussed, our CIA includes, in principle, contributions that in nonrelativistic frameworks appear as interaction currents, so care has to be taken when results of different formalisms are compared.

It is well known that interaction currents can have a significant influence on the electromagnetic 3N form factors. Therefore, a calculation in CIA cannot be expected to provide a quantitative description of the data over the entire range of the transferred momentum. Nevertheless, there is a number of important aspects that can be learned from a CST calculation in CIA: 

(a) The first aspect is technical in nature. The calculation of the form factors requires the numerical computation of rather envolved expressions, some of which have a complicated singularity structure. Since this is the first time these form factors are calculated in CST, it is important to learn how this can be done in an efficient and numerically stable way. In the same category belongs the study of the convergence of the employed expansion into helicity partial waves. 

(b) For a given 3N vertex function obtained by solving the CST 3N bound state equation with a chosen NN interaction model, we study the relative importance of the various different contributions to the total result. For instance, there are six Feynman diagrams that define the CIA, and it is useful to determine whether they are of comparable magnitude or if one may neglect some of them. Another interesting question is the importance of negative-energy states. The Feynman propagator of an off-shell nucleon can be decomposed into positive- and negative-energy parts, or equivalently into parts with positive or negative $\rho$-spin. The latter are related to the ``pair terms'' or ``Z-graphs'' in the language of time-ordered perturbation theory. It is known that, in nonrelativistic calculations with relativistic corrections
the Z-graphs connected to one-pion exchange have a sizeable effect. While a direct quantitative comparison of our negative-energy-state contributions to these pion Z-graphs is not easy because the decomposition into positive- and negative-energy states is frame dependent, one may still get a general idea about their size.

(c) Still operating with one particular 3N vertex function, we perform sensitivity studies with respect to the parameterization of the single-nucleon current. In addition to changes in the Dirac and Pauli nucleon form factors, off-shell nucleon currents also have off-shell terms and their associated form factors. We investigate the sensitivity of the results on changes in these off-shell currents. 

(d) We investigate the model dependence of the form factors by varying the NN interaction model from which the 3N vertex function was computed. This is done by using various members of a family of potentials that were fitted with different, fixed values of the scalar off-shell coupling. These models produce rather different 3N binding energies while the $\chi^2$ to the data varies only moderately, which allows one to assess also binding energy effects on the form factors in a systematic way.

(e) Since the calculations in the present work use NN interaction models in which the pion-nucleon coupling is purely of pseudo-vector form, the coupling to negative-energy nucleon states (or Z-graphs) is expected to be suppressed. Therefore it makes sense to compare our results to the ones in impulse approximation obtained in a nonrelativistic framework \cite{Mar98}, where relativistic effects are calculated to order $(v/c)^2$, while in CST they are included in all orders. The comparison may thus help to decide whether the $(v/c)$ expansions of relativistic corrections are justified.

(f) Last, but certainly not least, the calculations in this work give an indication of whether or not the CST calculations of three-body form factors will require any new physics.  For example, when we first calculated the three-nucleon binding energies in the CST we obtained about $-6$ MeV, {\it much\/} too underbound.  In trying to understand this result we discovered that the binding energy was very sensitive to the off-shell couplings of the scalar mesons, and that the introduction of such couplings not only  ``corrected'' the binding energy but also improved the fits to the two-body data.  We might not have realized that such previously not considered couplings were {\it needed\/} to give a good fit to the two-body data if we had not been initially confronted with such a large discrepancy in the binding energy.  In short, solving the binding energy ``problem'' lead to the discovery of a new interaction that, in CST, is also {\it necessary\/} for an efficient description of two-body data.  In this context, study of the form factors not only serves as a stringent test of these off-shell scalar couplings, but also exposes us to the possibility that another new mechanism will be required for the understanding of the form factors.  We return to this issue at the end of the paper.


This paper is organized in the following way: after the introduction in section I, section II provides a brief overview of the formalism in which the electromagnetic 3N form factors were calculated. The numerical results are presented and discussed in section III, and section IV contains our conclusions. Appendices A and B show some details about how the calculations were carried out.

\section{Formalism}

\subsection{Charge and magnetic form factors of composite spin-1/2 particles}

We start by reminding the reader that the most general form of the spin 1/2 current of the 3N bound states (with mass $M_t$) depends on two form factors
\begin{eqnarray}
\left\langle M'| J^\mu_{3N} |M \right\rangle &=&\bar u(P'_t,M')\Big[F^{3N}_1(Q^2)\,\gamma^\mu 
\nonumber\\
&&\quad\quad+F^{3N}_2(Q^2)\,\frac{i\sigma^{\mu\nu}\qph_\nu}{2M_t}\Big]\, u(P_t,M)\, ,\qquad \label{j3}
\end{eqnarray}
where $\qph=P'_t-P_t$ is the photon momentum, $P_t$ and $P'_t$ are the initial and final trinucleon momenta, $M$ and $M'$ the respective 
spin projections along the $z$-axis, , and we use the convention $Q^2=-\qph^2$.  (Note the use of the roman $\qph$ to represent the photon four-momentum, instead of the more customary $q$.)

It is convenient to rewrite this current using the Gordon decomposition (valid between on-shell spinors),
\begin{equation}
 \frac{i\sigma^{\mu\nu}\qph_\nu}{2M_t} = \gamma^\mu -\frac{(P'_t+P_t)^\mu}{2M_t} \, ,
 \label{gordon}
\end{equation} 
and to evaluate it in the Breit frame, where $\qph^0=0$ and ${\bf P}'_t={\bf q}/2=-{\bf P}_t$. We obtain
\bea
 \left\langle M'| J^0_{3N} |M \right\rangle & = &\bar u(P'_t,M')\Big[F^{3N}_M(Q^2)\,\gamma^0 \nonumber\\
&&-F_2^{3N}(Q^2) \frac{E}{M_t} 
\Big]\, u(P_t,M)\, ,\label{j30}\\
\left\langle M'| J^i_{3N} |M \right\rangle &=&F^{3N}_M(Q^2)\,\bar u(P'_t,M')\,\frac{i\sigma^{i\,\nu}\qph_\nu}{2M_t}\, u(P_t,M)\, ,\nonumber\\
\label{j3i}
\eea
where $E=P_t^{\prime 0}=P_t^0$ and $F_M=F_1+F_2$ is the familiar magnetic form factor.  Equation (\ref{j3i}) shows immediately that, if we choose $\qph$ to lie in the $+\hat z$ direction,  $J^3_{3N}=0$ and the spatial components of the current depend only on the magnetic form factor.

To evaluate these matrix elements we first start with the spinors for the three-body states when they are at rest:
\begin{equation}
u(0,M)=\left({\begin{array}{c} 1 \cr 
0 \end{array}}  \right) \otimes\chi_{_M}
\end{equation}
where $M=\pm\frac{1}{2}$ with
\begin{equation}
\chi_{\frac{1}{2}}= \left({\begin{array}{c} 1\cr 
0\end{array}}  \right)\qquad
\chi_{-\frac{1}{2}}= \left({\begin{array}{c} 0\cr 
1\end{array}}  \right)\, . \label{2spinor}
\end{equation}
Then, the initial and final three-body states will  be obtained by boosting the rest spinors to the correct frame.  The Dirac operator for an active boost to a frame with velocity
$\sinh \xi=Q/2M_t$ in the $+\hat z$ direction (we use $Q \equiv |{\bf q}|$) is denoted $S(B(\xi \hat{e}^3))$ and defined in Eq.~(\ref{eq:boostrot}).
 
With this notation for the Breit frame matrix elements, where the initial state is boosted in the $-\hat z$ direction, the matrix element (\ref{j30}) becomes
\begin{align}
&\left\langle M'| J^0_{3N} |M \right\rangle =\bar u(0,M')S^{-1}(B(\xi \hat{e}^3))\Big[F_M(Q^2)\,\gamma^0 
\nonumber\\
&\qquad\qquad-F_2(Q^2)\,\frac{E_Q}{M_t}\Big]\,S(B(-\xi\hat{e}^3))\, u(0,M) \nonumber\\
&\qquad=\bar u(0,M')\,\Bigg[F_M(Q^2) 
-F_2(Q^2)\,\frac{E^2_Q}{M_t^2}\Bigg]\, u(0,M)\nonumber\\
&\qquad=\delta_{M,M'}\,\Big[F_1(Q^2)-\frac{Q^2}{4M_t^2}F_2(Q^2)\Big]\, ,\label{j0a}
\end{align}
where $E_Q=M_t(\cosh^2\xi/2+\sinh^2\xi/2)=M_t\cosh\xi=\sqrt{M_t^2+Q^2/4}$.  The 0th component of the current in the Breit frame conserves spin (or flips helicity) and equals the charge form factor $F_C=F_1-\tau F_2$ with $\tau=Q^2/4M_t^2$.

Similarly, the $x$ component of the current is
\bea
\left\langle M'| J^1_{3N} |M \right\rangle &=&\frac{Q\,F^{3N}_M(Q^2)}{2M_t}\,\bar u(0,M')\,S^{-1}(B(\xi\hat{e}^3))
\nonumber\\
&&\qquad\qquad\times
i\Sigma_2\,S(B(-\xi\hat{e}^3))\, u(0,M) \qquad
\nonumber\\
&=&\frac{Q\,F^{3N}_M(Q^2)}{2M_t}\,\bar u(0,M')\,
i\Sigma_2\, u(0,M)\nonumber\\
&=&2M'\,\delta_{M',\,-M}\frac{Q\,F^{3N}_M(Q^2)}{2M_t}\, , \label{j3xa}
\label{j3x}
\eea
where we replaced $-i\sigma^{13}=i\Sigma_2$ [with the $\Sigma_i$ matrices in Dirac space defined in Eq.~(\ref{eq:sigmaalpha})] and the phase $2M'$ arises from the matrix element of $i\Sigma_2$. 

The $y$ component of the current is obtained in the same way, yielding
\begin{equation}
 \left\langle M'| J^2_{3N} |M \right\rangle =i 2M'\,\delta_{M',\,-M}\frac{Q\,F^{3N}_M(Q^2)}{2M_t}\, . \label{j3ya}
\end{equation} 
In this work, we actually calculate the 3N current matrix elements in the Lab frame. Since the current matrix elements transform like a four-vector, we can apply a boost from the Breit to the Lab frame,
\begin{equation}
\left\langle M'| J^{\mu}_\mathrm{Lab} | M \right\rangle  = \left[B(\xi \hat{e}^3)\right]^{\mu}_{\phantom{\mu}\nu} \left\langle M'| J^{\nu}_\mathrm{Breit}  | M \right\rangle \, ,
\end{equation} 
in order to determine the relation between the charge and magnetic form factors and the Lab frame current matrix elements.
\begin{align}
F^{3N}_{C}(Q^{2}) &= \frac{\langle M|J^{0}_{3N,\mathrm{Lab}}| M \rangle}{\sqrt{1+ \frac{Q^{2}}{4M_{t}^{2}}  }}  \, , \\
F^{3N}_{M}(Q^{2}) &= - (2M) \frac{2M_{t}}{Q} \langle -M| J^{1}_{3N,\mathrm{Lab}} | M \rangle \\
&= i (2 M) \frac{2M_{t}}{Q} \langle -M | J^{2}_{3N,\mathrm{Lab}} | M \rangle\, .
\end{align}
We conclude that the charge and magnetic 3N form factors can be obtained from the matrix elements of the time and space components of the 3N electromagnetic current in any convenient frame.

\subsection{The three-nucleon vertex functions and the relativistic wave function}

The three-body form factors in the spectator theory are expressed in terms of the bound-state Faddeev vertex function $\Gamma_{\lambda_1\lambda_2\alpha}(k_1,k_2,k_3)$ with (by convention) particles 1 and 2 on mass shell with helicities $\lambda_1$ and $\lambda_2$, and particle 3 off mass shell with the associated Dirac index $\alpha$ ($k_i$ is the four-momentum of nucleon $i$ with mass $m$). The total four-momentum (which is conserved) is $P=k_1+k_2+k_3$ and $k_1^2=k_2^2=m^2$. The on-shell energy of a nucleon with momentum $k_i$ is denoted $E(k_i)\equiv \sqrt{m^2+{\bf k}_i^2}$. In this paper we adopt the alternative notation for $\Gamma=\Gamma(k_1,k_2;P)$, convenient because $k_3=P-k_1-k_2$ is a dependent variable. This vertex function is related to the relativistic wave function $\Psi$, {\it defined\/} by the relation
\begin{eqnarray}
\Psi_{\lambda_1\lambda_2\alpha}(k_1,k_2;P)=
G_{\alpha\alpha'}(k_3)
\Gamma_{\lambda_1\lambda_2\alpha'}(k_1,k_2;P)
\, ,
\label{wavefunction}
\end{eqnarray}
where we continue to use $k_3$ as the argument of $G$ to make the notation more compact.  The propagator $G$ is
\begin{align}
G_{\alpha\alpha'}(k_3)&=\left[\frac{1}{(m-\slashed{k}_3)}\right]_{\alpha'\alpha}
\nonumber\\
&= \frac{m}{E(k_3)}\sum_\lambda\left[\frac{u_{\alpha'}({\bf k}_3,\lambda)
\bar u_{\alpha}({\bf k}_3,\lambda)}{k^0_{3}-E(k_3)-i\epsilon} \right.
\nonumber\\
&\left.
\qquad\qquad\qquad-\frac{v_{\alpha'}(-{\bf k}_3,\lambda)
\bar v_{\alpha}(-{\bf k}_3,\lambda)}{k^0_{3}+E(k_3)-i\epsilon}\right]
\nonumber\\
&=\frac{m}{E(k_3)}\sum_{\lambda\rho}\left[\frac{\rho\,u_{\alpha'}^\rho({\bf k}_3,\lambda)
\bar u_{\alpha}^\rho({\bf k}_3,\lambda)}{k^0_{3}-\rho E(k_3)-i\epsilon}\right]\, ,
\label{oneprop}
\end{align} 
where $E(k_3)$ is the {\it on-shell\/} energy of particle 3, $k_3^0$ its {\it off-shell\/} energy, and the $\rho$-spin is either + or $-$ with the convention
\bea
u_{\alpha}^+({\bf k}_3,\lambda)&=&u_{\alpha}({\bf k}_3,\lambda)\nonumber\\
u_{\alpha}^-({\bf k}_3,\lambda)&=&v_{\alpha}(-{\bf k}_3,\lambda)\, .
\eea
 
\begin{figure}
%
\centerline{\hspace*{-0.1in}
\mbox{
\includegraphics[width=2.in]{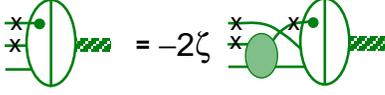}}}
\caption{\footnotesize\baselineskip=10pt  (Color online) 
Diagrammatic representation of the Faddeev equation (\ref{eq:3NFad}).}
\label{fig:faddeev}
\end{figure}  


\begin{figure}
\vspace*{0.2in}
\centerline{
\mbox{
\includegraphics[width=2.5in]{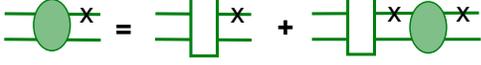}
}}
\caption{\footnotesize\baselineskip=10pt  (Color online) 
Diagrammatic representation of the Eq.~(\ref{twobodyoff}) for the
symmetrized two-body scattering subamplitude $M_{\beta\alpha,\lambda'_2\alpha'}$ with
{\it both\/} final particles off shell.}
\label{fig:Meqoff}
\end{figure}  

The vertex function satisfies the following Faddeev equation (shown diagrammatically in Fig.~\ref{fig:faddeev}, with $\zeta=+1$ for bosons and $\zeta=-1$ for fermions)
\begin{eqnarray}
\lefteqn{\Gamma_{\lambda_1\lambda_2\alpha}(k_1,k_2;P)=-\int
\frac{m\,d^3k'_2}{E(k'_2)\,(2\pi)^3}} \nonumber \\
& & \times
\sum_{\lambda'_2}
M_{\lambda_2\alpha,\lambda'_2\alpha'}(k_2,k_2';P_{23})\,
2\,\zeta\,{\cal P}_{12}\,
\Psi_{\lambda_1\lambda'_2\alpha'}(k_1,k'_2;P)
\, , \nonumber\\
\label{eq:3NFad}
\end{eqnarray}
where $P_{23}=P-k_1$ is the total four-momentum of the pair, $M$ is the two-body scattering amplitude satisfying
\begin{eqnarray}
\lefteqn{M_{\lambda_2\alpha,\lambda'_2\alpha'}(k_2,k'_2;P_{23}) = V_{\lambda_2\alpha,\lambda'_2\alpha'}(k_2,k'_2;P_{23})} \nonumber\\
&&
-\int
\frac{m\,d^3k''_2}{E(k''_2)\,(2\pi)^3}
\sum_{\lambda''_2}
V_{\lambda_2\alpha,\lambda''_2\beta}(k_2,k''_2;P_{23}) \nonumber\\
&&\times 
G_{\beta\beta'}(P_{23}-k''_2)M_{\lambda''_2\beta',\lambda'_2\alpha'}(k''_2,k'_2;P_{23})\, ,
\label{twobody}
\end{eqnarray}
${\cal P}_{12}$ is the permutation operator that interchanges particles 1 and 2, and $V$ is the two body interaction kernel (which we will frequently call the ``potential'' because of its close connection with the nonrelativistic potential).
For convenience we
have adopted the notation
\begin{eqnarray}
\lefteqn{V_{\lambda_2\alpha,\lambda'_2\alpha'}(k_2,k'_2;P_{23})} \nonumber \\
&&
\equiv 
\bar u_{\beta}(k_2,\lambda_2)V_{\beta\alpha,\beta'\alpha'}(k_2,k'_2;P_{23})\,
u_{\beta'}(k_2',\lambda'_2)
\label{vnotation}
\end{eqnarray}
so care must be taken to distinguish Dirac indicies from helicity indices. 
Whenever a Dirac index is replaced by a helicity index, a contraction with an
on-shell, positive energy spinor, such is shown in Eq.~(\ref{vnotation}), is
implied, {\it and unless otherwise stated\/}, it is assumed that the particle is on-shell.  The on-shell Dirac spinors are normalized to $\bar u u=1$.

Calculations of the form factors requires knowledge of the vertex
function with the {\it two interacting\/} nucleons
off-shell.   This vertex function was defined in Ref.~\cite{Gro04} and can be obtained using the Faddeev Eq.~(\ref{eq:3NFad}), generalized to the case when {\it both\/} of the final state interacting nucleons are off-shell
\begin{eqnarray}
\lefteqn{
\Gamma_{\lambda_1\beta\alpha}(k_1,k_2;P)=-\int
\frac{m\,d^3k'_2}{E(k'_2)\,(2\pi)^3}
}\nonumber \\
&&
\times
\sum_{\lambda'_2}
M_{\beta\alpha,\lambda'_2\alpha'}(k_2,k_2';P_{23})
2\,\zeta\,{\cal P}_{12}\,
\Psi_{\lambda_1\lambda'_2\alpha'}(k_1,k'_2;P)
\, , \nonumber \\
\label{eq:3NFadoff}
\end{eqnarray}
where now $k_2^2\ne m^2$. 
The off-shell scattering amplitude is obtained by quadratures from the off-shell kernel $V_{\beta\alpha,\lambda'_2\alpha'}$ and the on-shell scattering amplitude $M_{\lambda''_2\beta',\lambda'_2\alpha'}$ using a generalization of the two-body equation (\ref{twobody})
\begin{eqnarray}
\lefteqn{
M_{\beta\alpha,\lambda'_2\alpha'}(k_2,k'_2;P_{23}) = V_{\beta\alpha,\lambda'_2\alpha'}(k_2,k'_2;P_{23})
} \nonumber\\
&&
-\int
\frac{m\,d^3k''_2}{E(k''_2)\,(2\pi)^3}
\sum_{\lambda''_2}
V_{\beta\alpha,\lambda''_2\beta}(k_2,k''_2;P_{23}) \nonumber \\
&& \times
G_{\beta\beta'}(P_{23}-k''_2)M_{\lambda''_2\beta',\lambda'_2\alpha'}(k''_2,k'_2;P_{23})\, .
\label{twobodyoff}
\end{eqnarray}
This equation is illustrated in Fig.~\ref{fig:Meqoff}.
The off-shell kernel $V_{\beta\alpha,\lambda'_2\alpha'}$ is  known (in principal), and is discussed in more detail in Appendix \ref{app:A}.

\subsection{The three-nucleon form factors}

\begin{figure*}
\includegraphics[width=6.4cm,angle=-90]{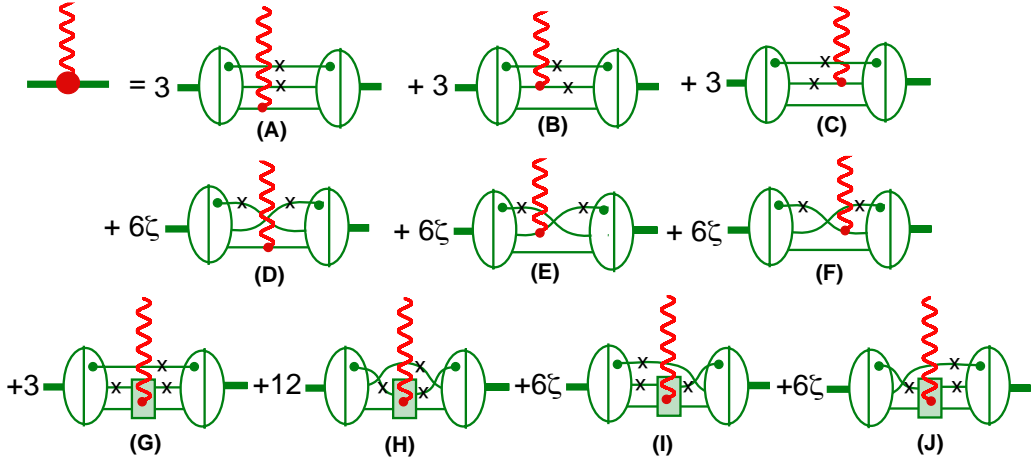}
\caption{\footnotesize\baselineskip=10pt  (Color online) 
The electromagnetic 3N current in CST for elastic electron scattering from the 3N bound state.  Diagrams (A) to (F) define the complete impulse approxiumation (CIA), in which the photon couples to single nucleons, which can be off-shell (A and D) or on-shell before or after the photon-nucleon vertex (B, C, E, and F). The interaction diagrams (G) to (J) describe processes in which gthe photon couples to two-body currents associated with the two-nucleon kernel.}
\label{fig:coreFF}
\end{figure*} 

The diagrams needed to calculate these form factors are displayed in Fig.~\ref{fig:coreFF}.  The
diagrams A--F are referred to as 
the complete impulse approximation (CIA).  Diagrams G--J are the interaction currents.  The algebraic result for
the six diagrams that make up the CIA can be written 
\begin{widetext}
\begin{eqnarray}
J^\mu_{\rm CIA}&=&3e\int\!\!\int
\frac{m^2\,d^3k_1d^3k_2}{E(k_1)E(k_2)\,(2\pi)^6}
\sum_{\lambda_1\lambda_2}\Bigl\{
\bar\Psi_{\lambda_1\lambda_2\alpha'}(k_1,k_2;P'_t)\,
[1+2\,\zeta{\cal P}_{12}]
\,j_{\alpha'\alpha}^\mu(k'_3,k_3)\,
\Psi_{\lambda_1\lambda_2\alpha}(k_1,k_2;P_t)\nonumber\\
& &+\bar\Gamma_{\lambda_1\beta'\alpha}(k_1,k'_2;P'_t)
\,G_{\beta'\beta}(k'_2)
\,j_{\beta\gamma}^\mu(k'_2,k_2)\,u_{\gamma}(k_2,\lambda_2)\,
[1+2\,\zeta{\cal P}_{12}]\,
\Psi_{\lambda_1\lambda_2\alpha}(k_1,k_2;P_t)\nonumber\\
& &+\bar\Psi_{\lambda_1\lambda_2\alpha}(k_1,k'_2;P'_t)\,
[1+2\,\zeta{\cal P}_{12}]\,\bar u_{\gamma}(k'_2,\lambda_2)
\,j_{\gamma\beta'}^\mu(k'_2,k_2)\,G_{\beta'\beta}(k_2)\,
\Gamma_{\lambda_1\beta\alpha}(k_1,k_2;P_t)
\Bigr\}\, , \label{CIA}
\end{eqnarray}
\end{widetext}

\noindent where
the doubly off-shell vertex functions are evaluated
using (\ref{eq:3NFadoff}),
$j_{\alpha'\alpha}(k',k)$ is the single nucleon current for
off-shell nucleons with incoming (outgoing) four-momenta
$k^\mu$ ($k'^\mu$), $P_t$ is the four-momentum of the incoming, and  $P'_t=P_t+q$ of the outgoing three-body nucleus, $k'_i=k_i + q$, and in every term
$k_1^2=k_2^2=m^2$.  Each diagram has two
off-shell propagators and three interactions, for a phase of
$i(-i)^5=1$.  Mathematical expressions for the diagrams with interaction currents are  given in \cite{Gro04} (similar expressions are also derived in \cite{Kvi97b} and \cite{Ada05}), and will not be needed here. Equation~(\ref{CIA}) with the
two particle off-shell vertex function defined by (\ref{eq:3NFadoff}) is evaluated numerically
in this paper.

\subsection{Three-nucleon partial wave helicity states}
\label{sec:pw}
In the 3N bound states, the three nucleons are treated as identical particles with an isospin degree of freedom. The expression for the 3N current should therefore be symmetric with respect to which nucleon the photon couples to, and which nucleons are on or off mass shell. However, permutation symmetry allows us to relate all possible diagrams that contribute to the current to ones in which nucleon 1 is a spectator and on mass-shell, while nucleons 2 and 3 form a pair that undergoes a two-body interaction before or after the photon couples to one of them (in CIA). The photon can also couple to the interacting pair directly, which produces a two-body current. The resulting total 3N current is displayed in Fig.~\ref{fig:coreFF}.

A closer inspection of this 3N current shows that particle 1 is always on mass-shell, particle 3 always off mass-shell, and particle 2 in some cases on and in others off mass-shell. It is therefore sufficient to introduce 3N basis states with these characteristics when we want to express amplitudes as matrix elements of operators between 3N states.

The propagator of an off-shell particle can be decomposed into positive- and negative-energy contributions, as described by Eq.~(\ref{oneprop}), where also the $\rho$-spin notation was introduced. In terms of the basis states, the presence of an off-shell particle reflects itself in the appearance of two $\rho$-spin states $\rho=\pm$, whereas an on-shell particle is described by a positive $\rho$-spin only. 

The 3N basis states with one nucleon off mass shell were defined in \cite{Sta97b}. The case of the two-nucleon system with both nucleons off mass shell was discussed in \cite{Gro08}. Here we only need to combine the results of \cite{Sta97b} and \cite{Gro08} in order to specify the more general case of 3N states with two nucleons off mass shell as required for the calculations of the 3N form factors.

We use a basis of 3N states that are tensor products of three one-nucleon states. The states are constructed in a sequence of steps. We start in the rest frame of the two-nucleon system composed of nucleons 2 and 3, and use a notation in which any variable with a ``$\tilde{\phantom{x}}$'' explicitly refers to this frame. Note that we use this notation only when the symbol without ``$\tilde{\phantom{x}}$'' is used to describe the same variable in a different frame.

Nucleons 2 and 3 have four-momenta $\tilde{k}_2=(\tilde{k}_{20},\tilde{\mathbf{k}}_2)$ and $\tilde{k}_3=(\tilde{k}_{30},\tilde{\mathbf{k}}_3)$. Their relative momentum is 
\begin{equation}
\tilde{p}=\frac{1}{2} (\tilde{k}_2-\tilde{k}_3) \, ,
\end{equation}
and, since $\tilde{\mathbf{k}}_2+\tilde{\mathbf{k}}_3=0$, we have $\tilde{\mathbf{k}}_2=-\tilde{\mathbf{k}}_3=\tilde{\mathbf{p}}$. The energy components are subject to the constraint $\tilde{k}_{20}+\tilde{k}_{30}=W$, where $W$ is the total energy in the two-nucleon rest frame.

A particular state of nucleon 2, which will serve as our reference state, has its three-momentum $\tilde{\mathbf p}$ aligned along the positive z-axis, and its helicity and $\rho$-spin are $\lambda_2$ and $\rho_2$. It is written as
\begin{equation}
 | (\tilde p,0,0) \lambda_2 \rho_2 \rangle \, ,
\label{eq:statek20}
\end{equation}
where $\tilde{\mathbf{p}}$ is specified through its magnitude and polar angles $\tilde{\theta}$ and $\phi$ in the form $(\tilde p,\tilde{\theta},\phi)$. We use the same symbol for the four-vector and the magnitude of the three-vector in order to avoid an awkward notation. Which one is meant is always clear from the context.
 
We denote a spatial rotation by an angle $\zeta$ about an axis $\hat{n}$ as $R(\zeta\hat{n})$, and a boost along the direction $\hat{n}$ with rapidity $\eta$ as $B(\eta\hat{n})$. The corresponding representations in 
Dirac space are
\begin{align}
S( R( \zeta \hat{n} ) ) & = \exp\left( {-i {\bf\Sigma}\cdot \hat{n} \frac{\zeta}{2}} \right) \nonumber\\
& = 
\mathbbm{1}\cos  \left ( \frac{\zeta}{2}  \right ) - i \sin  \left ( \frac{\zeta}{2} \right ) 
\sum_{i=1}^{3} 
\left( \hat{n} \cdot \hat{e}^{i} \right) \Sigma_i
 \, , \\
S( B(\eta \hat{n} ) ) & = \exp\left( {\bf\alpha}\cdot \hat{n} \frac{\eta}{2}\right)  \nonumber\\
&= 
\mathbbm{1} \cosh  \left ( \frac{\eta}{2}  \right ) + \sinh  \left ( \frac{\eta}{2} \right ) 
\sum_{i=1}^{3} 
 \left( \hat{n} \cdot \hat{e}^{i}\right) \alpha_i
 \, ,
\label{eq:boostrot}
\end{align}
where $\sigma^{1}$, $\sigma^{2}$, $\sigma^{3}$ are the Pauli matrices, and we introduce the Dirac matrices
\begin{equation}
 \Sigma_i \equiv \left ( 
\begin{array}{cc} 
\sigma^{i} & 0 \\ 
0 & \sigma^{i} 
\end{array} 
\right ) \, ,
\qquad
\alpha_i = \left ( 
\begin{array}{cc}  
0 & \sigma^{i} \\
\sigma^{i} & 0
\end{array} 
\right ) 
 \, .  \label{eq:sigmaalpha} 
\end{equation} 
The unit vectors $\hat{e}^{1}$, $\hat{e}^{2}$, and $\hat{e}^{3}$ point in the $x$, $y$, and $z$ direction, respectively. The special combination of rotations $ R( \alpha \hat{e}^{3}) R(\beta \hat{e}^{2}) R(\gamma \hat{e}^{3})$ is abbreviated as $R_{ \alpha, \beta,\gamma}$.

With these conventions, a general state in which the three-momentum points in the direction of the polar angles $\tilde{\theta}$ and $\phi$ is 
\begin{equation}
|(\tilde p,\tilde \theta,\phi)\lambda_2 \rho_2\rangle  =
S(R_{\phi,\tilde \theta,0}) |(\tilde{p},0,0)\lambda_2 \rho_2\rangle\, .
\label{eq:statek2}
\end{equation}

The three-momentum of nucleon 3 points into the opposite direction of the one of nucleon 2. In the phase convention by Wick \cite{Wic62}, its state is defined by starting from the reference state (\ref{eq:statek20}) with helicity $\lambda_3$ and $\rho$-spin $\rho_3$, applying a rotation $R_{\pi,\pi,0}$, which aligns its momentum along the negative $z$-axis, followed by a rotation in a general direction,
\begin{equation}
|(\tilde{p},\tilde \theta,\phi)\lambda_3\rho_3\rangle  =
e^{-i\pi s_3} S(R_{\phi,\tilde \theta,0}) |(\tilde{p},\pi,\pi)\lambda_3 \rho_3 \rangle\, ,
\label{eq:statek3}
\end{equation}
where $s_3=1/2$ is the spin of particle 3. The reason for introducing the extra phase factor $e^{-i\pi s_3}$ is discussed in \cite{Wic62} and \cite{Sta97b}.

The two-nucleon state can now be written as
\begin{eqnarray}
\lefteqn{
 |(\tilde{p},\tilde \theta,\phi)\lambda_2 \lambda_3;\rho_2 \rho_3\rangle  =
e^{-i\pi s_3} 
} \nonumber\\
&&\times S(R_{\phi,\tilde \theta,0}) \left\lbrace |(\tilde{p},0,0),\lambda_2 \rho_2\rangle
\otimes
|(\tilde{p},\pi,\pi),\lambda_3 \rho_3\rangle \right\rbrace  \, ,
\end{eqnarray}
where the common rotation $R_{\phi,\tilde \theta,0}$ acts on the spaces of both particles simultaneously.

Two-nucleon states with definite total angular momentum $j$ and total helicity $m$ are obtained through
\begin{eqnarray}
\lefteqn{
| \tilde{p} j m, \lambda_2 \lambda_3 ;\rho_2 \rho_3 \rangle
= \eta_j
\int_{0}^{2\pi} d \phi
\int_{0}^{\pi} d \tilde{\theta} \sin \tilde{\theta} 
} \nonumber \\
&&\times
{\cal D}^{(j)*}_{m,\lambda_2 -\lambda_3}(\phi,\tilde \theta,0) \,
|(\tilde{p},\tilde \theta,\phi)\lambda_2 \lambda_3;\rho_2 \rho_3\rangle \, ,
\label{eq:state2pw}
\end{eqnarray}
where we use the familiar Wigner ${\cal D}$-functions and the abbreviation
\begin{equation}
\eta_j  \equiv  \left( \frac{2j+1}{4\pi} \right)^{1/2} \, .\label{Eq7a}
\end{equation}
We can treat these composite two-body states as if they belonged to a single particle with spin $j$ and helicity $m$. Its total momentum is zero, because we are still in the two-body rest frame.
Applying the boost $Z(q) \equiv B(\eta_q \hat{e}^3)$ takes it to a state with three-momentum $q$ in $z$-direction.\footnote{Note that this ``$q$'' is totally unrelated to the photon momentum ``$\qph$''. It is unfortunate that both variables are traditionally given the same symbol, and we try to facilitate their distinction by using different font styles.} 
The rapidity of the boost is determined by $\sinh \eta_q = q/W(q)$, where the invariant mass of the nucleon pair is $W(q)=\sqrt{\left[ M_t-E(q)\right] ^2-q^2}$. 

We proceed then by introducing the state of particle 1, which in the three-body rest frame has its three-momentum in the opposite direction of the (23) pair, in exactly the same manner as the state of particle 3 was defined in the two-body rest frame,
\begin{equation}
 | (q,\pi,\pi) \lambda_1; \rho_1 \rangle = 
e^{-i\pi s_1}  S(R_{\pi,\pi,0}) |(q,0,0)\lambda_1 ;\rho_1\rangle\, ,
\label{eq:statek1}
\end{equation}
where the spin of particle 1 is $s_1=1/2$. In this paper, nucleon 1 is always on mass shell. Its $\rho$-spin is therefore always $\rho_1=+$ and will be suppressed from here on.

A tensor product of this state with the two-body state defined in Eq.~(\ref{eq:state2pw}) yields 
a three-body state with zero total three-momentum 
and the momentum of the (23) pair, which has spin $j$ and helicity $m$, in the positive $z$-direction, while nucleon 1 has helicity $\lambda_1$ and its three-momentum in opposite direction to the pair. A general 3N state is obtained by rotating the state aligned along the $z$-axis into a general direction by performing a rotation $R_{\Phi,\Theta,0}$. 

Repeating the procedure of Eq.~(\ref{eq:state2pw}), and exploiting the fact that a rotation about the $z$-axis commutes with a boost along the same axis, we can write the 3N partial wave helicity states with total angular momentum $J$ and total helicity $M$ in the form
\begin{eqnarray}
 \lefteqn{|q J M, \tilde p j m,\lambda_1 (\lambda_2 \lambda_3); \rho_2 \rho_3 \rangle 
} \nonumber \\
&& 
=
\eta_J \eta_j \int dS 
\;{\cal D}^{(J)*}_{M,m-\lambda_1}(S)
\int_0^{\pi} d\tilde{\theta} \sin\tilde{\theta}\;
d^{(j)}_{m,\lambda_2-\lambda_3}(\tilde{\theta})
\nonumber\\
&&\times
 S(R_{S})\,
\Bigl\lbrace | (q,\pi,\pi),\lambda_1 \rangle
\otimes 
S(Z(q)) S(R_{0,\tilde\theta,0})  \nonumber\\
&&
 \times 
| (\tilde{p},0,0), \lambda_2 \lambda_3 ;\rho_2 \rho_3 \rangle 
\Bigr\rbrace 
\end{eqnarray}
where we have used the abbreviations
\begin{eqnarray}
&&R_{S} = R_{\Phi,\Theta,\phi}\nonumber\\
&&{\cal D}^{(J)*}_{M,m-\lambda_1}(S) =
{\cal D}^{(J)*}_{M,m-\lambda_1}(\Phi,\Theta,\phi) \nonumber\\
&&\int dS =\int_{0}^{2\pi} d\Phi\int_{0}^{\pi} d\Theta \sin \Theta
\int_{0}^{2\pi} d\phi \, . 
\end{eqnarray}
Note that so far we have suppressed the energy components $q^0$ and $\tilde p^0$ of the four-vectors, which don't play any role in the definition of the partial wave basis. While nucleon 1 is on mass shell and $q^0$ given by $q^0=E(q)$, $\tilde p^0$ is not determined through $\bm p$ when both pair nucleons are off mass shell, and $\tilde p^0$ has to be included in the specification of the state. 

Our isospin states are constructed as in \cite{Sta97b},
\begin{equation}
 |\left[ (t_2 t_3)T t_1 \right] {\cal T} {\cal T}_z \rangle \, ,
\end{equation} 
where the isospins of nucleons 2 and 3 are first coupled to the pair isospin $T$, which is then in turn coupled with the isospin of nucleon 1 to yield the total 3N isospin $\cal T$ and its projection ${\cal T}_z$. Since for the 3N bound states $t_1=t_2=t_3={\cal T} = 1/2$, we will write the isospin states simply as 
$ |T {\cal T}_z \rangle$.

The complete specification of the helicity basis states is then
\begin{eqnarray}
 \lefteqn{|q J M, \tilde{p}^0 \tilde p j m,\lambda_1 (\lambda_2 \lambda_3); \rho_2 \rho_3 ; T {\cal T}_z \rangle 
} \nonumber \\
&& 
=
\eta_J \eta_j \int dS 
\;{\cal D}^{(J)*}_{M,m-\lambda_1}(S)
\int_0^{\pi} d\tilde{\theta} \sin\tilde{\theta}\;
d^{(j)}_{m,\lambda_2-\lambda_3}(\tilde{\theta})
\nonumber\\
&&\times
 S(R_{S})\,
\Bigl\{ | (q,\pi,\pi),\lambda_1 \rangle
\otimes 
S(Z(q)) S(R_{0,\tilde\theta,0})  \nonumber\\
&&
 \times 
| (\tilde p^0 \tilde{p},0,0), \lambda_2 \lambda_3 ;\rho_2 \rho_3 \rangle 
\Bigr\}
\otimes
|\left[ (t_2 t_3)T t_1 \right] {\cal T} {\cal T}_z \rangle \, .
\label{eq:3Npwhelbasis}
\end{eqnarray}

For the numerical solution of the Faddeev equation (\ref{eq:3NFad}) for the 3N vertex function, we use a basis of states with good parity and particle exchange symmetry, which is obtained \cite{Sta97b} through linear combinations of the states (\ref{eq:3Npwhelbasis}).

\section{Results and Discussion}

In this section, we present the results of our calculations of the elastic electromagnetic form factors of $^3$He and $^3$H in the complete impulse approximation (CIA). The underlying 3N vertex functions in (\ref{CIA}) were obtained by solving the corresponding Faddeev-type 3N bound state equation (\ref{eq:3NFad}) which depends on NN scattering amplitudes as dynamical input. The NN amplitudes are solutions of the CST two-body equation (\ref{twobody}) for given NN interaction models. We begin by a brief description of the interaction models used in our form factor calculations.

\subsection{NN interaction models}

In order to study the model dependence of the form factors, we used several members of a family of one-boson exchange (OBE) potentials that were first used in Ref.\ \cite{Sta97}, in the first CST calculation of the triton binding energy.   These models are based on the exchange of 6 bosons, namely the $\pi, \eta, \sigma, \delta, \omega,$
and $\rho$. The free potential parameters were determined by fitting to the $NN$ phase shifts below 350 MeV and to deuteron properties.  

A distinctive feature of these models is that they
include off-shell couplings of the scalar mesons
$\sigma$ (isoscalar) and $\delta$ (isovector) to nucleons, of the form
\begin{equation}
g_s\Lambda_s(p',p)= g_s \left[ 1 -\displaystyle{\frac{\nu_s}{2m}}
 \left(m-\slashed{p}' +m -\slashed{p} \right) \right]  \, , \label{eq:sNN}
\end{equation}
where $\Lambda_s(p',p)$ is the meson-nucleon vertex, and $p$ and $p'$ are the four-momenta of the incoming and
outgoing nucleons. The couplings proportional to $\nu_s$ do not
contribute if the nucleons are on-shell, hence the name ``off-shell couplings''.  (Note that the definition of $\nu_s$ used in Eq.~(\ref{eq:sNN}) differs from that used in the recent work of Ref.\ \cite{Gro08}.)

It turned out that the triton binding energy is very sensitive to the strength of the scalar off-shell coupling. In the first exploratory calculations, it was convenient to vary $\nu_\sigma$ and $\nu_\delta$ not independently, but keeping their ratio fixed. This was done by expressing each coupling constant through a common scale factor $\nu$, according to
\begin{equation}
\nu_\sigma=-0.75\,\nu \, ,\qquad\quad
\nu_\delta=2.60\,\nu\, . \label{eq2}
\end{equation}
The family of models discussed here has values of $\nu$ varying from 0 to 2.6.

Table~\ref{Tab:NNpar} lists the parameters that determine the scalar off-shell couplings. The table also shows that, as $\nu$ is increased from 0 to 2.6, the triton binding energy varies over a large range. At the value $\nu=1.6$ (model W16), the experimental binding energy is crossed, and simultaneously the best fit to the NN data (lowest $\chi^2/\mathrm{N_{data}}$) is achieved.  
Clearly, this singles out W16 as the most realistic model. While $\chi^2/\mathrm{N_{data}}$ deteriorates when $\nu$ moves away from 1.6, it does increase only moderately over the considerable range. We have therefore a convenient method at our disposal to generate a family of potentials that differ considerably in the 3N binding energy, but yield roughly equivalent fits to the NN data. This is an almost ideal situation to study the model dependence in our 3N electromagnetic form factor calculations.

A more detailed description of these potentials and the complete list of parameters can be found in \cite{Gro98P}. In all of these models, the pion-nucleon coupling is of pure pseudovector form.  (Recent fits to the data described in Ref.\ \cite{Gro08} give $\chi^2$/N $\simeq$ 1 and still show the same correlation between $E_t$ and $\chi^2$.  Results from these new models will be discussed elsewhere.)

\begin{table}
\caption{Scalar meson parameters of the NN potential models used in the calculations of the 3N vertex functions. Also given are the $\chi^2/\mathrm{N_{data}}$ of the models obtained in fits to a NN data base of 1994, and the corresponding triton binding energies $E_t$. All masses and binding energies are in MeV.}
 \begin{tabular}{cccccc}
\hline
\hline
  & W00 & W10 & W16 & W19 & W26 \\
\hline
$\nu$ & 0.0 & 1.0 & 1.6 & 1.9 & 2.6 \\
$g_\sigma^2/4\pi$ &   5.84067 $\,$  &$\,$  5.50753 $\,$  & $\,$  4.99887 $\,$  & $\,$  4.67948 $\,$  & $\,$  4.05718  \\
$\nu_\sigma$ & 0.0 & -0.75 & -1.2 & -1.425 & -1.95 \\
$m_\sigma$ & 525 & 515 & 506 & 501 & 491 \\
$g_\delta^2/4\pi$ & 0.14812 $\,$  & 0.69046 & 0.62818 & 0.47598 & $\,$ 0.25045 \\
$\nu_\delta$ & 0.0 & 2.6 & 4.16 & 4.94 & 6.76 \\
$m_\delta$ & 390 & 540 & 512 & 474 & 399 \\
\hline
$\chi^2/\mathrm{N_{data}}$  &  3.00 & 2.45 & 2.25 & 2.27 & 2.56 \\
$E_t$ & 6.217 & 7.411 & 8.489 & 9.072 & 10.533\\
\hline\hline
 \end{tabular} 
\label{Tab:NNpar}
\end{table} 

\subsection{Partial wave convergence}
\begin{table}[!htbp]
\caption{Partial wave convergence of the charge form factor of $^3$He at selected values of the momentum transfer $Q$. The first column shows the maximum total pair angular momentum included in the 3N partial waves. Columns two to six are the form factor calculated in the corresponding truncated partial wave basis at Q=1.0, 3.0, 5.0, 7.0, and 9.0 fm$^{-1}$. Except for those in column two, the values of the form factor are multiplied by a power of ten indicated in the last line. All results are for model W16 with the MMD nucleon form factor.}
\begin{tabular}{cccccc}
\hline\hline
$ j_\mathrm{max}$ &\multicolumn{5}{c}{ $Q$ (fm$^{-1}$) } \\
   & 1.0 & 3.0 & 5.0 & 7.0 & 9.0 \\
\hline
1 & $\,$ 0.5585 $\,$ & $\,$ 1.7603 $\,$ & $\,$ -1.5546 $\,$ & $\,$ -0.9903 $\,$ & $\,$ 1.039 $\,$ \\
2 & 0.5716 & 1.8772 & -1.5314 & -0.9815 & 1.201  \\
3 & 0.5733 & 1.8866 & -1.5519 & -1.0471 & 1.174  \\
4 & 0.5744 & 1.8903 & -1.5511 & -1.0618 & 1.119  \\
5 & 0.5743 & 1.8904 & -1.5515 & -1.0623 & 1.128  \\
6 & 0.5744 & 1.8896 & -1.5518 & -1.0644 & 1.119 \\
 & & $\times 10^{-2}$  & $\times 10^{-3}$ & $\times 10^{-4}$ & $\times 10^{-5}$ \\
 \hline \hline
\end{tabular}
\label{Tab:FC3Hepw}
\end{table}

The electromagnetic form factors of the 3N bound states have been calculated in the basis of partial wave helicity states described in Section \ref{sec:pw}. With an increasing number of partial waves, the calculations become more and more time consuming. It is important to have an idea about the accuracy of the results that can be achieved with a limited number of partial waves. It is costumary to define a truncated basis of three-body states through the maximum value of the two-body total angular momentum, $j_\mathrm{max}$, of the included basis states. Table \ref{Tab:FC3Hepw} shows the charge form factor of $^3$He, for a particular model calculation which will be described later, at 5 different values of the momentum transfer $Q$, depending on $j_\mathrm{max}$. Table \ref{Tab:FM3Hepw} shows the same for the magnetic form factor of $^3$He. 
\begin{table}[!htbp]
\caption{Same as Table \ref{Tab:FC3Hepw}, but for the magnetic form factor of $^3$He.}
\begin{tabular}{cccccc}
 \hline\hline
 $ j_\mathrm{max}$ &\multicolumn{5}{c}{ $Q$ (fm$^{-1}$) } \\
  & 1.0 & 3.0 & 5.0 & 7.0 & 9.0 \\
\hline
1 & $\,$ 0.5063 $\,$ & $\,$ -0.6586 $\,$ & $\,$ -3.4357 $\,$ & $\,$ -0.6531 $\,$ & $\,$ 4.3456 $\,$ \\
2 & 0.5202 & -0.6207 & -3.6999 & -1.0678 & 4.4701 \\
3 & 0.5217 & -0.6181 & -3.6482 & -0.8808 & 4.7954 \\
4 & 0.5227 & -0.6200 & -3.6497 & -0.9454 & 4.5796 \\
5 & 0.5226 & -0.6190 & -3.6489 & -0.9376 & 4.6228 \\
6 & 0.5226 & -0.6206 & -3.6479 & -0.9413 & 4.5910 \\
 & & $\times 10^{-2}$  & $\times 10^{-3}$ & $\times 10^{-4}$ & $\times 10^{-5}$ \\
\hline \hline
\end{tabular}
\label{Tab:FM3Hepw}
\end{table}

What both tables indicate is that at lower $Q$ a good stability of the results is reached already with $j_\mathrm{max}=2$. Not surprisingly, the results are less stable for higher $Q$, although uncertainties of the order of a few percent are perfectly satisfactory. 

Diagrams A and the sum of B and C require numerical integrations over two momenta and two angles, while in diagrams D and the sum of E and F two momenta and three angles are integrated. In principle, the precision of the numerical integrations can be improved upon by increasing the number of integration points. However, we feel that the current precision, which we estimate to be about three significant figures, is sufficient for the purpose of this work. Small changes of the order of one percent would not be visible in the figures presented.

 \subsection{Model dependence of the CIA}

 \begin{figure*}[tbh]
 \centering
 \includegraphics[width=17cm,bb=49 380 530 716]{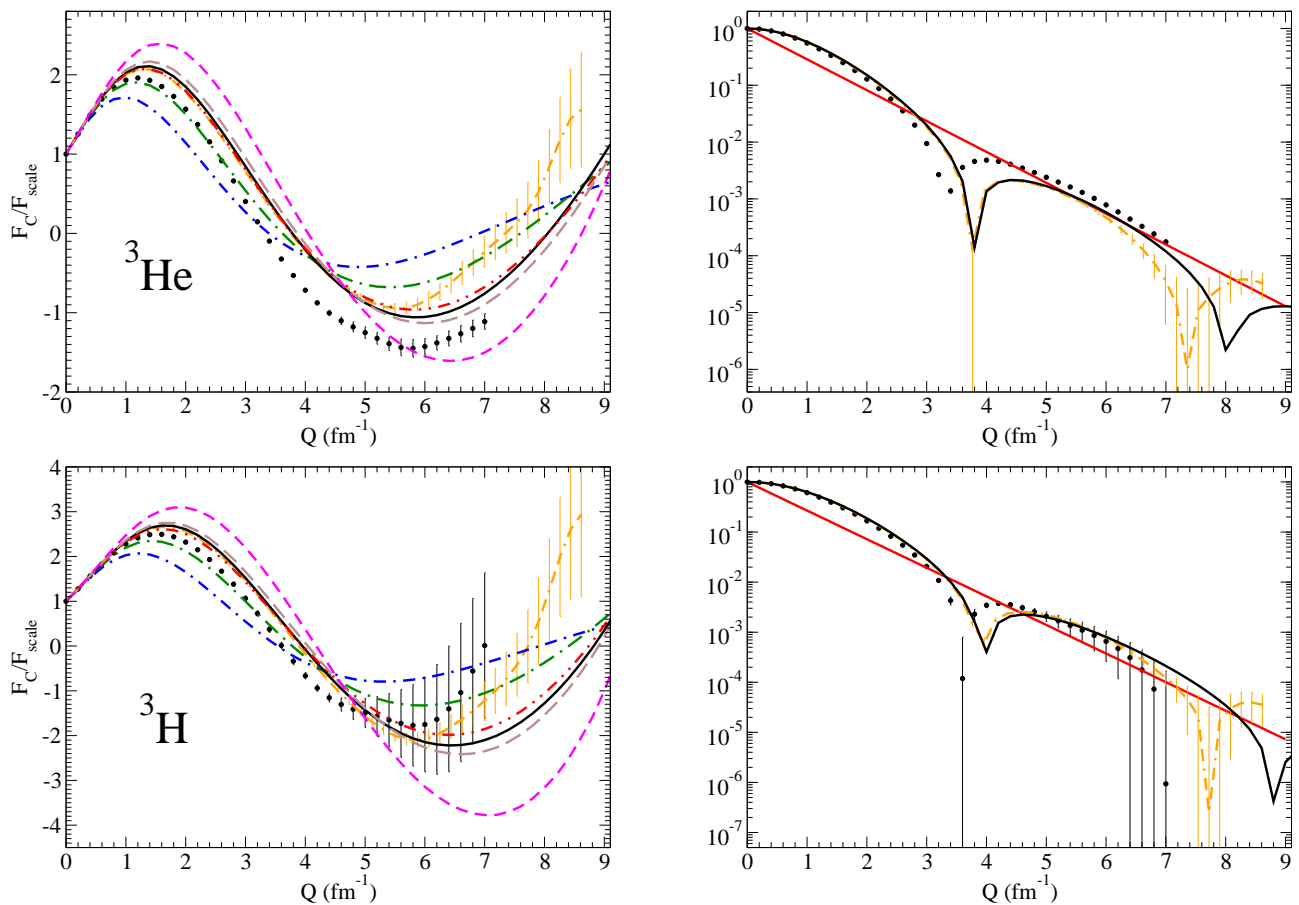}
\caption{(Color online) Charge form factors of $^3$He and $^3$H for different NN interaction models in CIA. The left panel shows the form factors divided by a scale function (solid straight line in the right panel). Shown are the results for models W00 (short dash-dotted), W10 (long dash-dotted), W16 (dash-double dotted), W19 (long dashed), and W26 (short dashed), all with the MMD nucleon form factor. In addition, W16 (solid line) is compared to IARC by Marcucci (double dash-dotted), where both calculations used the Galster nucleon form factor. The theoretical error bars of the Greens function Monte Carlo IARC calculations are also given. The right panel shows the W16/Galster and IARC/Galster results together with the scale function (solid straight line) in the traditional semi-log plot. Both panels also show the experimental data (full circles)\cite{Sic01}.}
 \label{fig:Wxx-CIA-3NFFch-scaled}
\end{figure*}

%

 \begin{figure*}[tbh]
 \centering
 \includegraphics[width=17cm,bb=49 379 530 716]{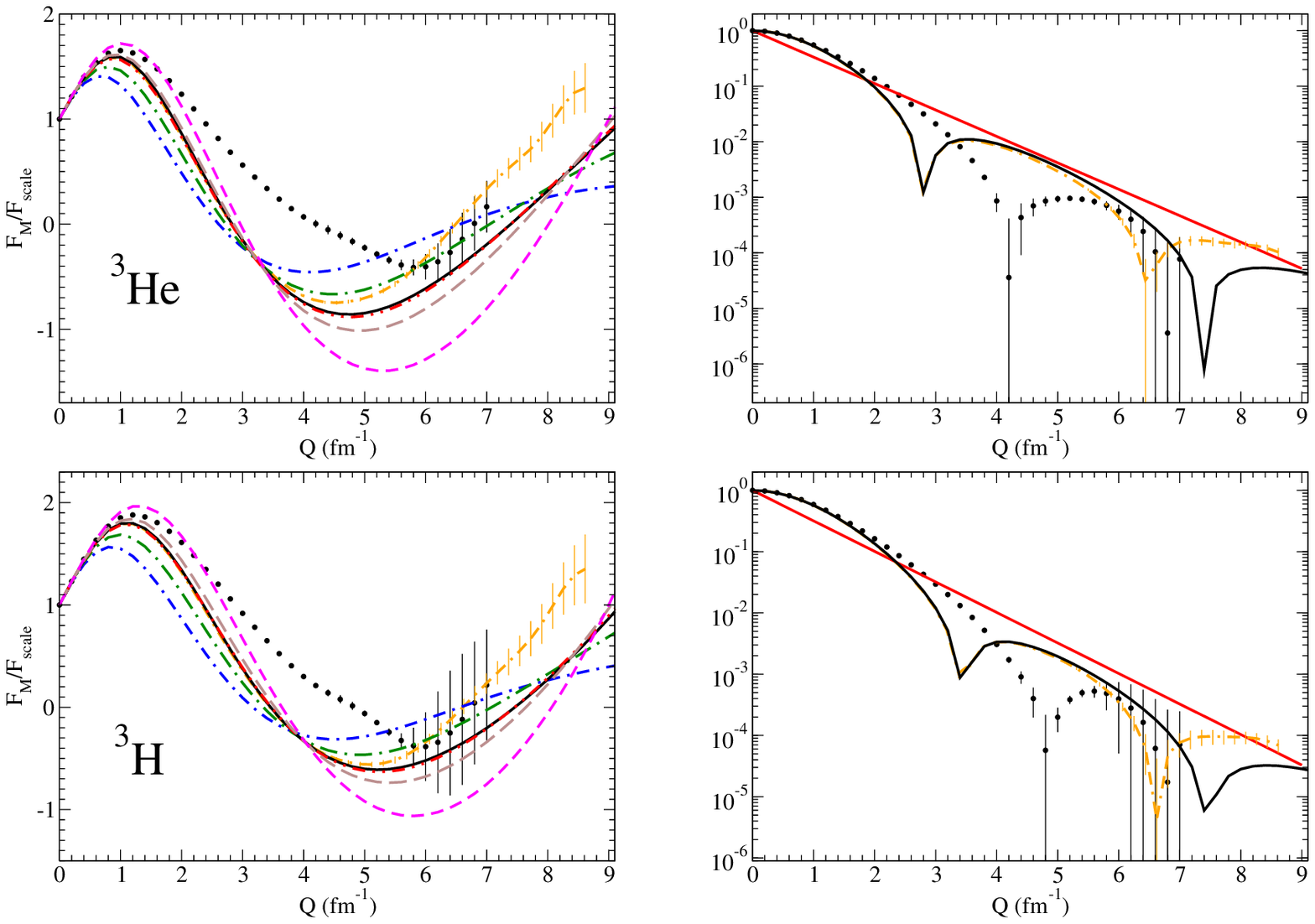}
\caption{(Color online) Magnetic form factors of $^3$He and $^3$H for different NN interaction models in CIA. The various curves are defined in the caption of Fig.\ \ref{fig:Wxx-CIA-3NFFch-scaled}.}
 \label{fig:Wxx-CIA-3NFFmag-scaled}
\end{figure*}

\begin{figure*}[tbh]
 \centering
 \includegraphics[width=17cm,bb=48 383 530 716]{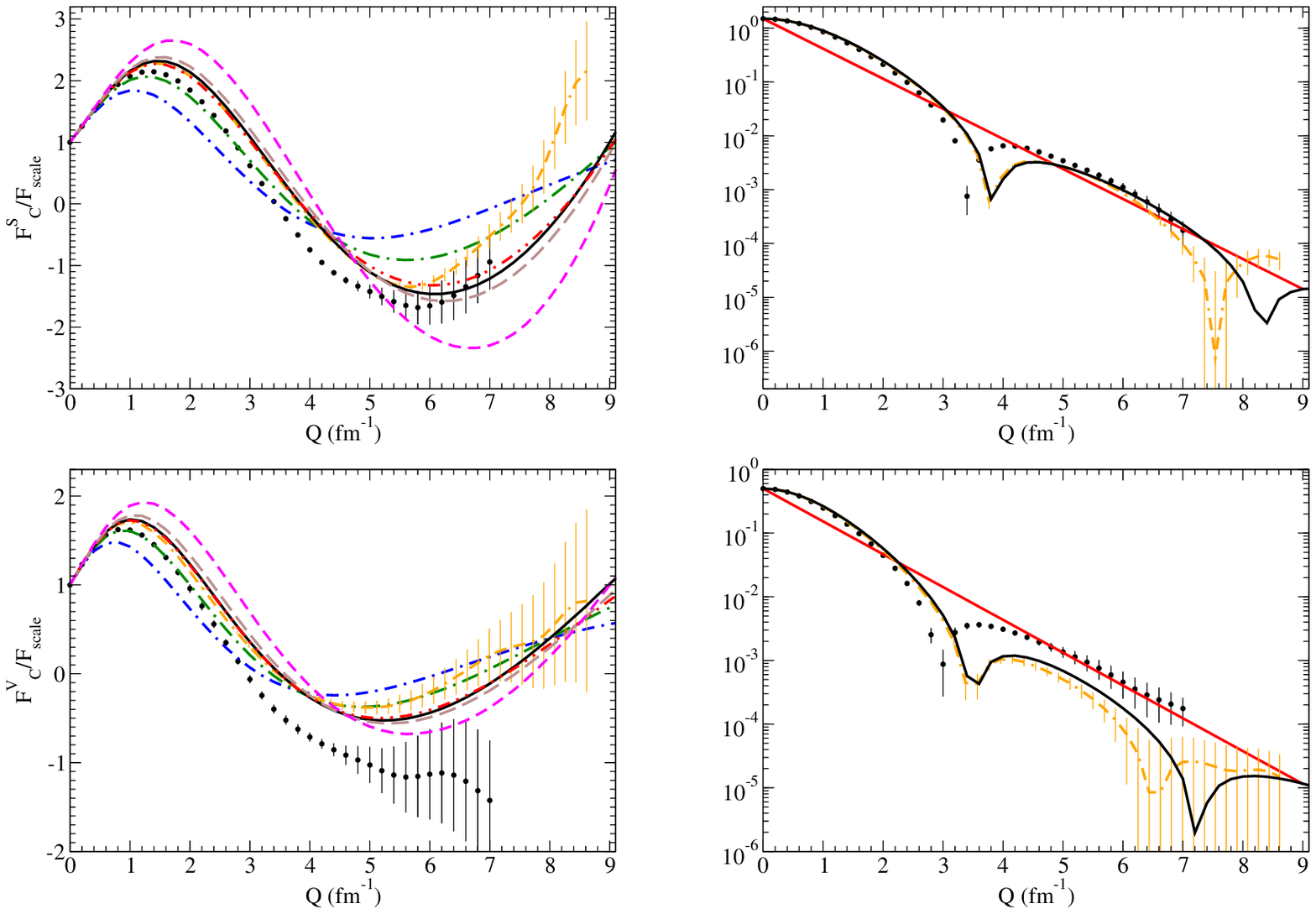}
\caption{(Color online) Isoscalar and isovector charge form factors of the 3N bound states for different NN interaction models in CIA. The various curves are defined in the caption of Fig.\ \ref{fig:Wxx-CIA-3NFFch-scaled}.}
 \label{fig:Wxx-CIA-3NISIVch}
\end{figure*}
\begin{figure}
 \centering
\end{figure}
\begin{figure*}[tbh]
 \centering
 \includegraphics[width=17cm,bb=48 384 530 716]{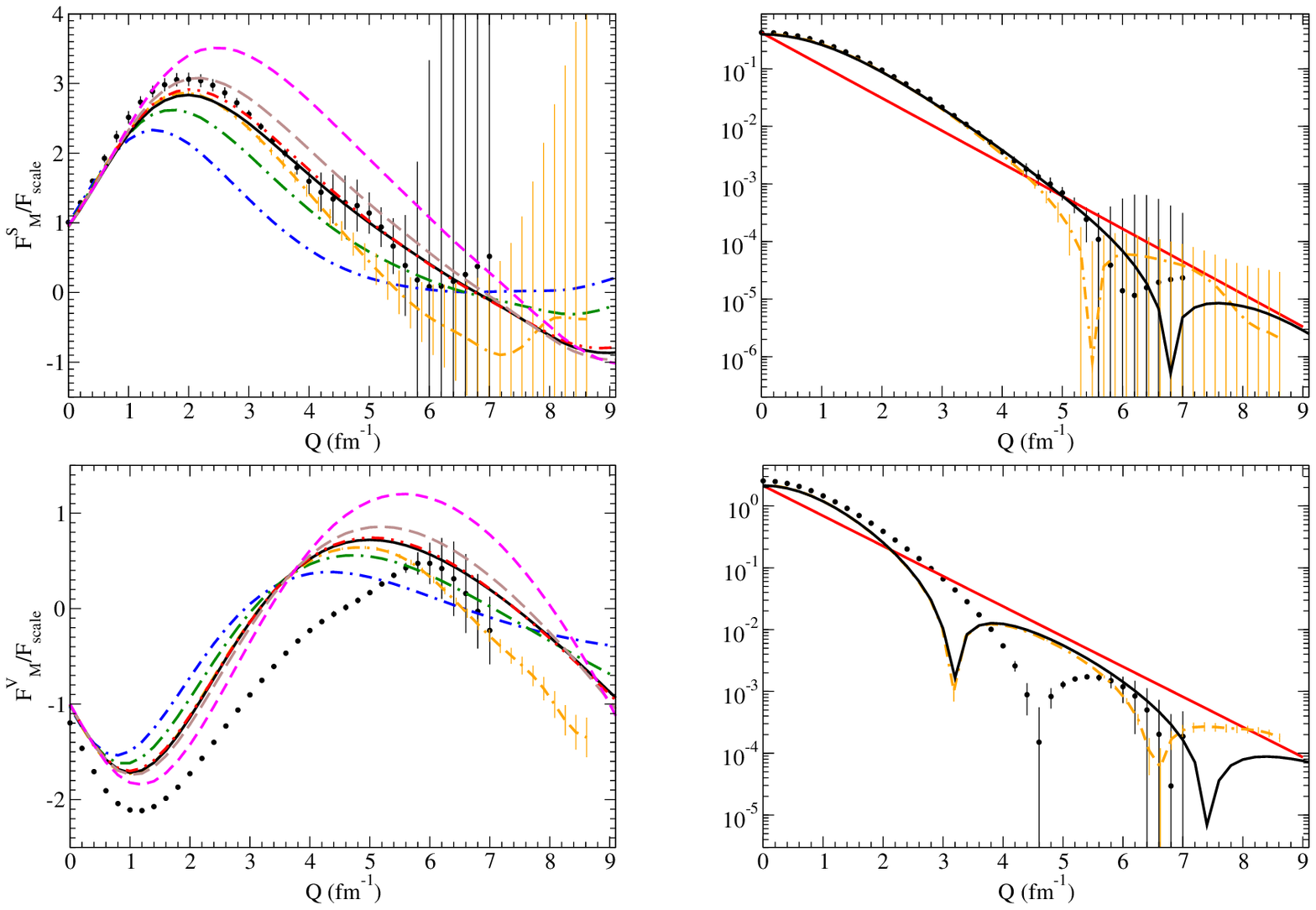}
\caption{(Color online) Isoscalar and isovector magnetic form factors of the 3N bound states for different NN interaction models in CIA. The various curves are defined in the caption of Fig.\ \ref{fig:Wxx-CIA-3NFFch-scaled}.}
 \label{fig:Wxx-CIA-3NISIVmag}
\end{figure*}

Figures \ref{fig:Wxx-CIA-3NFFch-scaled}--\ref{fig:Wxx-CIA-3NISIVmag}  compare our CIA results for the selected models listed in Table \ref{Tab:NNpar} to IARC (Impulse Approximation with Relativistic Corrections) calculations within the framework described in Refs.~\cite{Mar98, Mar05}. The IARC calculations use a  one-nucleon-current with wave functions obtained from the Argonne AV18 NN and Urbana IX 3N potentials, and include first-order relativistic corrections. Our models do not include the strong Coulomb corrections for the $pp$  interaction, and hence give the same binding energies for both $^3$H and $^3$He. To help in the comparison of our results with the IARC, L.~Marcucci provided us with IARC calculations in which the $pp$ Coulomb interaction was excluded \cite{Mar08pc}, such that the used $^3$H and $^3$He wave functions also had the same binding energies.

Figures\ \ref{fig:Wxx-CIA-3NFFch-scaled} and \ref{fig:Wxx-CIA-3NFFmag-scaled} show the charge and magnetic form factors of $^3$H and $^3$He, while Figs.\ \ref{fig:Wxx-CIA-3NISIVch} and \ref{fig:Wxx-CIA-3NISIVmag} show the isoscalar and isovector combinations of the charge and magnetic form factors.  Each figure has four panels.  The right hand panels give a log plot of the absolute value of the form factors vs.\ $Q$.  Note that they all fall off by about 6 orders of magnitude over the range in $Q$ shown in the plot.  In each of these right hand panels an exponential scale function is shown (the straight line).  To get a better idea of the relative differences in the models, the left hand panels show the form factors {\it divided\/} by the scale function.  This removes the strong exponential dependence and permits us to show both the sign and the relative value of each curve on a linear scale.

The first conclusion that can be drawn from these graphs is that, in every case, model W16 is remarkably close to the IARC calculation at low $Q$, confirming that these two calculations are in essential agreement with one another.  This is a very pleasing result: it confirms for the first time that the CST not only yields the correct 3N binding energy, but that it is also capable of a good description of the electromagnetic structure of the 3N bound states. The differences in the underlying dynamics between IARC (the nonrelativistic AV18 NN potential and an irreducible 3N force) and the CST (relativistic NN kernel with off-shell couplings, no irreducible 3N force) seem to be less important at small $Q$ than the fact that they yield the same nuclear binding energy. As will be discussed in more detail in Sec.~\ref{Sec:3E}, in the case of the family of CST interaction models used here, the CST single-nucleon current contains approximately the same physics as the one of IARC. The close agreement of the results is therefore perfectly understandable.

\begin{figure}[hbt]
 \centering
 \includegraphics[width=5cm,angle=-90]{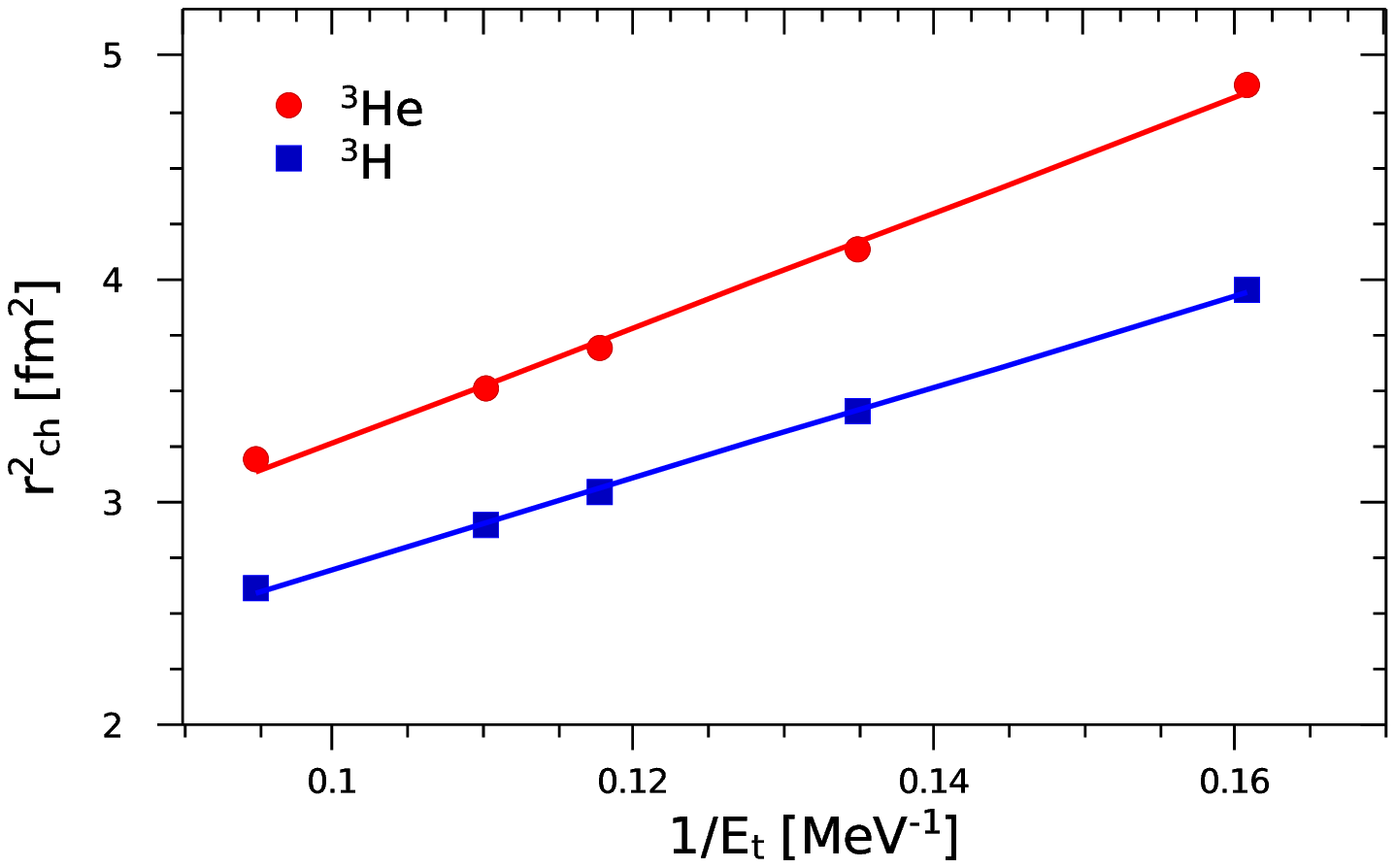}
\caption{(Color online) Mean square charge radii of $^3$He and $^3$H versus inverse binding energy for models W00, W10, W16, W19, and W26, in order from right to left. The straight lines are linear fits through the points and show that the expected linear relation between $\langle r^2 \rangle$ and $1/E_t$ is indeed well satisfied for these models.}
 \label{fig8}
\end{figure}

Next, note that model W16 departs from the IARC at a $Q$ of about 4 to 6 fm$^{-1}$.  The Monte Carlo calculations of the IARC begin to show large errors for $Q\ge 6$ fm$^{-1}$, so it is difficult to make a precise comparison above this point, but the trends are still clear.  The charge form factors agree to $Q\simeq 6$ fm$^{-1}$ beyond which the IARC models oscillate with a shorter wavelength than does W16.  A similar thing happens with the magnetic form factors, but the IARC and W16 break away at a lower $Q\simeq 4$ fm$^{-1}$.  The isoscalar and isovector combinations show the same behavior.

The variation in the CIA models is remarkably smooth, and seems to be explained almost entirely by the different three-body binding energies predicted by each of these models.  Recall that the low momentum behavior of the charge form factors can be written
\bea
F_C(Q^2)\simeq 1-\sfrac16\, Q^2 \left<r^2\right> +\cdots \, .
\eea
Using the values of the root mean square charge radii in Table \ref{Tab:rms-radii}, which also shows the magnetic radii for completeness, it can be verified directly in Fig.~\ref{fig8} that the well-known relation $ \left<r^2\right> \propto 1/E_t$ between the mean square radius of the bound state and the binding energy $E_t$ holds very well for the CST models. 

This behavior predicts model W00 with the smallest binding energy will have the greatest curvature at $Q=0$, and that W26 with the largest binding energy will have the smallest curvature, a behavior confirmed by the plots.  The surprising fact is that this behavior seems to persist on to higher $Q$, with W00 oscillating with the shortest  wave length and W26 the longest.

\begin{table}
\caption{Charge and magnetic root mean square radii of $^3$He and $^3$H for the five models of Table \ref{Tab:NNpar}, calculated in CIA with the MMD parametrization of the nucleon electromagnetic form factors. The corresponding 3N binding energies are also listed.}
 \begin{tabular}{lccccc}
\hline\hline
 & & \multicolumn{4}{c}{r.m.s.\ radius (fm)}  \\
 & & \multicolumn{2}{c}{charge} &  \multicolumn{2}{c}{magnetic}  \\
Model & $E_t$ (MeV) & $^3$He & $^3$H  & $^3$He & $^3$H  \\
\hline
W00 & $\,$ 6.217  $\,$  &  $\,$  2.204  $\,$  &  $\,$  1.986  $\,$  &  $\,$  2.361  $\,$  &  $\,$  2.172  $\,$   \\
W10 & 7.411 & 2.030 & 1.842 & 2.195 & 2.027 \\
W16 & 8.489 & 1.917 & 1.742 & 2.061 & 1.946 \\
W19 & 9.072 & 1.869 & 1.698 & 1.991 & 1.914 \\
W26 & 10.533 & 1.783 & 1.613 & 1.819 & 1.862 \\
\hline\hline
 \end{tabular} 
\label{Tab:rms-radii}
\end{table}

Finally, note that the comparison with the data shows significant discrepancies beyond $Q\simeq1$ fm$^{-1}$, particularly for the isovector magnetic combination. In the case of the IARC calculations, it has been shown \cite{Mar98} that the inclusion of large exchange-current contributions brings the theoretical calculations into good agreement with the data. While we are not yet ready to calculate the interaction currents for our CST models, we believe that they will be large also. We reemphasize that the observed discrepancy between the data and the theoretical results is due to known physics not included in the impulse approximations in both cases.
This aspect is discussed further below in Sec.\ \ref{Sec:3E}.

\begin{table}
\caption{Magnetic moments of $^3$He, $^3$H, as well as their isoscalar and isovector combinations $\mu_S$ and $\mu_V$ in nuclear magnetons. The first two lines are calculated with model W16 in CIA, in combination with the MMD and the Galster parametrization of the nucleon electromagnetic form factors. The third line is a IARC/Galster calculation by Marcucci. The last line shows the experimental values.}

\begin{tabular}{lcccc}
\hline 
\hline 
 & $\mu(^3\mathrm{He})$ & $\mu(^3\mathrm{H})$ & $\mu_S$ & $\mu_V$ \\
\hline
W16/MMD      &  $\,$  -1.747  $\,$  &  $\,$  2.550  $\,$  &  $\,$  0.402  $\,$  &  $\,$  -2.149  $\,$  \\
W16/Galster  & -1.749 & 2.546 & 0.398 & -2.147 \\
IARC/Galster & -1.763 & 2.572 & 0.404 & -2.168 \\
\hline
Experiment   & -2.127 & 2.979 & 0.426 & -2.553 \\
\hline
\hline
\end{tabular}  
\label{tab:2}
\end{table} 

 \begin{figure*}[tbh]
\centering
 \includegraphics[width=17cm,bb=48 392 530 716]{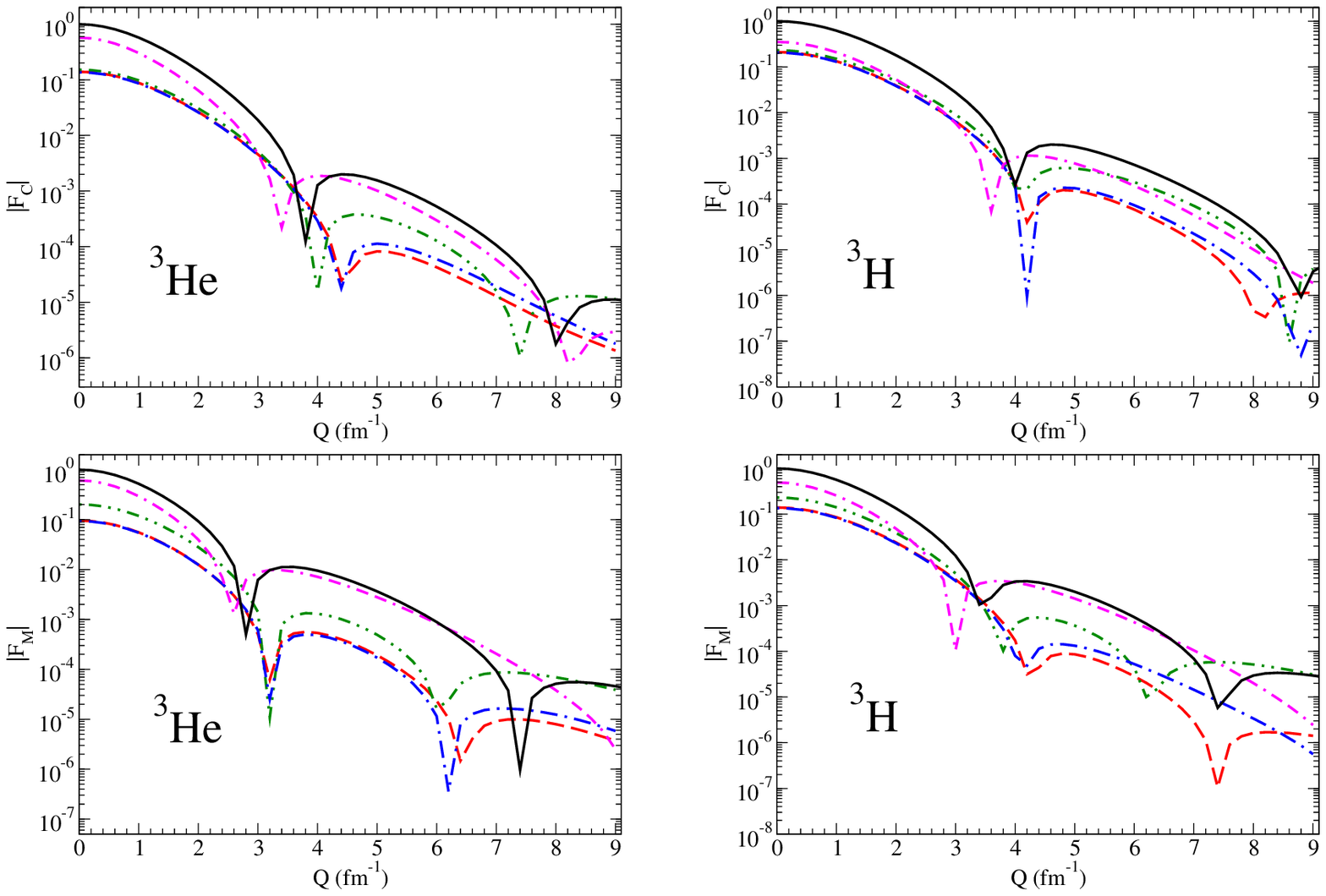}
\caption{(Color online) Contributions of the various diagrams to the charge (upper panel) and magnetic (lower panel) form factors of $^3$He and $^3$H. The solid line is the total result in CIA (sum of diagrams A -- F) for the NN interaction model W16 and the MMD nucleon electromagnetic form factors. The other lines are the partial results for diagrams A (dashed), B+C (dash-dotted), D (double dot-dashed), and E+F (double dash-dotted).}
\label{fig:W16-CIA-diag}
\end{figure*}

Finally, in Table \ref{tab:2} we compare calculations of the magnetic moments of the three-body nuclei.  Note the reasonably close agreement between the calculations, and the well known discrepancy with the data (due to missing exchange current contributions).

\subsection{Contributions of the six diagrams}

The contributions of each of the six diagrams that make up the CIA are shown in Fig.~\ref{fig:W16-CIA-diag}.  Note that all give comparable results; none can be neglected.  

It is particularly interesting to see that the sum of diagrams B and C (each contains an integrable singularity, but their sum has no singularity) is almost precisely equal to diagram A (in all cases).  This was expected, but  a similar result does not hold for the sum of diagrams E and F.  Each of these is singular, their sum is finite, but it is not equal to diagram D, as originally expected.  Note also that the contributions of diagrams A and D to the charge form factors are very similar, at least at low $Q^2$, but that they give substantially different results for the magnetic form factors.  These relationships will be studied in a future paper. 

 \begin{figure*}[tbh]
 \centering
 \includegraphics[width=17cm,bb=49 376 530 715]{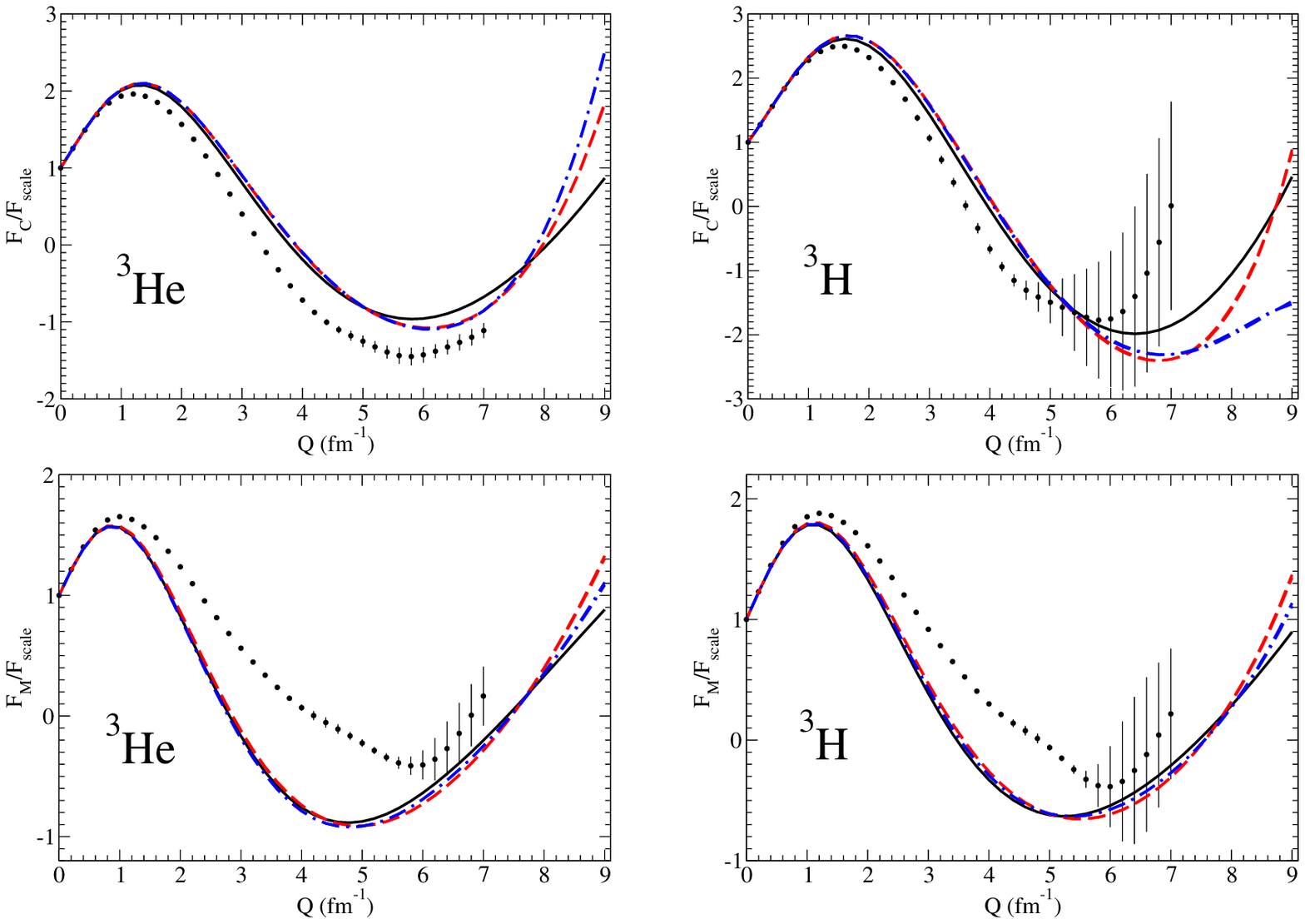}
\caption{(Color online) Charge (upper panels) and magnetic (lower panels) form factors of $^3$He and $^3$H in CIA, divided by the respective scale functions of Fig.~\ref{fig:Wxx-CIA-3NFFch-scaled}, with model W16 and different nucleon electromagnetic form factors. The MMD parametrization is used for the on-shell part of the nucleon current in all cases. The solid line is obtained using only the on-shell nucleon current NCI (\textit{i.e.}, $f_0=f'_0=1$ and $g_0=0$). The dashed line uses the full current (\ref{d15a}) in the version NCII (with $f'_0=f_0$ and $F_{3N}=G_{EN}$).  Finally,  the dash-dotted line shows the full current NCIV (with $f'_0=1$).  We also tried replacing $F_{3N}=G_{EN}$ by $F_{3N}=F_{1N}$ for each case (currents NCIII and NCV); the effect is too small to be seen on the plot. }
 \label{fig:W16-CIA-jNvar}
\end{figure*}

\subsection{Dependence on the nucleon form factors}

In Ref.\ \cite{Gro87} it was shown how to construct gauge invariant interactions using composite particles.  Subsequently \cite{Gro96,Ada02} the general form of the one nucleon current operator for use in such a program was derived.  The one used most frequently is
\begin{eqnarray}
\lefteqn{j_N^\mu(p',p) =  f_0(p'^2,p^2)\Big\{ (F_1(Q^2)-1) \tilde{\gamma}^\mu + \gamma^\mu \Big\}}
\nonumber\\
&&+ g_0(p'^2,p^2) \Lambda_- (p') \Big\{
(F_3(Q^2)-1) \,\tilde{\gamma}^\mu 
 +  \gamma^\mu \Big\}
\Lambda_- (p)\nonumber\\
&&+f'_0(p'^2,p^2)\,F_2(Q^2)\,\frac{i \,\sigma^{\mu \nu}
q_\nu }{2m}   \,  ,  \label{d15a}
\end{eqnarray}
where $F_{1,2}(Q^2)$ are the on-shell nucleon form factors, $F_3(Q^2)$
is a completely unknown form factor describing the off-shell
structure of the nucleon (subject to the constraint that $F_3(0)=1$),
$\Lambda_- (p) =(m-\slashed{p})/2m$, 
$\tilde{\gamma}^\mu=\gamma^\mu-q^\mu \rlap/q /q^2$, $f'_0$ is a completely undetermined function of $p'^2$ and $p^2$ (subject to the constraint $f'_0(m^2,m^2)=1$),  and $f_0$ and
$g_0$ are functions of $p^2$ and $p'^2$ completely determined by the
Ward-Takahashi identity for a dressed nucleon
\bea
q_\mu j_N^\mu(p',p)=S^{-1}_N(p)-S^{-1}_N(p')
\eea
where $S_N(p)$ is the dressed propagator for the nucleon
\bea
S_N(p)=\frac{[h(p^2)]^2}{m-\slashed{p}}\, .
\eea
In the applications discussed in this paper, the nucleon form factor is 
\bea
h(p^2)=\left[\frac{(\Lambda_N^2-m^2)^2}{(\Lambda_N^2-m^2)^2+(m^2-p^2)^2}\right]^2
\eea
and $f_0$ and $g_0$ are given in terms of $h$
\begin{eqnarray}
f_0(p'^2,p^2)=&& \frac{h}{h'} \;\frac{m^2 - p'^2}{p^2 - p'^2} +
\frac{h'}{h}\;
\frac{m^2 - p^2}{p'^2 - p^2} \nonumber\\
g_0(p'^2,p^2)=&&\left( \frac{h}{h'} - \frac{h'}{h} \right) 
\frac{4 m^2}{p'^2 -p^2} \,  .  \label{d17}
\end{eqnarray}
with $h\equiv h(p^2)$ and $h'\equiv h(p'^2)$.

In order to include the isospin dependence of the current correctly, care has to be taken to maintain the Ward-Takahashi identity satisfied. The following expression meets this requirement:
\begin{align}
&j_N^{\mu}(p',p) =
\frac{1+\tau^{3}}{2} f_{0}(p^{\prime 2} , p^{2} ) \gamma^{\mu} \nonumber \\
& +
\left [ \left ( F_{1p}(Q^{2}) - 1 \right ) \frac{1+\tau^{3}}{2} + F_{1n}(Q^{2}) \frac{1-\tau^{3}}{2} \right ]
f_{0}(p^{\prime 2} , p^{2} ) \tilde{\gamma}^\mu
\nonumber \\
& +
\left [ F_{2p}(Q^{2}) \frac{1+\tau^{3}}{2} +F_{2n}(Q^{2}) \frac{1-\tau^{3}}{2} \right ]
f_{0}(p^{\prime 2} , p^{2} ) 
\frac{i \sigma^{\mu \nu} q_{\nu}}{2m} \nonumber \\
& +
\frac{1+\tau^{3}}{2}
g_{0}(p^{\prime 2} , p^{2} )
\Lambda_- (p')
\gamma^{\mu}
\Lambda_- (p) \nonumber \\
& +
\left [ \left ( F_{3p}(Q^{2}) - 1 \right ) \frac{1+\tau^{3}}{2} + F_{3n}(Q^{2}) \frac{1-\tau^{3}}{2} \right ] \nonumber \\
& \quad \times
g_{0}(p^{\prime 2} , p^{2} )
\Lambda_- (p')
\tilde{\gamma}^\mu
\Lambda_- (p) \, . \label{cap3_fv76y}
\end{align}

Note that the nucleon current is always used in conjunction with a {\it conserved\/} electron current, and hence the terms in $q^\mu$ vanish.  Making this simplification in Eq.~(\ref{cap3_fv76y}), and combining the proton and neutron current into a single, isospin dependent expression, gives
\begin{align}
j_N^\mu(p',p) = &f_0(p'^2,p^2)\,\,F_{1N}(Q^2)\, \gamma^\mu \nonumber\\
+&f'_0(p'^2,p^2)\,F_{2N}(Q^2)\,\frac{i \,\sigma^{\mu \nu}
q_\nu }{2m}  \nonumber\\
+& g_0(p'^2,p^2) F_{3N}(Q^2) \Lambda_- (p')  \gamma^\mu 
\Lambda_- (p)\,  ,  \label{d15b}
\end{align}
where, for $i=\{1,2,3\}$ and the nucleon isospin projection $\tau^3$,
\begin{align}
F_{iN}(Q^2)=F_{ip}(Q^2) \frac{1+\tau^3}{2}+F_{in}(Q^2) \frac{1-\tau^3}{2}\, .
\end{align}

In previous applications, $f'_0$ has been taken to be equal to $f_0$, and usually $F_{3N}=G_{EN}$ (where $N$ stands for $p$ or $n$), but no systematic study of the dependence of these factors has been completed.  

In this work, we performed 3N form factor calculations with 5 different single-nucleon currents, which can be characterized by the different choices of the off-shell form factors $f_0$, $f'_0$, $g_0$, as well as of $F_{3N}$.

The first nucleon current, labeled NCI, is the usual on-shell current, with $f_0=f'_0=1$ and $g_0=0$. Next we consider a genuine full off-shell current, NCII, with $f_0$ and $g_0$ given by Eq.\ (\ref{d17}), $f'_0=f_0$, and $F_{3N}=G_{EN}$. Current NCIII differs from NCII only by the choice $F_{3N}=F_{1N}$. To study the sensitivity of the results on $f'_0$ we constructed current NCIV, which is equal to NCII except for $f'_0=1$. Finally, current NCV equals NCIV apart from $F_{3N}=F_{1N}$.

Figure \ref{fig:W16-CIA-jNvar} shows the effect of some variations in the choice of the single-nucleon form factors. In general, the changes are very small for $Q\alt5$ fm$^{-1}$. Only at larger values of $Q$ the various cases begin to diverge from each other, particularly in the charge form factors. The three displayed lines correspond to NCI, NCII, and NCIV. The results for NCIII are indistinguishable from NCII in the figure, as are those for NCV from NCIV. We conclude that the inclusion of off-shell form factors, as well as the choice of $f'_0(p'^2,p^2)$, has only marginal effects at low $Q$, but becomes relevant for larger values of $Q$. On the other hand, the particular parametrization of $F_{3N}(Q)$ is much less important. 

 \begin{figure}
 \centering
  \includegraphics[width=8cm]{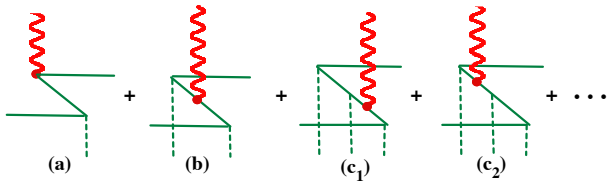}
\caption{(Color online) Examples of Z-graphs in time-ordered perturbation theory, which are automatically included through the coupling of the photon to negative-energy nucleon states in the full CIA calculations. In the 3N system, the meson lines (dashed) can couple to any of the remaining two nucleons (not shown).}
 \label{fig:pion-current}
\end{figure}

\subsection{Negative energy states as relativistic corrections vs.\ interaction currents} \label{Sec:3E}

We have seen that the CIA models and the nonrelativistic IARC agree quite well at low $Q$, so that (i) differences in the physics they describe shows up only at higher $Q$, and (ii) their failure to explain the low $Q$ data is probably due to the omission of the same physics.   Here we will look at both of these questions.

One must be cautious in comparing these two approaches -- the language used to describe the physics differs and the same processes can be called by two different names.   Negative energy states are automatically included in the CST calculations, and are considered relativistic corrections,  but in the IARC formalism they are considered to be interaction currents.  

The negative energy (or interaction current) terms we are discussing are illustrated in Fig.~\ref{fig:pion-current}.  The figure shows a selected set of $Z$-diagrams involving one, two and three pion exchanges.  These are ``time-ordered'' diagrams with time flowing from left to right, so the diagonal lines (flowing backward in time) represent negative energy contributions, which can be reinterpreted as contributions from the production of virtual nucleon-anticuclean pairs.  In the Feynman diagram formalism, more natural to the CST, these $Z$-diagrams are automatically included as part of the off-shell nucleon propagator.  The CIA  automatically includes contributions from the {\it infinite\/} sum of all of these diagrams. 
In the IARC formalism, only diagram \ref{fig:pion-current}(a) is included.  The other processes, thought to be small, are never included in IARC.  Furthermore, in the nonrelativistic language of IARC, some of these higher order diagrams generate three-body currents, so, in the language of IARC, our calculation also includes some  three-body currents.  However, in the language of the CST, none of these contributions are three-body currents. 

\begin{figure}
 \centering
  \includegraphics[width=5cm]{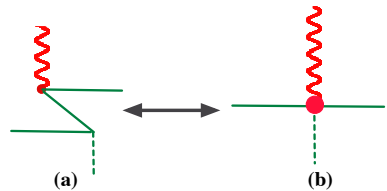}
\caption{(Color online) The pion interaction current.  With ps coupling the  contact current (b) is zero and the $Z$-diagram (a) is large.  With pv coupling, the $Z$-diagram (a) is small, and the contact term (b) is large.  There is an approximate equivalence between (a) calculated using ps coupling and (b) calculated with pv coupling.}
 \label{fig:pion-eq}
\end{figure}

There is another ambiguity that adds to the confusion.  If pseudoscalar (ps) pion coupling is used to calculate the pion exchange currents (the choice made by IARC), then there is no four-point $\gamma\pi NN$ current, but when pseudovector (pv) coupling is used, there is a (large) four-point $\gamma\pi NN$ current.   This is illustrated in Fig.~\ref{fig:pion-eq}.   It has been known for a long time that the leading order contribution from the $Z$-diagram (a) in ps coupling is equivalent to the contact current (b) in pv coupling.  In the CST models used here, a pure pv pion coupling was used, so the $Z$-diagrams are small and the interaction current of Fig.~\ref{fig:pion-eq}(b) must be added (it would be part of the contributions coming from Figs.~\ref{fig:coreFF}G -- J).  Hence, our results are similar to the IARC calculation because, even though we include the $Z$-diagram \ref{fig:pion-eq}(a) and IARC do not, ours is small.  The large part of the pion interaction current, \ref{fig:pion-eq}(a) for IARC and \ref{fig:pion-eq}(b) for CST, is omitted by both calculations.

 \begin{figure*}[tbh]
 \centering
 \includegraphics[width=17cm,bb=49 376 530 715]{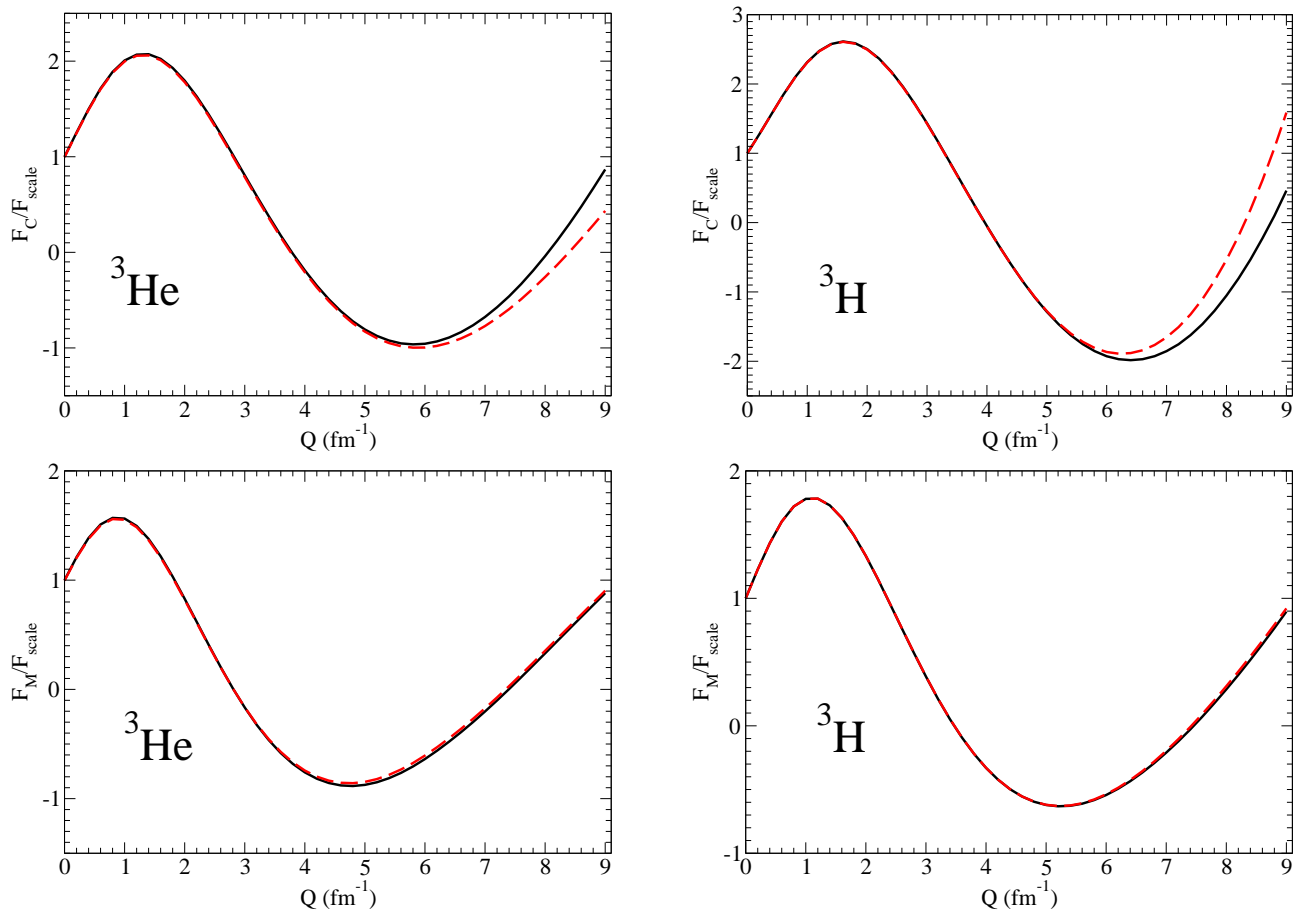}
\caption{(Color online) Charge (upper panels) and magnetic (lower panels) form factors of $^3$He and $^3$H divided by the respective scale functions of Fig.~\ref{fig:Wxx-CIA-3NFFch-scaled}. The solid line is the full result, the dashed line excludes all three-body negative-energy states in the 3N vertex function. The results are obtained with the NN model W16 and the MMD nucleon form factor in CIA.}
 \label{fig:W16-CIA-neg}
\end{figure*}

How big are the negative energy contributions shown in Fig.~\ref{fig:pion-current}?  We have said that they are small, because we use pv coupling for the pion.  A more precise picture is given 
 Fig.~\ref{fig:W16-CIA-neg}, which shows the influence of negative-energy states in three-body channels on the charge and magnetic form factors of $^3$He and $^3$H. It compares the full calculation with W16/MMD with another calculation in which only positive-energy three-body channels are taken into account. Note that this does not mean that negative-energy states are completely eliminated from these calculations, because the three-nucleon vertex functions were calculated from all channels.  But this comparison does show the effect of removing all of the diagrams in Fig.~ \ref{fig:pion-current}.  
The figure shows that the charge form factors of $^3$He and $^3$H are slightly changed, mainly for momentum transfer larger than about 6 fm$^{-1}$, while the magnetic form factors remain essentially unchanged. 

\subsection{Critique of the physics missing from the CIA}

We conclude with a brief discussion of the physics missing from the CIA.  It is useful to look at the isoscalar and isovector magnetic and charge combinations separately.


Careful examination of the  {\it isoscalar magnetic\/} form factor given in Fig.~\ref{fig:Wxx-CIA-3NISIVmag} shows that the CIA is very close to the data, except possibly in the region near $Q$  from about 2 to 3 fm$^{-1}$ (where the discrepancy is about $\sim 20$\%).  This is confirmed by the IARC studies \cite{Mar98}, which show that interaction currents in this channel are quite small (the largest ``corrections'' to the one-body current are the spin-orbit relativistic corrections, included in both the IARC and in the CST models).  The model dependence shown in the figure suggests that the discrepancy in the region $Q$ from  2 to 3 fm$^{-1}$ (not visible on the logarithmic scale shown in the right hand panel and in Ref.~\cite{Mar98})  might very well be corrected by use of more accurate wave functions.  Accurate $^3$H wave functions were produced after this work was finished, and a better $^3$He wave function (with the correct binding energy) can be produced once we develop an accurate $pp$ interaction model.

The {\it isovector magnetic\/} form factor (Fig.~\ref{fig:Wxx-CIA-3NISIVmag}) and both the {\it isoscalar\/} and {\it isovector charge\/} form factors (Fig.~\ref{fig:Wxx-CIA-3NISIVch}) show large discrepancies that are explained in Ref.~\cite{Mar98} by large pion and rho interaction currents missing from both the IARC and the CIA results presented here.  It remains to be seen whether these corrections, when added to the CIA, will bring the CST predictions into agreement with the data.

Finally, as observed in all of the figures above, relativistic effects grow significantly with increasing $Q^2$, and the new Jefferson Laboratory high $Q^2$ measurements of the $^3$He form factors, to be released soon, will provide a strong test of the relativistic theory.  The existence or non-existence of a second minimum will be a particularly interesting signal.  Our CIA calculations show a second minimum, but until we have added the missing interaction currents we are unable to predict its location.

\section{Conclusions}
We have performed the first numerical calculations of the electromagnetic three-nucleon form factors in the Covariant Spectator Theory. This framework is manifestly covariant and includes relativistic effects such as boosts, Wigner spin rotations, and negative-energy states exactly, without resorting to expansions in orders of $v/c$. These calculations were done in Complete Impulse Approximation for a number of relativistic three-nucleon bound-state vertex functions, derived from a family of two-nucleon interaction models that produce different three-nucleon binding energies while maintaining a roughly equivalent good fit to the nucleon-nucleon data. From the findings of these calculations, we want to highlight the following:

The first observation is that the CST description of the three-nucleon electromagnetic form factors yields very reasonable results, confirming the validity of this formalism. The calculations are demanding, but numerically stable.

Our study of the model dependence of the results indicates that the three-nucleon binding energy determines their behavior to larger momentum transfer than previously expected. This is also evident when we compare to nonrelativistic impulse approximation calculations with relativistic corrections (IARC), which are based on the Argonne AV18/UIX two- and three-nucleon potentials. 
The three-nucleon form factors obtained with model W16, which reproduces the experimental triton binding energy and comes also closest to the one predicted by the nonrelativistic AV18/UIX potentials, are remarkably similar to the IARC form factors up to moderate values of the momentum transfer (around 4 to 6 fm$^{-1}$).  

We find that each of the six diagrams of CIA is significant, so none can be omitted.

Variations in the parametrization of the single-nucleon electromagnetic form factors, as well as the inclusion of off-shell form factors, have little effect on the results.

A similar observation can be made about the influence of negative-energy states in the three-nucleon channels, that give rise to Z-graph-type contributions to the form factors. We explain their smallness by the use of pure pseudo-vector $\pi NN$ coupling in our two-nucleon potentials, which suppresses Z-graphs. The same reason makes it perfectly understandable why our results are so similar to the ones in IARC, that do not include any Z-graphs. 

Since the CIA calculations do not contain interaction currents, which we believe to be large, a good agreement with the data over the whole range of the considered momentum transfer cannot yet be expected. An execption is the isoscalar magnetic form factor, where interaction currents are small, and indeed we find a good agreement of our results with the data. In all other cases, interaction currents will have to be calculated when a very close description of the data for all values of $Q$ is the objective.

\begin{acknowledgments}
We thank L.\ Marcucci for providing the results of IARC calculations that were specially taylored for comparison with our results, and R.\ Schiavilla and J.\ Adam, Jr., for helpful discussions.
S.\ A.\ P.\ and A.\ S.\ received support from FEDER and FCT under grant Nos.\ SFRH/BD/8432/2002 and POCTI/ISFL/2/275. F.\ G.\ was supported by Jefferson Science Associates, LLC under U.S. DOE Contract No.~DE-AC05-06OR23177.  F.G. wishes to thank the Centro de F\'isica Nuclear da Universidade de Lisboa for its hospitality.
\end{acknowledgments}
 
\appendix

\section{Calculation of the 3N vertex function with two off-shell nucleons}
\label{app:A}

As discussed above, the contributions when two nucleons are off-shell require knowledge of the two-body off-shell vertex function defined in (\ref{eq:3NFadoff}).  This is determined by quadratures from the two-particle off-shell scattering amplitude, which is in turn determined by quadratures from the two-body off-shell kernel $V_{\beta\alpha,\lambda'_2\alpha'}(k_2,k'_2;P_{23})$, as illustrated in Fig.~\ref{fig:Meqoff}.

The off-shell kernel enlarges the number of degrees of freedom in two different ways.  

First, since the kernel connects to the off-shell propagator for particle 2, its expansion [similar to the expansion (\ref{oneprop}) for particle 3] requires knowledge of both the positive and negative energy projections of the kernel:
\bea
&&V^{\rho_2,+}_{\lambda_2 \alpha,\lambda'_2\alpha'}(k_2,k'_2;P_{23})
\nonumber\\
&&\qquad\qquad=\overline{u}^{\rho_2}_\beta(k_2,\lambda_2) V_{\beta\alpha,\lambda'_2\alpha'}(k_2,k'_2;P_{23})\, .\qquad
\eea
The $\rho_2=-$ projections, not needed before, are defined in Ref.~\cite{Gro08}, and can be computed straightforwardly from knowledge of the Dirac structure of the OBE model.  

Second, the relative energy of the final state is now no longer fixed.  In general, this relative energy is 
\bea
k_{23}^0\equiv\sfrac12(k_{2}^0-k_{3}^0)=k_2^0-\sfrac12W_{23}\, .
\eea
When particle 2 is on shell, $k_2^0=E_{k_2}$ and the relative energy depends on $|{\bf k}_2|$ and is not an independent variable.  When both particles are off-shell, $k_{23}^0$ is an independent variable.  In this work we found it convenient to write $k_{23}^0$ in terms of the new variable $x_0$, where
\bea
k_{23}^0=x_0\left(E_{k_2}-\sfrac12W_{23}\right)\, ,
\label{eq:A3}
\eea
so that $x_0=1$ when particle 2 is on shell.

To have the behavior required by the (generalized) Pauli principal, the potential must exhibit a Pauli exchange symmetry (meaning that, depending on the isospin, it must be either symmetric or antisymmetric) when particles 2 and 3 are exchanged.   However, when particles 2 and 3 are exchanged, $k_{23}^0\to-k_{23}^0$, so that $x_0\to-x_0$.  We conclude that the kernel must have the Pauli exchange symmetry when $x_0\to-x_0$.  For the on-shell case ($x_0=1$), this symmetry is discussed in detail in Refs.~\cite{Gro92,Gro08}.  

This symmetry is imposed in practice by explicitly symmetrizing the OBE kernels.  The symmetrized meson propagators are
\bea
\overline{\Delta}_{m_v}\simeq\frac12\left(\frac{N(q_+)}{m_v^2-t_+(x_0)}\pm\frac{N(q_-)}{m_v^2-t_-(x_0)} \right)
\label{eq:A4}
\eea 
where $q_\pm=\pm k_{23}-k'_{23}$ and $t_\pm=q_\pm^2$.  In Refs.~\cite{Gro92,Gro08} the first term in (\ref{eq:A4}) is referred to as the direct term; the second as the exchange or ``alternating'' term.
 Since the initial particle 2 is on-shell, the momentum transfers are
\bea
t_\pm(x_0)&=&\Big[E'-\sfrac12W_{23}\mp x_0(E-\sfrac12W_{23})\Big]^2\nonumber\\
&&-({\bf k}\mp{\bf k}')^2\, ,
\eea
where, for simplicity, we use the notation $k\equiv k_{23}$ and $k'\equiv k'_{23}$.  Near the physical scattering region, when $W_{23}\simeq 2E\simeq 2E'$, the momentum transfers $t_\pm(x_0)\simeq-({\bf k}\mp{\bf k}')^2$  are negative.  However, when the momenta are very large, $t_\pm$ can become positive, and the denominators $m_v^2-t_\pm$ can be zero.   To see that the momentum transfers can become positive, consider the case when ${\bf k}-{\bf k}'=0$ and $|{\bf k}|\to\infty$.  Then
\bea
\lim_{{\bf k}\to\infty}\;t_\pm(x_0)\to{\bf k}^2(1\pm x_0)^2\, .
\eea
Hence the momentum transfer always becomes positive, and the kernel has a singularity, unless $x_0=1$ and the same particle is on-shell both before and after the interaction (the direct term with $t=t_+$).  In this case it is easy to show the $t_+(1)<0$ for all ${\bf k}$ and  ${\bf k}'$.  

A detailed analysis shows that some of these singularities arise from physical particle production; others are spurious singularities that are cancelled by higher order kernels neglected in the OBE approximation.  In either case these singularities should be removed, and prescriptions for removing them are discussed in detail elsewhere (the interested reader should study Ref.~\cite{Gro08}).  The work presented in this paper uses prescription A for the on-shell amplitudes.  This prescription can be summarized by the replacements $t_\pm(x_0)\to t_\pm(\pm1)$ which insures that the meson propagators for both the direct (same particle on-shell) and  ``alternating''  (different particles on-shell) have no singularities.  The exact $x_0$ dependence is retained in the numerators $N(q)$ so that some differences between direct and alternating terms are preserved.  Hence, under prescription A all symmetries are preserved,  but  some terms (those that would arise from antisymmetrizing the denominators) are zero.  In choosing prescription A  we lose some of the off-shell dependence.  A more complete picture can be obtained from prescription C (described in Ref.~\cite{Gro08}), and the consequences of this choice  will be described elsewhere.

\section{Calculation of the three-nucleon electromagnetic current in CIA}

\noindent
In this appendix we show some details of the calculation of the 3N electromagnetic current in CIA, \textit{i.e.}, of diagrams (A) to (F) of Fig.\ \ref{fig:coreFF}. The calculations are carried out in the Lab frame. Of course, since we are working in a covariant framework, the final result does not depend on this particular choice of reference frame.

First, we write the helicity partial wave basis states (\ref{eq:3Npwhelbasis}) in a way that shows the involved transformations on the individual particle states in more detail:
\begin{widetext}
\begin{align}
\vert  q \tilde{p}^{0} \tilde{p} M j m & \lambda_{1} \lambda_{2} \lambda_{3}  \rho_{2} \rho_{3} T \mathcal{T}_{z} \rangle
=
\sqrt{\frac{2j+1}{8\pi^{2}}} \int_{0}^{2\pi} d \varPhi \int_{0}^{\pi} d \varTheta \sin \varTheta \int_{0}^{2\pi} d \tilde{\phi}
\int_{0}^{\pi} d \tilde{\theta} \sin \tilde{\theta} \,
{\cal D}^{(1/2)\ast}_{M , m-\lambda_{1}} ( \varPhi , \varTheta , 0    ) {\cal D}^{(j)\ast}_{m , \lambda_{2} - \lambda_{3} } ( \tilde{\phi} , 
\tilde{\theta} , 0    ) 
 \nonumber \\
 \times & \biggl\{ \Bigl[ \vert p_{1}  \rangle  \otimes S ( R_{ \varPhi , \varTheta , 0} ) S ( R_{ \pi , \pi , 0 } ) u^{+}( q , \lambda_{1} )\Bigr] 
\otimes
\Bigl[ \vert p_{2} \rangle \otimes
S ( R_{ \varPhi , \varTheta , 0} ) S ( Z( q ) ) S ( R_{ \tilde{\phi} , \tilde{\theta} , 0} )
u^{\rho_{2}}( \tilde{p} , \lambda_{2} )\Bigr] 
\nonumber \\ 
& \otimes 
\Bigl[ \vert p_{3} \rangle \otimes S ( R_{ \varPhi , \varTheta , 0} ) S ( Z( q ) ) S ( R_{ \tilde{\phi} , \tilde{\theta} , 0} )
S ( R_{ \pi , \pi , 0 } )
u^{\rho_{3}}( \tilde{p} , \lambda_{3} )\Bigr] \biggr\}
\otimes \vert T {\mathcal{T}}_{z} \rangle \, . \label{cap3_++54tr}
\end{align}
\end{widetext}
The total angular momentum is always $J=1/2$ for the 3N bound states and is from here on suppressed in the state kets. The possible values of the other discrete quantum numbers are $ M = \pm 1/2   $, $j=0,1,\dotsc$, $ m = - j , -j + 1 , \dotsc , j    $,
$ \lambda_{1} $, $ \lambda_{2} $, $ \lambda_{3} = \pm 1/2 $, subject to the constraints
$ \abs{m-\lambda_{1}} \le 1/2 $,
$ \abs{\lambda_{2}-\lambda_{3}} \le j $, and
$ \rho_{2} $, $ \rho_{3} = \pm $. The four-momenta 
$ p_{1} $, $ p_{2} $, $ p_{3} $ of the three nucleons are obtained by the space-time transformations 
\begin{align}
p_{1} &= R_{ \varPhi , \varTheta , 0} R_{ \pi , \pi , 0 } 
\bigl( E( q ) , 0 , 0 , q \bigr) \, , \label{cap3_+45tr} \\
p_{2} &= R_{ \varPhi , \varTheta , 0} Z(  q ) R_{ \tilde{\phi} , \tilde{\theta} , 0}
\bigl(  \tilde{p}^{0} , 0 ,0 , \tilde{p}  \bigr) \, , \label{cap3_+45tr1} \\
p_{3} &= R_{ \varPhi , \varTheta , 0} Z(  q ) R_{ \tilde{\phi} , \tilde{\theta} , 0} R_{ \pi , \pi , 0 }
\bigl(  W( q ) - \tilde{p}^{0} , 0 , 0 ,\tilde{p}  \bigr) \, , \label{cap3_+45tr2}
\end{align}
and for single-particle state vectors we use the normalization $ \langle p' \vert p \rangle = ( 2 \pi )^{4} \delta^{4} ( p' - p ) $.

The helicity spinors in (\ref{cap3_++54tr}) are
\begin{equation}
\begin{split}
u^{+}( p , \lambda ) 
& = 
\left ( 
\begin{array}{c} 
\cosh \frac{\eta_p}{2} \\
2\lambda\sinh \frac{\eta_p}{2}  
\end{array} 
\right ) 
\otimes
\chi ( \lambda ) \, ,\\
u^{-}( p , \lambda ) & =  
\left ( 
\begin{array}{c} -2\lambda \sinh \frac{\eta_p}{2} \\
 \cosh \frac{\eta_p}{2}
\end{array} 
\right ) 
\otimes
\chi ( \lambda ) \, , 
\end{split}
\end{equation}
with the rapidity $\eta_p$ given by $\tanh \eta_p = p/E(p)$, and the two-component spinors
\begin{equation}
\chi ( 1/2 ) =  
\left ( 
\begin{array}{c} 
1 \\ 0
\end{array} 
\right )  \, , \quad 
\chi ( - 1/2 ) =  
\left ( 
\begin{array}{c} 
0 \\ 1
\end{array} 
\right ) \, .
\end{equation}
The state (\ref{cap3_++54tr}) has
total angular momentum $1/2$ and helicity $M$. As long as the 3N bound state remains at rest or is boosted along the positive $z$-axis, $M$ is also the $z$-projection of the total angular momentum.

The total four-momentum of the 3N bound state at rest is $P_t=(M_t,0,0,0)$. Its vertex state with nucleon 1 on mass shell, helicity $M$, and isospin projection ${\mathcal T}_z$, can be written as a linear combination of the basis states (\ref{cap3_++54tr}):
\begin{widetext}
\begin{align}
& \vert \Gamma^{1} ( P_t,M,{\mathcal T}_{z} ) \rangle =
\sum_{\substack{jmT 
\\ \lambda_{1} \lambda_{2} \lambda_{3}  }} 
\sum_{\substack{ 
  \rho_{2} \rho_{3} 
\\  \rho'_{2} \rho'_{3} }}
\int_{0}^{q_{s}} \frac{d q\,  q^{2}}{(2\pi)^{3}2 E(q)} 
\int_{- \infty}^{+ \infty} \frac{d \tilde{p}^{0}}{2\pi} 
\int_{0}^{+ \infty} \frac{d \tilde{p} \, \tilde{p}^{2}}{(2\pi)^{3}  }   
\left ( \frac{m}{E(\tilde{p})} \right )^{4}
\notag \\
& \quad \times
C(  q \tilde{p}^{0} \tilde{p} M j m \lambda_{1} \lambda_{2} \lambda_{3}  \rho'_{2} \rho'_{3} T \mathcal{T}_{z}  )
O_{\rho_{2}\rho'_{2}} (\tilde{p},\lambda_{2})
O_{\rho_{3}\rho'_{3}} (\tilde{p},\lambda_{3})
\vert  q \tilde{p}^{0} \tilde{p} M j m \lambda_{1} \lambda_{2} \lambda_{3}  \rho_{2} \rho_{3} T \mathcal{T}_{z} \rangle \, , \label{qw5}
\end{align}
\end{widetext}
where the functions $C$ are real because of time reversal invariance, and the matrix $O$ is defined as
\begin{equation}
O_{{\rho}' \rho} ( p , \lambda )  =  
{\bar{u}}^{\rho'} ( p , \lambda ) u^{\rho} ( p , \lambda ) =
\left ( 
\begin{array}{cc} 
1 & -2\lambda \frac{p}{m} \\
-2\lambda \frac{p}{m} & -1 
\end{array} 
\right )_{\rho' \rho} \, .
\end{equation}
The integration over $q$ in (\ref{qw5}) extends up to the finite value $q_s=(M_t^2-m^2)/2M_t$ for which the mass of the 2N subsystem becomes zero and the limit of the region of timelike 2N states is reached. As was shown in \cite{Sta97b}, the functions $C$ go to zero smoothly as $q$ approaches $q_s$, which makes it possible to treat $q_s$ as a natural cut-off momentum without making the 3N vertex functions discontinuous.

When the Feynman diagrams of the CIA are evaluated, one encounters frequently the particular sequence of Lorentz transformations of a boost in $z$-direction, followed by a rotation about the $y$-axis and then by another boost in $z$-direction. It turns out to be very useful that this is equivalent to one boost in $z$-direction between two rotations about the $y$-axis, in the following way:
\begin{equation}
B( \eta_{1} \hat{e}^{3}  ) R( \theta_{1} \hat{e}^{2}  ) B( \eta_{2} \hat{e}^{3}  ) =
R( \theta_{2} \hat{e}^{2}  ) B( \eta_{3} \hat{e}^{3}  ) R( \theta_{3} \hat{e}^{2}  ) \, .
\label{prop_34er}
\end{equation}
For given rapidities $\eta_{1}$, $\eta_{2}$, and rotation angle $\theta_{1}$, the corresponding rapidity $\eta_{3}$ and rotation angles $\theta_{2}$ and $\theta_{3}$ can be found from:
\begin{equation}
\sinh \eta_{3} = \sqrt{( \cosh \eta_{1} \cosh \eta_{2} +  \sinh \eta_{1} \sinh \eta_{2} \cos \theta_{1} )^{2}-1} \, .
\end{equation}

If $ \eta_{3} > 0  $ then 
\begin{align} 
\sin \theta_{2} &= \frac{ \sin \theta_{1} \sinh \eta_{2}  }{ \sinh \eta_{3} } \, , \\
\cos \theta_{2} &= \frac{ \sinh \eta_{1} \cosh \eta_{2} + \cosh \eta_{1} \sinh \eta_{2} \cos \theta_{1}}{ \sinh \eta_{3} } \, , \\
\sin \theta_{3} &= \frac{ \sin \theta_{1} \sinh \eta_{1}  }{ \sinh \eta_{3} } \, , \\
\cos \theta_{3} &= \frac{ \cosh \eta_{1} \sinh \eta_{2} + \sinh \eta_{1} \cosh \eta_{2} \cos \theta_{1}}{ \sinh \eta_{3} } \, ,
\end{align}
with $ 0 \le \theta_{2} , \theta_{3} < 2 \pi $. 
If $ \eta_{3} = 0  $ then $\theta_{3} =0$ and
\begin{align} 
\sin \theta_{2} &= \sin \theta_{1} \, , \\
\cos \theta_{2} &= \cos \theta_{1} \, ,
\end{align} 
where $ 0 \le \theta_{2} < 2 \pi $.

Similarly useful, a rotation about the $z$-axis between two rotations about the $y$-axis can be replaced by a rotation about the $y$-axis between two rotations about the $z$-axis:
\begin{equation}
R( \theta_{1} \hat{e}^{2}  ) R( \phi_{1} \hat{e}^{3}  ) R( \theta_{2} \hat{e}^{2}  ) =
R( \phi_{2} \hat{e}^{3}  ) R( \theta_{3} \hat{e}^{2}  ) R( \phi_{3} \hat{e}^{3}  ) \, .
\label{cap3_34rt78}
\end{equation}
From the given angles $\theta_{1}$, $\theta_{2}$, and $\phi_{1}$ one obtains first
\begin{equation}
\theta_{3} = \arccos ( \cos \theta_{1} \cos \theta_{2} - \sin \theta_{1} \sin \theta_{2} \cos \phi_{1} ) \, .
\end{equation}
If $ \theta_{3} \ne 0   $ and $ \theta_{3} \ne \pi   $ then 
\begin{align} 
\sin \phi_{2} &= \frac{ \sin \theta_{2} \sin \phi_{1}  }{ \sin \theta_{3} } \, , \\
\cos \phi_{2} &= \frac{ \sin \theta_{1} \cos \theta_{2} + \cos \theta_{1} \sin \theta_{2} \cos \phi_{1}}{ \sin \theta_{3} } \, , \\
\sin \phi_{3} &= \frac{ \sin \theta_{1} \sin \phi_{1}  }{ \sin \theta_{3} } \, , \\
\cos \phi_{3} &= \frac{ \cos \theta_{1} \sin \theta_{2} + \sin \theta_{1} \cos \theta_{2} \cos \phi_{1}}{ \sin \theta_{3} } \, ,
\end{align}
with $ 0 \le \phi_{2} , \phi_{3} < 2 \pi $.
On the other hand, if $ \cos \theta_{3} = \pm 1   $ then $\phi_{3} =0$, and
\begin{align} 
\sin \phi_{2} &= \frac{1}{2} \sin \phi_{1} ( \cos \theta_{1} \pm \cos \theta_{2} ) \, , \\
\cos \phi_{2} &= \frac{1}{2} [ \cos \phi_{1} ( 1 \pm \cos \theta_{1} \cos \theta_{2} ) \mp \sin \theta_{1} \sin \theta_{2} ]  \, ,
\end{align} 
with $ 0 \le \phi_{2} < 2 \pi $.


The 3N vertex functions were calculated in a helicity partial wave basis, which means that we have to calculate the matrix elements of the electromagnetic 3N current in terms of the functions
\begin{multline}
 C(  q \tilde{p}^{0} \tilde{p} M j m \lambda_{1} \lambda_{2} \lambda_{3}  \rho_{2} \rho_{3} T \mathcal{T}_{z}  ) = \\
\langle  q \tilde{p}^{0} \tilde{p} M j m  \lambda_{1} \lambda_{2} \lambda_{3}  \rho_{2} \rho_{3} T \mathcal{T}_{z} \vert 
\Gamma^{1} ( P_t,M,{\mathcal T}_{z} ) \rangle \, .
\end{multline}
Formally, we can write each of the diagrams of the CIA in the general operator form
\begin{equation}
 \langle \Gamma^{1} ( P'_t,M',{\mathcal T}_{z} ) \vert A j^\mu_N B \vert \Gamma^{1} ( P_t,M,{\mathcal T}_{z} ) \rangle \, ,
\label{eq:Joperator}
\end{equation} 
where $j^\mu_N$ is the operator of a single-nucleon current, and $A$ and $B$ are operators composed of off-shell propagators, on-shell projectors, and, in the cases of diagrams (D), (E), and (F), permutation operators. 
We can now insert a complete set of partial wave states (\ref{cap3_++54tr}) before  $A$ and after $B$ in (\ref{eq:Joperator}) to obtain the desired current matrix elements in partial wave form.

Note that in all diagrams, as part of the states (\ref{cap3_++54tr}), single-nucleon momentum kets $\vert p_i \rangle$ that are inserted on the right of $B$ are contracted to corresponding bras from the states inserted on the left of $A$. These inner products lead to Dirac delta functions that insure momentum conservation along the nucleon lines in each diagram. In the following sections, we show how these delta functions are evaluated, and we display the final form of the various diagrams in partial wave form.

\subsection{Calculation of diagram A}

\indent

The bound state vertex vector represented on right (left) side of diagram A
is an integral over momentum variables $q$, $\tilde{p}$ ($q'$, $\tilde{p}'$) and angular variables
$ \varPhi $, $ \varTheta $, $ \tilde{\phi}    $, $ \tilde{\theta}    $
($ \varPhi' $, $ \varTheta' $, $ \tilde{\phi}'    $, $ \tilde{\theta}'    $), and its total momentum is 
$P_{t}$ ($P'_{t}=B ( \eta \hat{e}^{3} )P_{t}$).

The product of momentum conserving Dirac delta functions in diagram A is:
\begin{widetext}
\begin{multline}  
\delta^{3}
\Bigl( 
B ( \eta \hat{e}^{3} ) 
R_{ \varPhi' , \varTheta' , 0 } 
R_{ \pi , \pi , 0 }
B ( \xi(q') \hat{e}^{3} )
v( m , 0 )  
-
R_{ \varPhi , \varTheta , 0 } 
R_{ \pi , \pi , 0 }
B ( \xi(q) \hat{e}^{3} )
v( m , 0 )  
\Bigr) 
\\ 
 \times
\delta^{3}
\Bigl( 
B ( \eta \hat{e}^{3} ) 
R_{ \varPhi' , \varTheta' , 0 } 
Z(  q' )
R_{ \tilde{\phi}' , \tilde{\theta}' , 0 } 
B ( \xi(\tilde{p}') \hat{e}^{3} ) 
v( m , 0 )  
-
R_{ \varPhi , \varTheta , 0 } 
Z(  q )
R_{ \tilde{\phi} , \tilde{\theta} , 0  } 
B ( \xi(\tilde{p}) \hat{e}^{3} ) 
v( m , 0 )  
\Bigr) \, , \label{cap3_45ty34}
\end{multline}
where we introduced the auxiliary four-vector $v(x,y)\equiv (x,0,0,y)$ and defined $\sinh \xi ( x ) = x/m$.

From the first  Dirac delta function in Eq.~\eqref{cap3_45ty34} we get 
$ \varPhi' = \varPhi $, $ \varTheta' = \pi - \varTheta_{1} $, $ q' = r_{1}   $, where
$ \varTheta_{1} $, $  r_{1} $ are obtained by applying (\ref{prop_34er}) to 
$ B ( -\eta \hat{e}^{3} )
R_{ 0 , \pi - \varTheta , 0 } 
B ( \xi(q) \hat{e}^{3} )$:
\begin{equation}
B ( -\eta \hat{e}^{3} )
R_{ 0 , \pi - \varTheta , 0 } 
B ( \xi(q) \hat{e}^{3} ) =
R_{ 0 ,  \varTheta_{1}  , 0 } 
B ( \xi ( r_{1} ) \hat{e}^{3} )
R_{ 0 ,  \varTheta_{2}  , 0 } \, .
\end{equation}
To evaluate the second delta function, we begin by applying rule (\ref{prop_34er}) twice,
\begin{align}
Z( - q ) R_{ 0 , - \varTheta , 0 } B ( \eta \hat{e}^{3} )  &=
R_{ 0 ,  \vartheta_{1} , 0 }
B ( \xi( s_{1} ) \hat{e}^{3} ) 
R_{ 0 ,  \vartheta_{2}  , 0 } \, , \label{zzx1} \\
B ( \xi( s_{1} ) \hat{e}^{3} ) 
R_{ 0 ,  \vartheta_{2} + \varTheta' , 0 }
Z(  q' ) &=
R_{ 0 ,   \vartheta_{3} , 0 }
B ( \xi( s_{2} ) \hat{e}^{3} ) 
R_{ 0 ,  \vartheta_{4}  , 0 } \, , \label{zzx2}
\end{align}
which defines $s_{1}$, $s_{2}$, $\vartheta_{1}$, $\vartheta_{2}$,
$\vartheta_{3}$, and $\vartheta_{4}$.

Next, we change the angular integration over $ \tilde{\phi}' $ and $ \tilde{\theta}' $ to the new angles $ \bar{\phi} $ and $\bar{\theta}$ (with $ 0 \le \bar{\phi} < 2 \pi $ and $0 \le \bar{\theta} \le \pi $), where
$ \tilde{\phi}' $ and $ \tilde{\theta}' $ are given in terms of $ \bar{\phi} $ and $\bar{\theta}$
through the application of (\ref{cap3_34rt78}),
\begin{equation} 
R_{ 0 , - \vartheta_{4} , 0  } R_{ \bar{\phi} , \bar{\theta} ,  0  } =
R_{ \tilde{\phi}' , \tilde{\theta}' , - \bar{\varphi}} \, ,  \label{zzx3}
\end{equation}
and we have
\begin{equation}
\int_{0}^{\pi} d \tilde{\theta}' \sin \tilde{\theta}' \int_{0}^{2 \pi} d \tilde{\phi}' =
\int_{0}^{\pi} d \bar{\theta} \sin \bar{\theta} \int_{0}^{2 \pi} d \bar{\phi} \, .
\end{equation}
Using (\ref{prop_34er}) in the form
\begin{equation}
B ( \xi( s_{2} ) \hat{e}^{3} ) 
R_{ 0 , \bar{\theta} , 0  }
B ( \xi(\tilde{p}') \hat{e}^{3} )
=
R_{ 0 , \vartheta_{5} , 0  }
B ( \xi( s_{3} ) \hat{e}^{3} ) 
R_{ 0 , \vartheta_{6} , 0  } \, ,
\end{equation}
determines  $ s_{3}$, and finally applying again  (\ref{cap3_34rt78}),
\begin{equation}
R_{ 0 ,  \vartheta_{1} + \vartheta_{3} , \bar{\phi} }
R_{ 0 , \vartheta_{5} , 0  }
=
R_{ \phi_{1} , \vartheta_{7} , \phi_{2}  } \, ,
\end{equation}
defines the angles $ \phi_{1} $ and $ \vartheta_{7}$.

What we have achieved is that the first momentum in the second delta function has been rewritten as
\begin{equation}
 B ( \eta \hat{e}^{3} ) 
R_{ \varPhi' , \varTheta' , 0 } 
Z(  q' )
R_{ \tilde{\phi}' , \tilde{\theta}' , 0 } 
B ( \xi(\tilde{p}') \hat{e}^{3} ) 
v( m , 0 ) = 
R_{ \varPhi , \varTheta , 0 } 
Z(  q )
R_{\phi_1 , \vartheta_7 , 0  } 
B ( \xi(s_3) \hat{e}^{3} ) 
v( m , 0 )  \, ,  \label{eq:DiagA-delta2}
\end{equation}
where we have also used the fact that rotations leave $v( m , 0 )$ unchanged. Comparing (\ref{eq:DiagA-delta2}) to the second momentum in the argument of the same delta function, we can read off $\tilde{\phi}=\phi_1$, $\tilde{\theta}=\vartheta_7$, and $\tilde{p}=s_3$.

Before writing down the final expression for diagram A, we have to address the issue of what consequences might arise from the limitation of the spectator momentum in the vertex function to values below the critical value $q_s$. 
The photon momentum absorbed by the 3N system in the lab frame is directed along the positive $z$-axis, which imparts on it a corresponding boost into the same direction. 
It is clear that even if the spectator momentum of the 3N system at rest lies below $q_s$, it can exceed the critical value after being boosted. Whether this is the case or not depends on the size of the boost, the initial spectator momentum, and its angle with respect to the direction of the boost.

We define $K$ to be the $z$-component of the photon momentum in the lab frame, which is related to the invariant square of the photon four-momentum, $-Q^2$, through $K=Q\sqrt{1+Q^2/4M^2_t}$. Then the boost rapidity $\eta$ is given through $ \sinh \eta =  K/M_{t}$. An analysis of the effect of this boost on the spectator momentum shows that, for a given $K$ and a fixed $q < q_s$, the magnitude of the boosted spectator momentum $q'$ is less than $q_s$ if and only if
$\varTheta \in I $, where 
$ I = [ 0 , \pi ]$ if $ 0 \le q < a $ and $ I = ] \arccos b  , \pi ]$ if
$a  \le q < q_s$, with
\begin{align}
a &= \frac{(M^{2}_{t}-m^2) \sqrt{M^{2}_{t}+K^{2}} - (M^{2}_{t}+m^2)   K} {2M^{2}_{t}} \, , \\
b &= \frac{ M_{t} \sqrt{ m^2 + q^{2}_{s} } - \sqrt{ (m^2 + q^{2})(M^{2}_{t}+K^{2})  } }{qK} \, .
\end{align}
These relations are valid as long as $0 < K < (M^2_t-m^2)/2m$, which is always the case in the calculations of this work. Note that $q' \to q_{s}$ as $ \varTheta \to \arccos b $ when $a \le q < q_{s}$.
We restrict the integration over $\varTheta$ to the interval $I$.

Diagram A can now be written as an integral over
$q$, $\varTheta$, $\tilde{p}'$, $\bar{\theta}$, $\bar{\phi}$, $\varPhi$:
\begin{align}
\langle M' | J^\mu_A | M \rangle &= \frac{3m^{2}}{2(2\pi)^{8}} \int_{0}^{q_{s}} dq \frac{q^{2}}{E(q)} 
\int_{I} d\varTheta \sin \varTheta  \int_{0}^{+\infty} d \tilde{p}' 
\frac{\tilde{p}^{\prime2}}{E(\tilde{p}')} \int_{0}^{\pi} d \bar{\theta} \sin \bar{\theta} \int_{0}^{2\pi} d \bar{\phi}
\int_{0}^{2\pi} d \varPhi
\sum_{ \substack{ T' j' m' \lambda_{1}'  \lambda_{2}' \lambda_{3}' \rho_{3}' \\  T j m \lambda_{1}  \lambda_{2} \lambda_{3} \rho_{3} } }
\notag \\
& \quad 
\sqrt{2j'+1}  \sqrt{2j+1}
\mathcal{D}^{(1/2)}_{M',m'-\lambda'_{1}}( \varPhi' , \varTheta' , 0 )
\mathcal{D}^{(1/2)\ast}_{M,m-\lambda_{1}}( \varPhi , \varTheta , 0 )
\mathcal{D}^{(j')}_{m',\lambda'_{2}-\lambda'_{3}}( \tilde{\phi}' , \tilde{\theta}' , 0 )
\notag \\
& \quad \times
\mathcal{D}^{(j)\ast}_{m,\lambda_{2}-\lambda_{3}}( \tilde{\phi} , \tilde{\theta} , 0 )
F^{(11)}_{\lambda'_{1}\lambda_{1}}
F^{(22)}_{\lambda'_{2}\lambda_{2}} 
\left(  R_{ \varPhi , 0 , 0 })\right) ^{\mu}_{\phantom{\mu}{\nu}}
\left(  F^{(1)}_{T'T\mathcal{T}_{z}} F^{(33)\nu}_{\rho'_{3}\lambda'_{3}\rho_{3}\lambda_{3}}
+ F^{(2)}_{T'T\mathcal{T}_{z}} F^{(33)\nu\tau}_{\rho'_{3}\lambda'_{3}\rho_{3}\lambda_{3}} \qph_{\tau}
+ F^{(3)}_{T'T\mathcal{T}_{z}} G^{(33)\nu}_{\rho'_{3}\lambda'_{3}\rho_{3}\lambda_{3}} \right)  
\notag \\
& \quad \times
C(  q' E(\tilde{p}^{\prime}) \tilde{p}' M' j' m' \lambda'_{1} \lambda'_{2} \lambda'_{3}  + \rho'_{3} T' \mathcal{T}_{z}  )
C(  q E(\tilde{p}) \tilde{p} M j m \lambda_{1} \lambda_{2} \lambda_{3}  + \rho_{3} T \mathcal{T}_{z}  )
\, , \label{cap3_789gfq}
\end{align}
where the following spinor matrix elements appear:
\begin{gather}
\begin{split}
F^{(11)}_{\lambda'_{1}\lambda_{1}} &= \bar{u}^{+}(0,\lambda'_{1}) S^{-1}( B ( \xi(q') \hat{e}^{3} ) )
S^{-1}(R_{\pi,\pi,0}) S^{-1} ( R_{0,\varTheta',0} ) S^{-1}( B ( \eta \hat{e}^{3} ) )
S ( R_{0,\varTheta,0} ) S(R_{\pi,\pi,0}) \\
& \quad \times S( B( \xi(q) \hat{e}^{3} ) ) u^{+}(0,\lambda_{1}) \, , \end{split} \label{cap3_4lpo}\\
\begin{split}
F^{(22)}_{\lambda'_{2}\lambda_{2}} &= \bar{u}^{+}(0,\lambda'_{2}) S^{-1}( B( \xi(\tilde{p}') \hat{e}^{3} ) )
S^{-1}(R_{\tilde{\phi}',\tilde{\theta}',0}) S^{-1}( Z(  q' ) )
S^{-1} ( R_{0,\varTheta',0} ) S^{-1}( B ( \eta \hat{e}^{3} ) )  \\
& \quad \times
S ( R_{0,\varTheta,0} )  S( Z(  q ) )
S(R_{\tilde{\phi},\tilde{\theta},0}) S( B( \xi(\tilde{p}) \hat{e}^{3} ) )
u^{+}(0,\lambda_{2}) \, , \end{split} \\
\begin{split}
F^{(33)\nu}_{\rho'_{3} \lambda'_{3} \rho_{3} \lambda_{3}} &= 
\frac{m}{E(\tilde{p}')} 
\frac{m}{E(\tilde{p})} 
\frac{1}{  [ ( \rho'_{3} + 1 )     E(\tilde{p}') - W (  q' ) ] [ ( \rho_{3} + 1 )     E(\tilde{p}) - W (  q ) ] }
\bar{u}^{\rho'_{3}}(\tilde{p}',\lambda'_{3}) \\
& \quad \times
S^{-1}(R_{ \pi , \pi ,0})
S^{-1}(R_{\tilde{\phi}',\tilde{\theta}',0}) 
S^{-1}( Z(  q' ) )
S^{-1} ( R_{0,\varTheta',0} ) S^{-1}( B ( \eta \hat{e}^{3} ) ) 
\gamma^{\nu}
S ( R_{0,\varTheta,0} )   \\
& \quad \times
 S( Z(  q ) )
S(R_{\tilde{\phi},\tilde{\theta},0}) 
S(R_{ \pi , \pi , 0  }) 
u^{\rho_{3}}(\tilde{p},\lambda_{3}) \, , \end{split} \label{cap3_12sd67_1} \\
\begin{split}
F^{(33)\nu\tau}_{\rho'_{3} \lambda'_{3} \rho_{3} \lambda_{3}} &= 
\frac{m}{E(\tilde{p}')} 
\frac{m}{E(\tilde{p})} 
\frac{1}{  [ ( \rho'_{3} + 1 )     E(\tilde{p}') - W (  q' ) ] [ ( \rho_{3} + 1 )     E(\tilde{p}) - W (  q ) ] }
\bar{u}^{\rho'_{3}}(\tilde{p}',\lambda'_{3}) \\
& \quad \times
S^{-1}(R_{ \pi , \pi ,0})
S^{-1}(R_{\tilde{\phi}',\tilde{\theta}',0}) 
S^{-1}( Z(  q' ) )
S^{-1} ( R_{0,\varTheta',0} ) S^{-1}( B ( \eta \hat{e}^{3} ) ) 
i\sigma^{\nu\tau}\\
& \quad \times
S ( R_{0,\varTheta,0} )   
 S( Z(  q ) )
S(R_{\tilde{\phi},\tilde{\theta},0}) 
S(R_{ \pi , \pi , 0  }) 
u^{\rho_{3}}(\tilde{p},\lambda_{3}) \, . \end{split} \label{cap3_12sd67_2} \\
\begin{split}
G^{(33)\nu}_{\rho'_{3} \lambda'_{3} \rho_{3} \lambda_{3}} &= 
\frac{1}{(2m)^{2}}
\left ( \frac{m}{E(\tilde{p}')} \right )^{2}
\left ( \frac{m}{E(\tilde{p})} \right )^{2}
\sum_{\sigma' \sigma = \pm}
O_{\rho'_{3} \sigma'}( \tilde{p}' , \lambda'_{3} )
O_{\rho_{3} \sigma}( \tilde{p} , \lambda_{3} )
\bar{u}^{\sigma'}(\tilde{p}',\lambda'_{3}) \\
& \quad \times
S^{-1}(R_{ \pi , \pi ,0})
S^{-1}(R_{\tilde{\phi}',\tilde{\theta}',0}) 
S^{-1}( Z(  q' ) )
S^{-1} ( R_{0,\varTheta',0} ) S^{-1}( B ( \eta \hat{e}^{3} ) ) 
\gamma^{\nu} \\
& \quad \times
S ( R_{0,\varTheta,0} ) S( Z(  q ) )
S(R_{\tilde{\phi},\tilde{\theta},0}) 
S(R_{ \pi , \pi , 0  }) 
u^{\sigma}(\tilde{p},\lambda_{3}) \, , \end{split} \label{cap3_qw12uj} \, .
\end{gather}
Expression (\ref{cap3_789gfq}) also contains the isospin-dependent functions
\begin{gather}
\begin{split}
F^{(1)}_{T'T\mathcal{T}_{z}} &= \tilde{f} ( q_{3}^{\prime 2} , q_{3}^{2} )
\left( 
F_{1p} ( Q^{2} )
\langle   T'
{\mathcal{T}}_{z} \vert
\frac{1+\tau^{3}_{3}}{2}
\vert   T
{\mathcal{T}}_{z} \rangle
+
F_{1n} ( Q^{2} )
\langle   T'
{\mathcal{T}}_{z} \vert
\frac{1-\tau^{3}_{3}}{2}
\vert T
{\mathcal{T}}_{z} \rangle
\right)  \, , \end{split} \label{eq:AF-1}\\ 
\begin{split}
F^{(2)}_{T'T\mathcal{T}_{z}} &= \tilde{g} ( q_{3}^{\prime 2} , q_{3}^{2} )
\left( 
F_{2p} ( Q^{2} )
\langle  T'
{\mathcal{T}}_{z} \vert
\frac{1+\tau^{3}_{3}}{2}
\vert  T
{\mathcal{T}}_{z} \rangle
+
F_{2n} ( Q^{2} )
\langle  T'
{\mathcal{T}}_{z} \vert
\frac{1-\tau^{3}_{3}}{2}
\vert  T
{\mathcal{T}}_{z} \rangle
\right)  \, , \end{split} \label{eq:AF-2}\\
\begin{split}
F^{(3)}_{T'T\mathcal{T}_{z}} &= \tilde{h} ( q_{3}^{\prime 2} , q_{3}^{2} )
\left( 
F_{3p} ( Q^{2} )
\langle  T'
{\mathcal{T}}_{z} \vert
\frac{1+\tau^{3}_{3}}{2}
\vert  T
{\mathcal{T}}_{z}\rangle
+
F_{3n} ( Q^{2} )
\langle  T'
{\mathcal{T}}_{z} \vert
\frac{1-\tau^{3}_{3}}{2}
\vert  T
{\mathcal{T}}_{z} \rangle
\right)  \, , \label{eq:AF-3}\end{split} \, ,
\end{gather}
where $ \tau^{3}_{3} $ is the isospin projection of nucleon 3, and the four-momenta $q'_{3}$ and $q_3$ are defined as
\begin{equation}
 q'_{3} = \bigl (  W (  q' ) - E(\tilde{p}') , 0 , 0 , \tilde{p}'  \bigr ) \, , \qquad
q_{3} = \bigl (  W (  q ) - E(\tilde{p}) , 0 , 0 , \tilde{p}  \bigr ) \, .
\label{plko}
\end{equation} 
We have constructed five different models of the electromagnetic nucleon current, which correspond to the different choices of the off-shell nucleon form factors given in Table \ref{tab:Nff}.
\begin{table}
 \begin{tabular}{lccccc}
\hline
\hline
  Model & $\tilde{f} ( p^{\prime 2} , p^{2} )$ & $\tilde{g} ( p^{\prime 2} , p^{2} )$ & $\tilde{h} ( p^{\prime 2} , p^{2} )$ &  $ F_{3p} ( Q^{2} ) $ & $ F_{3n} ( Q^{2} )$ \\
\hline
NCI & 1 &   1 & 0 \\
NCII & $f_{0} ( p^{\prime 2} , p^{2} ) $ & $f_{0} ( p^{\prime 2} , p^{2} ) $ & $g_{0} ( p^{\prime 2} , p^{2} ) $ &
$G_{Ep} ( Q^{2} )   $ & $G_{En} ( Q^{2} )   $ \\
NCIII & $f_{0} ( p^{\prime 2} , p^{2} ) $ & $f_{0} ( p^{\prime 2} , p^{2} ) $ & $g_{0} ( p^{\prime 2} , p^{2} ) $ & $F_{1p} ( Q^{2} )$ & $F_{1n} ( Q^{2} )   $ \\
NCIV & $f_{0} ( p^{\prime 2} , p^{2} ) $ & 1 & $g_{0} ( p^{\prime 2} , p^{2} ) $ &
$G_{Ep} ( Q^{2} )   $ & $G_{En} ( Q^{2} )   $\\
NCV & $f_{0} ( p^{\prime 2} , p^{2} ) $ & 1 & $g_{0} ( p^{\prime 2} , p^{2} ) $ & $F_{1p} ( Q^{2} )$ & $F_{1n} ( Q^{2} )   $ \\
\hline
\hline
 \end{tabular} 
\caption{The off-shell nucleon form factors $\tilde{f}$, $\tilde{g}$,  $\tilde{h}$, and the electromagnetic nucleon form factors $F_{3p}$ and  $F_{3n}$ in (\ref{eq:AF-1}) to (\ref{eq:AF-3}) for the models of the nucleon current used in the numerical calculations of this work.}
\label{tab:Nff}
\end{table} 

The isospin matrix elements needed in (\ref{eq:AF-1}) to (\ref{eq:AF-3}) are
\begin{equation}
 \langle  0
{\mathcal{T}}_{z} \vert
\frac{1\pm\tau^{3}_{3}}{2}
\vert  0
{\mathcal{T}}_{z} \rangle = \frac{1}{2} \, , \quad
\langle  1
{\mathcal{T}}_{z} \vert
\frac{1\pm\tau^{3}_{3}}{2}
\vert  1
{\mathcal{T}}_{z} \rangle = \frac{1}{2} \pm \frac{2}{3} {\mathcal{T}}_{z} \, , \quad
\langle  1
{\mathcal{T}}_{z} \vert
\frac{1\pm\tau^{3}_{3}}{2}
\vert  0
{\mathcal{T}}_{z} \rangle =  \pm \frac{1}{\sqrt{3}} {\mathcal{T}}_{z} \, , \quad
\langle  0
{\mathcal{T}}_{z} \vert
\frac{1\pm\tau^{3}_{3}}{2}
\vert  1
{\mathcal{T}}_{z} \rangle =  \pm \frac{1}{\sqrt{3}} {\mathcal{T}}_{z} \, . 
\end{equation}

$F^{(33)\nu}_{\rho'_{3} \lambda'_{3} \rho_{3} \lambda_{3}}$ in (\ref{cap3_12sd67_1}) and $F^{(33)\nu\tau}_{\rho'_{3} \lambda'_{3} \rho_{3} \lambda_{3}}$ in (\ref{cap3_12sd67_2}) contain factors of the form $\frac{m}{E(\tilde{p})}\frac{1}{( \rho_3 + 1 ) E(\tilde{p})-W(q)}$ (and the analogous expression with primed variables). These factors originate from the $\rho$-spin decomposition (\ref{oneprop}) of off-shell nucleon propagators. We introduce
\begin{equation}
\frac{m}{E(\tilde{p})} g^\rho (q,\tilde{p}) \equiv \frac{m}{E(\tilde{p})}\frac{1}{( \rho + 1 ) E(\tilde{p})-W(q)}
\end{equation}
as the $\rho$-spin propagator of nucleon 3.

The positive $\rho$-spin propagators 
$g^{+} (q,\tilde{p})$ and 
$g^{+} (q',\tilde{p}')$ are finite since  $2 E(\tilde{p}') - W (  q ) \ge 3m - M_{t} > 0$.
On the other hand, the negative $\rho$-spin propagator develops a singularity as $q$ approaches $q_s$. In the vicinity of $q_s$, 
\begin{equation}
g^{-} (q,\tilde{p})= -\frac{1}{W (  q ) } \sim -\frac{1}{  \sqrt{\frac{2M_{t}q_{s}}{E(q_{s})} (q_{s} - q ) } } \, . \label{uikj3}
\end{equation}
Similarly,
\begin{equation}
g^{-} (q',\tilde{p}')=-\frac{1}{W (  q' )} \sim -\frac{1}{  \sqrt{ 2 K q ( b - \cos \varTheta ) } }  \label{uikj4} 
\end{equation}
as $ \varTheta \to \arccos b$ and $ a \le q < q_{s} $. These propagators are only weakly singular, and they are still multiplied by the vertex function which goes to zero at the singularity. The total result is finite.

The integrations over $\varPhi$ and $\bar{\phi}$ in Eq. \eqref{cap3_789gfq} can be done analytically.
The only $\varPhi$-dependence in the integrand of \eqref{cap3_789gfq} resides in the rotation matrix 
$( R_{\varPhi , 0 , 0} )^{\mu}_{\phantom{\mu}{\nu}} $ and in a factor $ e^{-i(M'-M)\varPhi}$ from $\mathcal{D}^{(1/2)}_{M',m'-\lambda'_{1}}( \varPhi' , \varTheta' , 0 )
\mathcal{D}^{(1/2)\ast}_{M,m-\lambda_{1}}( \varPhi , \varTheta , 0 )$. The product is easily integrated:
\begin{align}
\int_{0}^{2\pi} & d \varPhi e^{-i(M'-M)\varPhi}  R_{\varPhi , 0 , 0}  =
( 2 - \abs{M'-M} ) \pi 
\left (
\begin{array}{cccc}
1 - \abs{M'-M} & 0 & 0 & 0 \\
0 & \abs{M'-M} & i ( M' - M ) & 0 \\
0 & - i ( M' - M ) & \abs{M'-M} & 0 \\
0 & 0 & 0 & 1 - \abs{M'-M}
\end{array}
\right ) \, . \label{cap3_qw56nm}
\end{align}
Next, the integrand of \eqref{cap3_789gfq} depends on 
$\bar{\phi}$ through $ \tilde{\phi}' $, $ \tilde{\theta}'$, $ \tilde{\phi} $, $ \tilde{\theta} $, which
appear in the factors 
$\mathcal{D}^{(j')}_{m',\lambda'_{2}-\lambda'_{3}}( \tilde{\phi}' , \tilde{\theta}' , 0 )$,
$\mathcal{D}^{(j)\ast}_{m,\lambda_{2}-\lambda_{3}}( \tilde{\phi} , \tilde{\theta} , 0 )$,
$ F^{(22)}_{\lambda'_{2}\lambda_{2}}    $,
$ F^{(33)\nu}_{\rho'_{3}\lambda'_{3}\rho_{3}\lambda_{3}} $,
$ F^{(33)\nu\tau}_{\rho'_{3}\lambda'_{3}\rho_{3}\lambda_{3}} $ and
$ G^{(33)\nu}_{\rho'_{3}\lambda'_{3}\rho_{3}\lambda_{3}} $.
We can write those spinor matrix elements as
\begin{gather}
\begin{split}
& F^{(22)}_{\lambda'_{2}\lambda_{2}} = 
\sum_{\xi'_{2} \xi_{2} = \pm 1/2}
\sum_{\tau'_{2} \tau_{2} = \pm }
\mathcal{D}^{(1/2)\ast}_{  \xi'_{2}  ,\lambda'_{2} }( \tilde{\phi}' , \tilde{\theta}' , 0 )
\mathcal{D}^{(1/2)}_{ \xi_{2}  ,\lambda_{2}    }( \tilde{\phi} , \tilde{\theta} , 0 )
\tau'_{2} \tau_{2}
[\bar{u}^{\tau'_{2}}(0,\lambda'_{2}) 
u^{+}(\tilde{p}',\lambda'_{2})]  \\
& \quad \times
[\bar{u}^{\tau_{2}}(0,\lambda_{2}) 
u^{+}(\tilde{p},\lambda_{2})]
\bar{u}^{\tau'_{2}}(0,\xi'_{2})  S^{-1}( Z(  q' ) )
S^{-1} ( R_{0,\varTheta',0} ) S^{-1}( B ( \eta \hat{e}^{3} ) )  
S ( R_{0,\varTheta,0} )   
S( Z(  q ) )
u^{\tau_{2}}(0,\xi_{2}) \, , \end{split} \label{cap3_12rtg_1} \\
\begin{split}
& F^{(33)\nu}_{\rho'_{3}\lambda'_{3}\rho_{3}\lambda_{3}} = 
\frac{m}{E(\tilde{p}')} g^{\rho'_3}(q',\tilde{p}')
\frac{m}{E(\tilde{p})} g^{\rho_3}(q,\tilde{p})
\sum_{\xi'_{3} \xi_{3} = \pm 1/2} 
\sum_{\tau'_{3} \tau_{3} = \pm }
\mathcal{D}^{(1/2)\ast}_{  \xi'_{3}  ,-\lambda'_{3} }( \tilde{\phi}' , \tilde{\theta}' , 0 )
\mathcal{D}^{(1/2)}_{ \xi_{3}  ,-\lambda_{3}    }( \tilde{\phi} , \tilde{\theta} , 0 )
 \tau'_{3}
[\bar{u}^{\rho'_{3}}(0,\lambda'_{3}) 
u^{\tau'_{3}}(\tilde{p}',\lambda'_{3})]  \\
& \quad \times 
\bar{u}^{\tau'_{3}}(0,\xi'_{3})  S^{-1}( Z(  q' ) ) 
S^{-1} ( R_{0,\varTheta',0} ) S^{-1}( B ( \eta \hat{e}^{3} ) )  \gamma^{\nu} 
S ( R_{0,\varTheta,0} )   
S( Z(  q ) )
u^{\tau_{3}}(0,\xi_{3}) 
\tau_{3}
[\bar{u}^{\rho_{3}}(0,\lambda_{3}) 
u^{\tau_{3}}(\tilde{p},\lambda_{3})]   
\, , \end{split} \\
\begin{split}
& F^{(33)\nu\tau}_{\rho'_{3}\lambda'_{3}\rho_{3}\lambda_{3}} = 
\frac{m}{E(\tilde{p}')} g^{\rho'_3}(q',\tilde{p}')
\frac{m}{E(\tilde{p})} g^{\rho_3}(q,\tilde{p})
\sum_{\xi'_{3} \xi_{3} = \pm 1/2} 
\sum_{\tau'_{3} \tau_{3} = \pm }
\mathcal{D}^{(1/2)\ast}_{  \xi'_{3}  ,-\lambda'_{3} }( \tilde{\phi}' , \tilde{\theta}' , 0 )
\mathcal{D}^{(1/2)}_{ \xi_{3}  ,-\lambda_{3}    }( \tilde{\phi} , \tilde{\theta} , 0 )
 \tau'_{3}
[\bar{u}^{\rho'_{3}}(0,\lambda'_{3}) 
u^{\tau'_{3}}(\tilde{p}',\lambda'_{3})]  \\
& \quad \times 
\bar{u}^{\tau'_{3}}(0,\xi'_{3})  S^{-1}( Z(  q' ) ) 
S^{-1} ( R_{0,\varTheta',0} ) S^{-1}( B ( \eta \hat{e}^{3} ) )  i \sigma^{\nu\tau}
S ( R_{0,\varTheta,0} )   
S( Z(  q ) )
u^{\tau_{3}}(0,\xi_{3}) 
\tau_{3}
[\bar{u}^{\rho_{3}}(0,\lambda_{3}) 
u^{\tau_{3}}(\tilde{p},\lambda_{3})]   
\, , \end{split} \\
\begin{split}
& G^{(33)\nu}_{\rho'_{3}\lambda'_{3}\rho_{3}\lambda_{3}} = 
\frac{m}{2E^{2}(\tilde{p}')}
\frac{m}{2E^{2}(\tilde{p})}
\sum_{\xi'_{3} \xi_{3} = \pm 1/2}
\sum_{\tau'_{3} \tau_{3} \omega'_{3} \omega_{3}= \pm }
\mathcal{D}^{(1/2)\ast}_{  \xi'_{3}  ,-\lambda'_{3} }( \tilde{\phi}' , \tilde{\theta}' , 0 )
\mathcal{D}^{(1/2)}_{ \xi_{3}  ,-\lambda_{3}    }( \tilde{\phi} , \tilde{\theta} , 0 )
\tau'_{3} 
O_{ \rho'_{3} \omega'_{3} }( \tilde{p}' , \lambda'_{3} ) 
[\bar{u}^{\omega'_{3}}(0,\lambda'_{3}) u^{\tau'_{3}}(\tilde{p}',\lambda'_{3})] \\
& \quad \times  
\bar{u}^{\tau'_{3}}(0,\xi'_{3})  S^{-1}( Z(  q' ) )
S^{-1} ( R_{0,\varTheta',0} ) S^{-1}( B ( \eta \hat{e}^{3} ) ) \gamma^{\nu} 
S ( R_{0,\varTheta,0} )  
S( Z(  q ) )
u^{\tau_{3}}(0,\xi_{3}) 
 \tau_{3}
O_{ \rho_{3} \omega_{3} }( \tilde{p} , \lambda_{3} )
[\bar{u}^{\omega_{3}}(0,\lambda_{3}) u^{\tau_{3}}(\tilde{p},\lambda_{3})]   
\, . \end{split} \label{cap3_12rtg_4}
\end{gather}
Hence the integration variable $\bar{\phi}$ appears only in the product
\begin{align}
& \mathcal{D}^{(j')}_{m',\lambda'_{2}-\lambda'_{3}}( \tilde{\phi}' , \tilde{\theta}' , 0 )
\mathcal{D}^{(j)\ast}_{m,\lambda_{2}-\lambda_{3}}( \tilde{\phi} , \tilde{\theta} , 0 )
\mathcal{D}^{(1/2)\ast}_{  \xi'_{2}  ,\lambda'_{2} }( \tilde{\phi}' , \tilde{\theta}' , 0 )
\mathcal{D}^{(1/2)}_{ \xi_{2}  ,\lambda_{2}    }( \tilde{\phi} , \tilde{\theta} , 0 )
\mathcal{D}^{(1/2)\ast}_{  \xi'_{3}  ,-\lambda'_{3} }( \tilde{\phi}' , \tilde{\theta}' , 0 ) 
\mathcal{D}^{(1/2)}_{ \xi_{3}  ,-\lambda_{3}    }( \tilde{\phi} , \tilde{\theta} , 0 )  \label{cap3_123d}
\end{align}
which can be written as
\begin{align}
& (-1)^{m-\lambda_{2}+\lambda_{3} -\lambda'_{2}+\lambda'_{3}+\xi'_{2}+\xi'_{3}}
\sum_{s'=-j'}^{j'}
\sum_{s=-j}^{j}
\sum_{\beta'_{2}\beta'_{3} \beta_{2}\beta_{3}=\pm 1/2}
e^{-i(s'+\beta'_{2}+\beta'_{3}+ s+\beta_{2}+\beta_{3})\bar{\phi}}
d^{(j')}_{m's'}(  - \vartheta_{4}  ) \notag \\
& \quad \times
d^{(j')}_{s',\lambda'_{2} - \lambda'_{3}} ( \bar{\theta} )
d^{(1/2)}_{ - \xi'_{2} \beta'_{2}   }( - \vartheta_{4} ) 
d^{(1/2)}_{ \beta'_{2}  - \lambda'_{2} }( \bar{\theta} )
d^{(1/2)}_{ - \xi'_{3} \beta'_{3}   }( - \vartheta_{4} ) 
d^{(1/2)}_{ \beta'_{3}   \lambda'_{3} }( \bar{\theta} )
d^{(j)}_{-m s}(  \vartheta_{1} + \vartheta_{3}  ) 
d^{(j)}_{s,-\lambda_{2}+\lambda_{3}}(  \vartheta_{5}  ) \notag \\
& \quad \times
d^{(1/2)}_{\xi_{2} \beta_{2}}(  \vartheta_{1} + \vartheta_{3})
d^{(1/2)}_{ \beta_{2}   \lambda_{2}}(  \vartheta_{5}  )
d^{(1/2)}_{ \xi_{3} \beta_{3}  }(  \vartheta_{1} + \vartheta_{3}  )
d^{(1/2)}_{ \beta_{3} - \lambda_{3}  }(  \vartheta_{5}  ) \, . \label{alex2001}
\end{align}
The $\bar{\phi}$ dependence is now isolated in the factor $e^{-i(s'+\beta'_{2}+\beta'_{3}+ s+\beta_{2}+\beta_{3})\bar{\phi}}$, and the analytical integration over $\bar{\phi}$ has become straightforward.

\subsection{Calculation of the sum of diagrams B and C}

\indent

Diagrams B and C are closely related by symmetry. We start with diagram C by evaluating its momentum-conserving Dirac delta functions,
\begin{multline} 
\delta^{3}
\Bigl( 
B ( \eta \hat{e}^{3} ) 
R_{ \varPhi' , \varTheta' , 0 } 
R_{ \pi , \pi , 0 }
B ( \xi(q') \hat{e}^{3} )
v( m , 0 )  
-
R_{ \varPhi , \varTheta , 0 } 
R_{ \pi , \pi , 0 }
B ( \xi(q) \hat{e}^{3} )
v( m , 0 )  
\Bigr)  \\
 \times
\delta^{4}
\Bigl( 
B ( \eta \hat{e}^{3} ) 
R_{ \varPhi' , \varTheta' , 0 } 
Z(  q' )
R_{ \tilde{\phi}' , \tilde{\theta}' , 0 }
R_{ \pi , \pi , 0 }  
v( W(q') - E(\tilde{p}') , \tilde{p}' ) 
-
R_{ \varPhi , \varTheta , 0 } 
Z(  q )
R_{ \tilde{\phi} , \tilde{\theta} , 0  } 
R_{ \pi , \pi , 0 } 
v( W(q) - \tilde{p}^{0} , \tilde{p} )  \Bigr) \, . \label{cap3_78tr09k}
\end{multline}
The first delta function in \eqref{cap3_78tr09k} is the same as the one in \eqref{cap3_45ty34} of diagram A. Therefore 
$ \varPhi' = \varPhi   $,
$ \varTheta' = \pi - \varTheta_{1}$, $q' = r_{1}$, and the interval $I$ for the $\varTheta$-integration is also the same as in the case of diagram A.

We use the variables $ \vartheta_{1} $, $\vartheta_{3}$, $s_{2}$ 
defined in Eqs.~\eqref{zzx1}, \eqref{zzx2},
and $\bar{\phi}$, $\bar{\theta}$, $ \tilde{\phi}'   $, $ \tilde{\theta}'   $ which are related through \eqref{zzx3}.
Now we introduce $s_{3}$ and $ \vartheta_{5} $ (with $ 0 \le \vartheta_{5} \le \pi $),
\begin{gather}
s_{3} = \sqrt{ ( \tilde{p}' \sin \bar{\theta} )^{2} + 
\left ( [ W(q') - E(\tilde{p}'] ) \frac{s_{2}}{m}    
-
\tilde{p}'\cos\bar{\theta} \frac{\sqrt{ m^{2} + s_{2}^{2} }}{m}    
\right )^{2}           
} \,  ,              \\
s_{3} \cos \vartheta_{5} 
= 
- 
\left[ W(q') - E(\tilde{p}') \right]  \frac{s_{2}}{m}
+
\tilde{p}'\cos\bar{\theta} \frac{\sqrt{ m^{2} + s_{2}^{2} }}{m}       
 \, . 
\end{gather} 
Next we define $ \phi_{1} $, $ \vartheta_{6} $, $ \phi_{2} $, $s_{4}$ by
\begin{gather}
R_{ 0 , \vartheta_{1} + \vartheta_{3} , \bar{\phi} }
R_{ 0 , \vartheta_{5} , 0 }
= R_{ \phi_{1} , \vartheta_{6} , \phi_{2} } \, , \\
s_{4} =  W(q) - [ W(q') - E(\tilde{p}') ] \frac{\sqrt{ m^{2}+s^{2}_{2} }}{m} 
+ \tilde{p}' \cos \bar{\theta}  \frac{s_{2}}{m}   \, .
\end{gather}
Evaluating the second Dirac delta function in Eq.~\eqref{cap3_78tr09k} in the way it was shown for diagram A,
we obtain $\tilde{p}^{0} = s_{4}$,
$ \tilde{p} = s_{3}    $, $ \tilde{\phi} = \phi_{1}      $,
$  \tilde{\theta} =  \vartheta_{6}       $.

Diagram C can now be written as an integral over
$q$, $ \varTheta $, $ \tilde{p}' $, $ \bar{\theta} $,
$\bar{\phi}$, $ \varPhi $:
\begin{align}
\langle M' | J^\mu_C | M \rangle &= 
\frac{3m^{2}}{2(2\pi)^{8}} \int_{0}^{q_{s}} dq \frac{q^{2}}{E(q)} 
\int_{I} d\varTheta \sin \varTheta  \int_{0}^{+\infty} d \tilde{p}' 
\frac{\tilde{p}^{\prime2}}{E(\tilde{p}')} \int_{0}^{\pi} d \bar{\theta} \sin \bar{\theta} \int_{0}^{2\pi} d \bar{\phi}
\int_{0}^{2\pi} d \varPhi
\sum_{ \substack{ T' j' m' \lambda_{1}'  \lambda_{2}' \lambda_{3}' \rho_{3}' \\  T j m \lambda_{1}  \lambda_{2} \lambda_{3} \rho_{2} \rho_{3} } }
\notag \\ & \quad
\sqrt{2j'+1}  \sqrt{2j+1}
\mathcal{D}^{(1/2)}_{M',m'-\lambda'_{1}}( \varPhi' , \varTheta' , 0 )
\mathcal{D}^{(1/2)\ast}_{M,m-\lambda_{1}}( \varPhi , \varTheta , 0 )
\mathcal{D}^{(j')}_{m',\lambda'_{2}-\lambda'_{3}}( \tilde{\phi}' , \tilde{\theta}' , 0 )
\notag \\ & \quad \times
\mathcal{D}^{(j)\ast}_{m,\lambda_{2}-\lambda_{3}}( \tilde{\phi} , \tilde{\theta} , 0 )
C(  q' E(\tilde{p}^{\prime}) \tilde{p}' M' j' m' \lambda'_{1} \lambda'_{2} \lambda'_{3}  + \rho'_{3} T' \mathcal{T}_{z}  )
C(  q \tilde{p}^{0} \tilde{p} M j m \lambda_{1} \lambda_{2} \lambda_{3}  \rho_{2} \rho_{3} T \mathcal{T}_{z}  )
\notag \\ & \quad \times
F^{(11)}_{\lambda'_{1}\lambda_{1}}
( R_{ \varPhi , 0 , 0 })^{\mu}_{\phantom{\mu}{\nu}}
\left( F^{(1)}_{T'T\mathcal{T}_{z}} F^{(22)\nu}_{\lambda'_{2}\rho_{2}\lambda_{2}}
+ F^{(2)}_{T'T\mathcal{T}_{z}} F^{(22)\nu\tau}_{\lambda'_{2}\rho_{2}\lambda_{2}} \qph_{\tau} \right) 
F^{(33)}_{\rho'_{3}\lambda'_{3}\rho_{3}\lambda_{3}}\, , \label{cap3_mk34f}
\end{align}
where $ F^{(11)}_{\lambda'_{1}\lambda_{1}} $ is defined in Eq. \eqref{cap3_4lpo}. The remaining spinor matrix elements are
\begin{gather}
\begin{split}
F^{(22)\nu}_{\lambda'_{2}\rho_{2} \lambda_{2}} &=
\frac{ \tilde{p}^{0} + \rho_{2} E(\tilde{p}) }{E^{2}(\tilde{p})-(\tilde{p}^0)^2} 
\frac{m}{E(\tilde{p})}  
\bar{u}^{+}(\tilde{p}',\lambda'_{2}) 
S^{-1}(R_{\tilde{\phi}',\tilde{\theta}',0}) S^{-1}( Z(  q' ) ) 
S^{-1} ( R_{0,\varTheta',0} )\\ 
& \quad \times 
S^{-1}( B ( \eta \hat{e}^{3} ) )   \gamma^{\nu}
S ( R_{0,\varTheta,0} )  S( Z(  q ) )
S(R_{\tilde{\phi},\tilde{\theta},0}) 
u^{\rho_{2}}(\tilde{p},\lambda_{2}) \, , \end{split} \label{cap3_23rt78_1} \\
\begin{split}
F^{(22)\nu\tau}_{\lambda'_{2}\rho_{2} \lambda_{2}} &=
\frac{ \tilde{p}^{0} + \rho_{2} E(\tilde{p})  }{E^{2}(\tilde{p})-(\tilde{p}^0)^2} 
\frac{m}{E(\tilde{p})}  
\bar{u}^{+}(\tilde{p}',\lambda'_{2}) 
S^{-1}(R_{\tilde{\phi}',\tilde{\theta}',0}) S^{-1}( Z(  q' ) ) 
S^{-1} ( R_{0,\varTheta',0} ) \\
& \quad \times
S^{-1}( B ( \eta \hat{e}^{3} ) )   i\sigma^{\nu\tau}
S ( R_{0,\varTheta,0} )  S( Z(  q ) )
S(R_{\tilde{\phi},\tilde{\theta},0})
u^{\rho_{2}}(\tilde{p},\lambda_{2}) \, , \end{split} \label{cap3_23rt78_2} \\
\begin{split}
F^{(33)}_{\rho'_{3} \lambda'_{3} \rho_{3} \lambda_{3}} &= 
 \frac{m}{E(\tilde{p}')} g^{\rho'_3}(q',\tilde{p}')
\left ( \frac{m}{E(\tilde{p})} \right )^{2}
\sum_{ \sigma_{3} = \pm}
O_{\rho_{3} \sigma_{3}}( \tilde{p} , \lambda_{3} )
\bar{u}^{\rho'_{3}}(\tilde{p}',\lambda'_{3})  \\
& \quad \times
S^{-1}(R_{ \pi , \pi ,0})
S^{-1}(R_{\tilde{\phi}',\tilde{\theta}',0}) 
S^{-1}( Z(  q' ) )
S^{-1} ( R_{0,\varTheta',0} ) S^{-1}( B ( \eta \hat{e}^{3} ) ) 
S ( R_{0,\varTheta,0} )  \\
& \quad \times  S( Z(  q ) )
S(R_{\tilde{\phi},\tilde{\theta},0}) 
S(R_{ \pi , \pi , 0  }) 
u^{\sigma_{3}}(\tilde{p},\lambda_{3}) \, , \end{split} \label{cap3_yyy7} \\ 
\end{gather}
and the isospin dependent factors are
\begin{align}
F^{(1)}_{T'T\mathcal{T}_{z}} &= \tilde{f} ( m^{2} , q_{2}^{2} )
\left(
F_{1p} ( Q^{2} )
\langle  T'
{\mathcal{T}}_{z} \vert
\frac{1+\tau^{3}_{2}}{2}
\vert  T
{\mathcal{T}}_{z} \rangle 
+
F_{1n} ( Q^{2} )
\langle  T'
{\mathcal{T}}_{z} \vert
\frac{1-\tau^{3}_{2}}{2}
\vert  T
{\mathcal{T}}_{z} \rangle
\right) \, , \\
F^{(2)}_{T'T\mathcal{T}_{z}} &= \tilde{g} ( m^{2} , q_{2}^{2} )
\left(
F_{2p} ( Q^{2} )
\langle  T'
{\mathcal{T}}_{z} \vert
\frac{1+\tau^{3}_{2}}{2}
\vert  T
{\mathcal{T}}_{z} \rangle 
+
F_{2n} ( Q^{2} )
\langle  T'
{\mathcal{T}}_{z} \vert
\frac{1-\tau^{3}_{2}}{2}
\vert  T
{\mathcal{T}}_{z} \rangle
\right) \, , 
\end{align}
with 
\begin{equation}
q_{2} = \bigl (   \tilde{p}^{0} , 0 , 0 , \tilde{p}  \bigr ) \, . \label{qwsa}
\end{equation}
The matrix elements involving the isospin $z$-projection of nucleon 2, $ \tau^{3}_{2}$, are
\begin{equation}
\langle  0
{\mathcal{T}}_{z} \vert
\frac{1\pm\tau^{3}_{2}}{2}
\vert  0
{\mathcal{T}}_{z} \rangle = \frac{1}{2} \, , \quad
\langle  1
{\mathcal{T}}_{z} \vert
\frac{1\pm\tau^{3}_{2}}{2}
\vert  1
{\mathcal{T}}_{z} \rangle = \frac{1}{2} \pm \frac{2}{3} {\mathcal{T}}_{z} \, , \quad
\langle  1
{\mathcal{T}}_{z} \vert
\frac{1\pm\tau^{3}_{2}}{2}
\vert  0
{\mathcal{T}}_{z} \rangle =  \mp \frac{1}{\sqrt{3}} {\mathcal{T}}_{z} \, , \quad
\langle  0
{\mathcal{T}}_{z} \vert
\frac{1\pm\tau^{3}_{2}}{2}
\vert  1
{\mathcal{T}}_{z} \rangle =  \mp \frac{1}{\sqrt{3}} {\mathcal{T}}_{z} \, .
\end{equation}
Just as it was shown for \eqref{cap3_789gfq}, the integrations over $\varPhi$ and $ \bar{\phi}$ in  \eqref{cap3_mk34f} can again be done analytically.

We turn our attention now to the integration over $\bar{\theta}$. This is complicated due to the moving singularity contained in the factor
$ ( \tilde{p}^{0}+\rho_{2} E(\tilde{p}) ) / ( E^{2}(\tilde{p}) - (\tilde{p}^0)^2 ) $ in 
\eqref{cap3_23rt78_1} and \eqref{cap3_23rt78_2}, which has its origin in the off-shell propagator of nucleon 2. 
In this work, we calculate diagram C separately from B, but eliminate the imaginary part of the propagator singularity, which cancels with the corresponding one of diagram B, and keep only the principal value integral.
The energy denominator can be written as
\begin{equation}
E^{2}(\tilde{p}) - (\tilde{p}^0)^2 =
A E( \tilde{p}' ) - B \tilde{p}' \cos \bar{\theta} - C \, ,
\end{equation}
where
\begin{equation}
A =  2 W(q') - 2 W(q) \frac{\sqrt{m^{2}+s_{2}^{2}}}{m} \, , \qquad 
B = 2 W(q) \frac{s_{2}}{m} \, , \qquad
C =   \frac{A^{2} - B^{2}}{4} \, .
\end{equation}
$A$, $B$, and $C$ do not depend on $\tilde{p}'$ nor on $ \bar{\theta} $, and $ B > \abs{A} $.
The location of the singularity depends on the relations between $A$, $B$, and $\tilde{p}'$, and we have to distinguish between several cases.
We define
\begin{align}
p_{\pm} &= \frac{ B C \pm \abs{A} \sqrt{ C^{2} + m^{2} ( B^{2} - A^{2} )  } }{A^{2}-B^{2}}  \\
\intertext{and, for $ \tilde{p}' \ne 0 $,}
D (\tilde{p}') &= \frac{-C + A E( \tilde{p}' ) }{B \tilde{p}' }  \, .
\end{align}
We find that
the equation $E^{2}(\tilde{p}) - (\tilde{p}^0)^2=0$ in the unknown $\bar{\theta}$ 
has at least one solution if and only if
one of the following conditions is satisfied:
\begin{align}
& A \ge 0, \: \tilde{p}' \ge p_{-} \, , \label{oigres_1} \\
& A < 0, \: B < \sqrt{A^{2} + 4m \abs{A} }, \: \tilde{p}' \ge - p_{+}  \, , \label{oigres_2} \\
& A < 0, \: B = \sqrt{A^{2} + 4m \abs{A} }, \: \tilde{p}' > 0 \, , \\
& A < 0, \: B > \sqrt{A^{2} + 4m \abs{A} }, \: \tilde{p}' \ge p_{+} \label{oigres_4} \, , \\
& \tilde{p}' = 0 \, . \label{oigres_5} 
\end{align}
The respective minimum values $p_{-}$, $- p_{+}$, and $p_{+}$ of $\tilde{p}'$ in \eqref{oigres_1}, \eqref{oigres_2}, and \eqref{oigres_4} are all positive.
In any of the cases \eqref{oigres_1}-\eqref{oigres_4}, the equation $E^{2}(\tilde{p}) - (\tilde{p}^0)^2=0$
has only one solution, namely $\bar{\theta} = \arccos D (\tilde{p}')$.
In the case of \eqref{oigres_5}, the set of solutions of $E^{2}(\tilde{p}) - (\tilde{p}^0)^2=0$
is the interval $[0,\pi]$.

Now that the location of the singularity has been determined, one can divide the integration region into subintervals according to the four cases (\ref{oigres_1}) -- (\ref{oigres_4}) and apply standard numerical techniques (we used a subtraction method) to perform the principal value integration.  

Diagram B does not need to be calculated once diagram C is known, since it can be obtained from the latter using time reversal and parity transformation properties. It can be shown that the components of diagram B are given by
\begin{gather}
\begin{split}
\langle M'  \vert J_{B}^{0} \vert  M  \rangle 
&= 
( B ( \eta \hat{e}^{3} ) )^{0}_{\phantom{0}0}
\langle M'  \vert J_{C}^{0} \vert  M  \rangle
-
( B ( \eta \hat{e}^{3} ) )^{0}_{\phantom{0}3}
\langle M'  \vert J_{C}^{3} \vert  M  \rangle
\, , \end{split} \\ 
\langle M'  \vert J_{B}^{1} \vert  M  \rangle
 = 
\langle M'  \vert J_{C}^{1} \vert  M  \rangle
\, , \\ 
\langle M'  \vert J_{B}^{2} \vert  M  \rangle
 =  
\langle M'  \vert J_{C}^{2} \vert  M  \rangle
\, , \\ 
\begin{split}
\langle M'  \vert J_{B}^{3} \vert  M  \rangle
& = 
- ( B ( \eta \hat{e}^{3} ) )^{3}_{\phantom{3}3} 
\langle M'  \vert J_{C}^{3} \vert  M  \rangle
+
( B ( \eta \hat{e}^{3} ) )^{3}_{\phantom{3}0}
\langle M'  \vert J_{C}^{0} \vert  M  \rangle
 \, . \end{split}
\end{gather}
%

\subsection{Calculation of diagram D}

\indent


The product of momentum conserving Dirac delta functions in diagram D is
\begin{align}
& 
\delta^{3}
\Bigl( 
B ( \eta \hat{e}^{3} ) 
R_{ \varPhi' , \varTheta' , 0 } 
Z(  q' )
R_{ \tilde{\phi}' , \tilde{\theta}' , 0 } 
B ( \xi(\tilde{p}') \hat{e}^{3} )
v( m , 0 )  
-
R_{ \varPhi , \varTheta , 0 } 
R_{ \pi , \pi , 0 }
B ( \xi(q) \hat{e}^{3} ) 
v( m , 0 )  
\Bigr)
\notag \\ & \quad \times
\delta^{3}
\Bigl( 
B ( \eta \hat{e}^{3} ) 
R_{ \varPhi' , \varTheta' , 0 } 
R_{ \pi , \pi , 0 }
B ( \xi(q') \hat{e}^{3} )
v( m , 0 )  
-
R_{ \varPhi , \varTheta , 0 } 
Z(  q ) 
R_{ \tilde{\phi} , \tilde{\theta} , 0  } 
B ( \xi(\tilde{p}) \hat{e}^{3} )
v( m , 0 )  
\Bigr) 
 \, . \label{cap3_56tr_b}
\end{align}
From the first  Dirac delta function in \eqref{cap3_56tr_b}
we obtain
$ \tilde{\phi}' = \varphi_{3}   $,
$ \tilde{\theta}' = \vartheta_{6}     $, and
$  \tilde{p}' = s_{2}     $, where
$  \varphi_{3}   $,
$  \vartheta_{6}     $,
$   s_{2}     $
are defined by applying
(\ref{prop_34er}) and (\ref{cap3_34rt78}):
\begin{align} 
Z(  -q' )
R_{ 0 , - \varTheta' , 0 }
B ( -\eta \hat{e}^{3} )
&=
R_{ 0 ,  \vartheta_{1} , 0 }
B ( \xi(s_{1}) \hat{e}^{3} )
R_{ 0 ,  \vartheta_{2} , 0 } \, , \\
R_{ 0 ,  \vartheta_{2} , 0 }
R_{ \varPhi -\varPhi' + \pi, \pi-\varTheta , 0 } 
&=
R_{ \varphi_{1} , \vartheta_{3} , \varphi_{2} }  \, , \\
B ( \xi(s_{1}) \hat{e}^{3} )
R_{ 0 ,  \vartheta_{3} , 0 }
B ( \xi(q) \hat{e}^{3} )
&=
R_{ 0 ,  \vartheta_{4} , 0 }
B ( \xi(s_{2}) \hat{e}^{3} )
R_{ 0 ,  \vartheta_{5} , 0 } \, , \\
R_{ 0 ,  \vartheta_{1} , \varphi_{1} }
R_{ 0 ,  \vartheta_{4} , 0 }
&=
R_{ \varphi_{3} ,  \vartheta_{6} , \varphi_{4} } \, .
\end{align}
The second delta function leads to
$ \tilde{\phi} = \phi_{3}   $,
$ \tilde{\theta} =  \theta_{6}    $, and
$ \tilde{p} = r_{2} $, 
where  --- again by using
(\ref{prop_34er}) and (\ref{cap3_34rt78}) ---
$  \phi_{3}   $,
$   \theta_{6}    $,
$  r_{2} $ are determined through
\begin{align} 
Z(  -q ) 
R_{ 0 , -\varTheta , 0 }  
B ( \eta \hat{e}^{3} ) 
&=
R_{ 0 , \theta_{1} , 0 }  
B ( \xi(r_{1}) \hat{e}^{3} )
R_{ 0 , \theta_{2} , 0 }  \, , \\
R_{ 0 , \theta_{2} , 0 }  
R_{ \varPhi' - \varPhi + \pi , \pi -\varTheta' , 0 } 
&=
R_{ \phi_{1} , \theta_{3} , \phi_{2} } \, , \\
B ( \xi(r_{1}) \hat{e}^{3} )
R_{ 0 , \theta_{3} , 0 }  
B ( \xi(q') \hat{e}^{3} )
&=
R_{ 0 , \theta_{4} , 0 }
B ( \xi(r_{2}) \hat{e}^{3} )
R_{ 0 , \theta_{5} , 0 } \, , \\
R_{ 0 , \theta_{1} , 0 }  
R_{ \phi_{1} , \theta_{4} , 0 }
&=
R_{ \phi_{3} , \theta_{6} , \phi_{4} }  \, .
\end{align}

We can write the current given by diagram D in the form
\begin{align}
\langle M' | J^\mu_D | M \rangle &= 
-\frac{3m^{2}}{(2\pi)^{8}} \int_{0}^{q_{s}} dq \frac{q^{2}}{E(q)}  
\int_{0}^{q_{s}} dq' \frac{q^{\prime2}}{E(q')} 
\int_{0}^{\pi} d\varTheta \sin \varTheta  \int_{0}^{\pi} d\varTheta' \sin \varTheta'
 \int_{0}^{2\pi} d \varPhi' \int_{0}^{2\pi} d \varPhi
\notag \\ & \quad
\sum_{ \substack{ T' j' m' \lambda_{1}'  \lambda_{2}' \lambda_{3}' \rho_{3}' \\  T j m \lambda_{1}  \lambda_{2} \lambda_{3} \rho_{3} } }
\sqrt{2j'+1}  \sqrt{2j+1}
\mathcal{D}^{(1/2)}_{M',m'-\lambda'_{1}}( \varPhi' , \varTheta' , 0 )
\mathcal{D}^{(1/2)\ast}_{M,m-\lambda_{1}}( \varPhi , \varTheta , 0 )
\mathcal{D}^{(j')}_{m',\lambda'_{2}-\lambda'_{3}}( \tilde{\phi}' , \tilde{\theta}' , 0 )
\notag \\ & \quad \times
\mathcal{D}^{(j)\ast}_{m,\lambda_{2}-\lambda_{3}}( \tilde{\phi} , \tilde{\theta} , 0 )
C(  q' E(\tilde{p}^{\prime}) \tilde{p}' M' j' m' \lambda'_{1} \lambda'_{2} \lambda'_{3}  + \rho'_{3} T' \mathcal{T}_{z}  )
C(  q E(\tilde{p}) \tilde{p} M j m \lambda_{1} \lambda_{2} \lambda_{3}  + \rho_{3} T \mathcal{T}_{z}  )
\notag \\ & \quad \times
F^{(12)}_{\lambda'_{1}\lambda_{2}} 
F^{(21)}_{\lambda'_{2}\lambda_{1}}
\left( R_{ \frac{\varPhi + \varPhi'}{2} , 0 , 0 }\right)^{\mu}_{\phantom{\mu}{\nu}}
\left( F^{(1)}_{T'T\mathcal{T}_{z}} F^{(33)\nu}_{\rho'_{3}\lambda'_{3}\rho_{3}\lambda_{3}}
+ F^{(2)}_{T'T\mathcal{T}_{z}} F^{(33)\nu\tau}_{\rho'_{3}\lambda'_{3}\rho_{3}\lambda_{3}} \qph_{\tau}
+ F^{(3)}_{T'T\mathcal{T}_{z}} G^{(33)\nu}_{\rho'_{3}\lambda'_{3}\rho_{3}\lambda_{3}} \right) \, , \label{cap3_aa12}
\end{align}
with the spinor matrix elements
\begin{gather}
\begin{split}
F^{(12)}_{\lambda'_{1}\lambda_{2}} &= \bar{u}^{+}(0,\lambda'_{1}) S^{-1}( B ( \xi(q') \hat{e}^{3} ) )
S^{-1}(R_{\pi,\pi,0}) S^{-1} ( R_{0,\varTheta',0} ) S^{-1}( B ( \eta \hat{e}^{3} ) )
S ( R_{\varPhi-\varPhi',\varTheta,0} ) \\
& \quad \times S( Z(  q ) )
S(R_{\tilde{\phi},\tilde{\theta},0}) S( B( \xi(\tilde{p}) \hat{e}^{3} ) )
u^{+}(0,\lambda_{2}) \, , \end{split} \\
\begin{split}
F^{(21)}_{\lambda'_{2}\lambda_{1}} &= \bar{u}^{+}(0,\lambda'_{2}) S^{-1}( B( \xi(\tilde{p}') \hat{e}^{3} ) )
S^{-1}(R_{\tilde{\phi}',\tilde{\theta}',0}) S^{-1}( Z(  q' ) )
S^{-1} ( R_{0,\varTheta',0} ) S^{-1}( B ( \eta \hat{e}^{3} ) )  \\
& \quad \times 
S ( R_{\varPhi-\varPhi',\varTheta,0} )S(R_{\pi,\pi,0})S( B ( \xi(q) \hat{e}^{3} ) )
u^{+}(0,\lambda_{1})
 \, , \end{split} \label{edfr} \\
\begin{split}
F^{(33)\nu}_{\rho'_{3} \lambda'_{3} \rho_{3} \lambda_{3}} &= 
 \frac{m}{E(\tilde{p}')} g^{\rho'_3}(q',\tilde{p}')
 \frac{m}{E(\tilde{p})} g^{\rho_3}(q,\tilde{p})
\bar{u}^{\rho'_{3}}(\tilde{p}',\lambda'_{3}) 
  \\
& \quad \times
S^{-1}(R_{ \pi , \pi ,0})
S^{-1}(R_{\tilde{\phi}',\tilde{\theta}',0}) 
S^{-1}( Z(  q' ) )
S^{-1} ( R_{0,\varTheta',0} ) S^{-1}( B ( \eta \hat{e}^{3} ) ) 
S ( R_{\frac{\varPhi-\varPhi'}{2},0,0} )   \\
& \quad \times
\gamma^{\nu}
S ( R_{\frac{\varPhi-\varPhi'}{2},0,0} )
S ( R_{0,\varTheta,0} )    S( Z(  q ) )
S(R_{\tilde{\phi},\tilde{\theta},0}) 
S(R_{ \pi , \pi , 0  }) 
u^{\rho_{3}}(\tilde{p},\lambda_{3}) \, , \end{split} \label{cap3_asd83u_1} \\
\begin{split}
F^{(33)\nu\tau}_{\rho'_{3} \lambda'_{3} \rho_{3} \lambda_{3}} &= 
 \frac{m}{E(\tilde{p}')}  g^{\rho'_3}(q',\tilde{p}')
 \frac{m}{E(\tilde{p})} g^{\rho_3}(q,\tilde{p})
\bar{u}^{\rho'_{3}}(\tilde{p}',\lambda'_{3}) 
  \\
& \quad \times
S^{-1}(R_{ \pi , \pi ,0})
S^{-1}(R_{\tilde{\phi}',\tilde{\theta}',0}) 
S^{-1}( Z(  q' ) )
S^{-1} ( R_{0,\varTheta',0} ) S^{-1}( B ( \eta \hat{e}^{3} ) ) 
S ( R_{\frac{\varPhi-\varPhi'}{2},0,0} )   \\
& \quad \times
i\sigma^{\nu\tau}
S ( R_{\frac{\varPhi-\varPhi'}{2},0,0} )
S ( R_{0,\varTheta,0} )    S( Z(  q ) )
S(R_{\tilde{\phi},\tilde{\theta},0}) 
S(R_{ \pi , \pi , 0  }) 
u^{\rho_{3}}(\tilde{p},\lambda_{3}) \, , \end{split} \label{cap3_asd83u_2} \\
\begin{split}
G^{(33)\nu}_{\rho'_{3} \lambda'_{3} \rho_{3} \lambda_{3}} &= 
\frac{1}{(2m)^{2}}
\left ( \frac{m}{E(\tilde{p}')} \right )^{2}
\left ( \frac{m}{E(\tilde{p})} \right )^{2}
\sum_{\sigma' \sigma = \pm}
O_{\rho'_{3} \sigma'}( \tilde{p}' , \lambda'_{3} )
O_{\rho_{3} \sigma}( \tilde{p} , \lambda_{3} )
\bar{u}^{\sigma'}(\tilde{p}',\lambda'_{3}) \\
& \quad \times
S^{-1}(R_{ \pi , \pi ,0})
S^{-1}(R_{\tilde{\phi}',\tilde{\theta}',0}) 
S^{-1}( Z(  q' ) )
S^{-1} ( R_{0,\varTheta',0} ) S^{-1}( B ( \eta \hat{e}^{3} ) ) \\
& \quad \times
S ( R_{\frac{\varPhi-\varPhi'}{2},0,0} ) \gamma^{\nu} S ( R_{\frac{\varPhi-\varPhi'}{2},0,0} )
S ( R_{0,\varTheta,0} ) S( Z(  q ) )
S(R_{\tilde{\phi},\tilde{\theta},0}) 
S(R_{ \pi , \pi , 0  }) 
u^{\sigma}(\tilde{p},\lambda_{3}) \, , \end{split} \label{cap3_cvb67j} 
\end{gather}
and the isospin-dependent factors
\begin{align}
F^{(1)}_{T'T\mathcal{T}_{z}} &= \tilde{f} ( q_{3}^{\prime 2} , q_{3}^{2} )
(
F_{1p} ( Q^{2} )
\langle  T'
{\mathcal{T}}_{z} \vert
P^{\mathrm{iso}}_{12} \frac{1+\tau^{3}_{3}}{2}
\vert  T
{\mathcal{T}}_{z} \rangle 
+
F_{1n} ( Q^{2} )
\langle  T'
{\mathcal{T}}_{z} \vert
P^{\mathrm{iso}}_{12}\frac{1-\tau^{3}_{3}}{2}
\vert  T
{\mathcal{T}}_{z} \rangle
) \, ,  \\
F^{(2)}_{T'T\mathcal{T}_{z}} &= \tilde{g} ( q_{3}^{\prime 2} , q_{3}^{2} )
(
F_{2p} ( Q^{2} )
\langle  T'
{\mathcal{T}}_{z} \vert
P^{\mathrm{iso}}_{12}\frac{1+\tau^{3}_{3}}{2}
\vert  T
{\mathcal{T}}_{z} \rangle 
+
F_{2n} ( Q^{2} )
\langle  T'
{\mathcal{T}}_{z} \vert
P^{\mathrm{iso}}_{12}\frac{1-\tau^{3}_{3}}{2}
\vert  T
{\mathcal{T}}_{z} \rangle
) \, ,  \\
F^{(3)}_{T'T\mathcal{T}_{z}} &= \tilde{h} ( q_{3}^{\prime 2} , q_{3}^{2} )
(
F_{3p} ( Q^{2} )
\langle  T'
{\mathcal{T}}_{z} \vert
P^{\mathrm{iso}}_{12}\frac{1+\tau^{3}_{3}}{2}
\vert  T
{\mathcal{T}}_{z} \rangle 
+
F_{3n} ( Q^{2} )
\langle  T'
{\mathcal{T}}_{z} \vert
P^{\mathrm{iso}}_{12}\frac{1-\tau^{3}_{3}}{2}
\vert  T
{\mathcal{T}}_{z} \rangle
) \, . 
\end{align}
The four-momenta $q'_{3}$ and $q_{3}$ are defined in \eqref{plko}, where now $q'$ is an independent variable,
and we have $\tilde{p}'=s_{2}$ and $\tilde{p}=r_{2}$. 
$P^{\mathrm{iso}}_{12}$ is the operator that interchanges the isospin states of nucleons 1 and 2, with the matrix elements
\begin{align}
\langle  0
{\mathcal{T}}_{z} \vert
P^{\mathrm{iso}}_{12}\frac{1\pm\tau^{3}_{3}}{2}
\vert  0
{\mathcal{T}}_{z} \rangle &=
\frac{1}{4} \mp \frac{1}{2} {\mathcal{T}}_{z} &  
\langle  1
{\mathcal{T}}_{z} \vert
P^{\mathrm{iso}}_{12}\frac{1\pm\tau^{3}_{3}}{2}
\vert  1
{\mathcal{T}}_{z} \rangle & =
-\frac{1}{4} \mp \frac{5}{6} {\mathcal{T}}_{z} \nonumber\\
\langle  1
{\mathcal{T}}_{z} \vert
P^{\mathrm{iso}}_{12}\frac{1\pm\tau^{3}_{3}}{2}
\vert  0
{\mathcal{T}}_{z} \rangle & =
-\frac{\sqrt{3}}{4} \mp \frac{\sqrt{3}}{6} {\mathcal{T}}_{z} &
\langle  0
{\mathcal{T}}_{z} \vert
P^{\mathrm{iso}}_{12}\frac{1\pm\tau^{3}_{3}}{2}
\vert  1
{\mathcal{T}}_{z} \rangle & =
-\frac{\sqrt{3}}{4} \mp \frac{\sqrt{3}}{6} {\mathcal{T}}_{z} \, .
\end{align}
In \eqref{cap3_aa12}, we change now integration variables from $\varPhi'$ and $\varPhi$ to $\phi$ and $\varphi$, with
\begin{equation} 
\phi = \frac{\varPhi - \varPhi'}{2} \, , \quad
\varphi = \frac{\varPhi + \varPhi'}{2} \,  , \label{cap3_gh_39}
\end{equation} 
which implies
\begin{equation}
\int_{0}^{2\pi} d \varPhi' \int_{0}^{2\pi} d \varPhi = 2 \int_{-\pi}^{\pi} d \phi \int_{\abs{\phi}}^{-\abs{\phi}+2\pi}
d \varphi \, .   \label{cap3_gh_39p}
\end{equation}
The only $\varphi$-dependence in the integrand comes from $\mathcal{D}^{(1/2)}_{M',m'-\lambda'_{1}}( \varPhi' , \varTheta' , 0 ) \mathcal{D}^{(1/2)\ast}_{M,m-\lambda_{1}}( \varPhi , \varTheta , 0 )$, which yields a factor
 $ e^{-i(M'-M)\varphi} ( R_{\varphi , 0 , 0} )^{\mu}_{\phantom{\mu}{\nu}} $ 
and can be integrated analytically. The three different cases are:

\noindent
if $  M' = M $
\begin{align}
\int_{\abs{\phi}}^{-\abs{\phi}+2\pi}
d \varphi
e^{-i(M'-M)\varphi}  R_{\varphi , 0 , 0}  =
\left (
\begin{array}{cccc}
2 \pi - 2 \abs{\phi} & 0 & 0 & 0 \\
0 & - 2 \sin \abs{\phi} & 0 & 0 \\
0 & 0 & - 2 \sin \abs{\phi} & 0 \\
0 & 0 & 0 & 2 \pi - 2 \abs{\phi}
\end{array}
\right ) \, , \label{cap3_zxc34_1}
\end{align}
if $  M' = - M = 1/2 $
\begin{align}
& \int_{\abs{\phi}}^{-\abs{\phi}+2\pi}
d \varphi
e^{-i(M'-M)\varphi}  R_{\varphi , 0 , 0}  = 
\left (
\begin{array}{cccc}
- 2 \sin \abs{\phi} & 0 & 0 & 0 \\
0 & - \abs{\phi} + \pi - \frac{1}{2} \sin ( 2 \abs{\phi} ) & i [  - \abs{\phi} + \pi + \frac{1}{2} \sin ( 2 \abs{\phi} )       ] & 0 \\
0 & - i  [  - \abs{\phi} + \pi + \frac{1}{2} \sin ( 2 \abs{\phi} )       ] & - \abs{\phi} + \pi - \frac{1}{2} \sin ( 2 \abs{\phi} ) & 0 \\
0 & 0 & 0 & - 2 \sin \abs{\phi}
\end{array}
\right ) \, , \label{cap3_zxc34_2}
\end{align}
if $  M' = - M = - 1/2 $
\begin{align}
& \int_{\abs{\phi}}^{-\abs{\phi}+2\pi}
d \varphi
e^{-i(M'-M)\varphi}  R_{\varphi , 0 , 0}  = 
\left (
\begin{array}{cccc}
- 2 \sin \abs{\phi} & 0 & 0 & 0 \\
0 & - \abs{\phi} + \pi - \frac{1}{2} \sin ( 2 \abs{\phi} ) & - i [  - \abs{\phi} + \pi + \frac{1}{2} \sin ( 2 \abs{\phi} )       ] & 0 \\
0 &  i  [  - \abs{\phi} + \pi + \frac{1}{2} \sin ( 2 \abs{\phi} )       ] & - \abs{\phi} + \pi - \frac{1}{2} \sin ( 2 \abs{\phi} ) & 0 \\
0 & 0 & 0 & - 2 \sin \abs{\phi}
\end{array}
\right ) \, . \label{cap3_zxc34_3}
\end{align}
As in the case of diagram A, there is no problem with propagator singularities and the numerical integrations can be performed in a straighforward manner.

\subsection{Calculation of the sum of diagrams E and F}

\noindent


In close analogy to diagrams B and C, also diagrams E and F are related by symmetry. We start with the calculation of diagram F and relate it later to diagram E.
Here, the product of momentum conserving delta functions is
\begin{align}
& 
\delta^{3}
\Bigl( 
B ( \eta \hat{e}^{3} ) 
R_{ \varPhi' , \varTheta' , 0 } 
Z(  q' )
R_{ \tilde{\phi}' , \tilde{\theta}' , 0 } 
B ( \xi(\tilde{p}') \hat{e}^{3} ) 
v( m , 0 )  
-
R_{ \varPhi , \varTheta , 0 } 
R_{ \pi , \pi , 0 }  
B ( \xi(q) \hat{e}^{3} ) 
v( m , 0 )  
\Bigr)
\notag \\ & \quad \times
\delta^{4}
\Bigl(
R_{ \varPhi , \varTheta , 0 } 
Z(  q ) 
R_{ \tilde{\phi} , \tilde{\theta} , 0  } 
v( \tilde{p}^{0} , \tilde{p} )  
- 
B ( \eta \hat{e}^{3} ) 
R_{ \varPhi' , \varTheta' , 0 } 
R_{ \pi , \pi , 0 } 
B ( \xi(q') \hat{e}^{3} )
v( m , 0 ) + \qph
\Bigr) \, . \label{cap3_34tryuj}
\end{align}
The first delta  functions is the same as the one in \eqref{cap3_34tryuj}, hence
$ \tilde{\phi}' = \varphi_{3}   $,
$ \tilde{\theta}' = \vartheta_{6}     $ and
$  \tilde{p}' = s_{2}     $. 

We define $r_{1}$ and $ \theta_{1} $ as
\begin{gather}
r_{1} = \sqrt{ q^{\prime 2} \sin^{2} \varTheta' + \left ( \frac{K}{M_{t}} E(q') - 
\frac{\sqrt{ M_{t}^{2} + K^{2} } }{ M_{t} } q' \cos \varTheta' - K           \right )^{2}        } \, ,\\
r_{1}  \cos \theta_{1} = 
\frac{K}{M_{t}} E(q') - 
\frac{\sqrt{ M_{t}^{2} + K^{2} } }{ M_{t} } q' \cos \varTheta' - K \, ,\qquad ( 0 \le \theta_{1} \le \pi)          \, ,
\end{gather}
and introduce $\phi_{1}$, $\phi_{2}$, $\theta_{2}$ by applying (\ref{cap3_34rt78}),
\begin{equation}
R_{ 0 , - \varTheta , -  \varPhi + \varPhi' + \pi  } 
R_{ 0 , \theta_{1} , 0 } 
=
R_{ \phi_{1} , \theta_{2} , \phi_{2} } \, .
\end{equation}
From the second Dirac delta  function in Eq. \eqref{cap3_34tryuj} we obtain
\begin{align}
\tilde{p}^{0} &=
\frac{\sqrt{ W^{2} (  q ) + q^{2} }}{W (  q )}
\Biggl (
E(q') \frac{\sqrt{ M_{t}^{2} + K^{2} } }{ M_{t} } - q' \cos \varTheta' \frac{K}{ M_{t} } - \sqrt{K^2-Q^2}
\Biggr ) 
-
\frac{q}{W (  q )}  r_{1} \cos \theta_{2} \, , \label{mkjn}
\end{align}
as well as $ \tilde{\phi} = \phi_{1}    $, $  \tilde{\theta}  = \theta_{3}   $, and
$\tilde{p} = r_{2}$, where
\begin{gather}
r_{2} = \Biggl [
r_{1}^{2} \sin^{2} \theta_{2} +
\Biggl (
- \frac{q}{W (  q )} 
\Biggl (
E(q') \frac{\sqrt{ M_{t}^{2} + K^{2} } }{ M_{t} } - q' \cos \varTheta' \frac{K}{ M_{t} } - \sqrt{K^2-Q^2}
\Biggr ) 
+
\frac{\sqrt{ W^{2} ( q ) + q^{2} }}{W ( q )} r_{1} \cos \theta_{2}
\Biggr )^{2}
\Biggr ]^{1/2} \, , \\
r_{2} \cos \theta_{3} =
- \frac{q}{W (  q )} 
\Biggl (
E(q') \frac{\sqrt{ M_{t}^{2} + K^{2} } }{ M_{t} } - q' \cos \varTheta' \frac{K}{ M_{t} } - \sqrt{K^2-Q^2}
\Biggr ) 
+
\frac{\sqrt{ W^{2} (  q ) + q^{2} }}{W (  q )} r_{1} \cos \theta_{2} \, , \qquad (0 \le \theta_{3} \le \pi) \, . 
\end{gather}
The current matrix elements of diagram F are given by
\begin{align}
\langle M' | J^\mu_F | M \rangle &= 
-\frac{3m^{2}}{ (2\pi)^{8}} \int_{0}^{q_{s}} dq \frac{q^{2}}{E(q)}  
\int_{0}^{\pi} d\varTheta \sin \varTheta 
\int_{0}^{2\pi} d \varPhi' \int_{0}^{2\pi} d \varPhi
\int_{0}^{q_{s}} dq' \frac{q^{\prime2}}{E(q')}
\int_{0}^{\pi} d\varTheta' \sin \varTheta'
\notag \\ & \quad
\sum_{ \substack{ T' j' m' \lambda_{1}'  \lambda_{2}' \lambda_{3}' \rho_{3}' \\  
T j m \lambda_{1}  \lambda_{2} \lambda_{3} \rho_{2} \rho_{3}} }
\sqrt{2j'+1}  \sqrt{2j+1}
\mathcal{D}^{(1/2)}_{M',m'-\lambda'_{1}}( \varPhi' , \varTheta' , 0 )
\mathcal{D}^{(1/2)\ast}_{M,m-\lambda_{1}}( \varPhi , \varTheta , 0 )
\mathcal{D}^{(j')}_{m',\lambda'_{2}-\lambda'_{3}}( \tilde{\phi}' , \tilde{\theta}' , 0 )
\notag \\ & \quad \times
\mathcal{D}^{(j)\ast}_{m,\lambda_{2}-\lambda_{3}}( \tilde{\phi} , \tilde{\theta} , 0 ) 
C(  q' E(\tilde{p}^{\prime}) \tilde{p}' M' j' m' \lambda'_{1} \lambda'_{2} \lambda'_{3} + \rho'_{3} T' \mathcal{T}_{z}  )
C(  q \tilde{p}^{0} \tilde{p} M j m \lambda_{1} \lambda_{2} \lambda_{3} \rho_{2} \rho_{3} T \mathcal{T}_{z}  )
\notag \\ & \quad \times
F^{(21)}_{\lambda'_{2}\lambda_{1}} 
\left( R_{ \frac{\varPhi + \varPhi'}{2} , 0 , 0 }\right)^{\mu}_{\phantom{\mu}{\nu}}
\left( F^{(1)}_{T'T\mathcal{T}_{z}} F^{(12)\nu}_{\lambda'_{1}\rho_{2}\lambda_{2}}
+ F^{(2)}_{T'T\mathcal{T}_{z}} F^{(12)\nu\tau}_{\lambda'_{1}\rho_{2}\lambda_{2}} \qph_{\tau} \right) 
F^{(33)}_{\rho'_{3}\lambda'_{3}\rho_{3}\lambda_{3}} \, , \label{cap3_lk_87_gt}
\end{align}
where $F^{(21)}_{\lambda'_{2}\lambda_{1}}$ is defined in \eqref{edfr}, and the remaining spinor matrix elements are
\begin{gather}
\begin{split}
F^{(12)\nu}_{\lambda'_{1}\rho_{2}\lambda_{2}} &= 
\frac{m}{E(\tilde{p})} \frac{\tilde{p}^{0}+\rho_{2}E(\tilde{p})}{E^{2}(\tilde{p})-(\tilde{p}^0)^2}
\bar{u}^{+}(q',\lambda'_{1}) 
S^{-1}(R_{\pi,\pi,0}) S^{-1} ( R_{ \frac{\varPhi'-\varPhi}{2} ,\varTheta',0} ) S^{-1}( B ( \eta \hat{e}^{3} ) )
\gamma^{\nu} \\ & \quad \times
S ( R_{ \frac{\varPhi-\varPhi'}{2} ,\varTheta,0} ) 
S( Z(  q ) )
S(R_{\tilde{\phi},\tilde{\theta},0}) 
u^{\rho_{2}}(\tilde{p},\lambda_{2}) \, ,  \end{split} \label{cap3_df345_1}\\
\begin{split}
F^{(12)\nu\tau}_{\lambda'_{1}\rho_{2}\lambda_{2}} &= 
\frac{m}{E(\tilde{p})} \frac{\tilde{p}^{0}+\rho_{2}E(\tilde{p})}{E^{2}(\tilde{p})-(\tilde{p}^0)^2}
\bar{u}^{+}(q',\lambda'_{1}) 
S^{-1}(R_{\pi,\pi,0}) S^{-1} ( R_{ \frac{\varPhi'-\varPhi}{2} ,\varTheta',0} ) S^{-1}( B ( \eta \hat{e}^{3} ) )
i\sigma^{\nu\tau} \\ & \quad \times
S ( R_{ \frac{\varPhi-\varPhi'}{2} ,\varTheta,0} ) 
S( Z(  q ) )
S(R_{\tilde{\phi},\tilde{\theta},0}) 
u^{\rho_{2}}(\tilde{p},\lambda_{2}) \, ,  \end{split} \label{cap3_df345_2}\\
\begin{split}
F^{(33)}_{\rho'_{3} \lambda'_{3} \rho_{3} \lambda_{3}} &= 
 \frac{m}{E(\tilde{p}')} g^{\rho'_3}(q',\tilde{p}')
\left ( \frac{m}{E(\tilde{p})} \right )^{2}
\sum_{ \sigma_{3} = \pm}
O_{\rho_{3} \sigma_{3}}( \tilde{p} , \lambda_{3} )
\bar{u}^{\rho'_{3}}(\tilde{p}',\lambda'_{3})  \\
& \quad \times
S^{-1}(R_{ \pi , \pi ,0})
S^{-1}(R_{\tilde{\phi}',\tilde{\theta}',0}) 
S^{-1}( Z(  q' ) )
S^{-1} ( R_{0,\varTheta',0} ) S^{-1}( B ( \eta \hat{e}^{3} ) )  \\
& \quad \times 
S ( R_{\varPhi-\varPhi',\varTheta,0} )  S( Z(  q ) )
S(R_{\tilde{\phi},\tilde{\theta},0}) 
S(R_{ \pi , \pi , 0  }) 
u^{\sigma_{3}}(\tilde{p},\lambda_{3}) \, . \end{split} \label{cap3_xxx_p} 
\end{gather}
The isospin dependent factors are given by
\begin{align}
F^{(1)}_{T'T\mathcal{T}_{z}} &= \tilde{f} ( m^{2} , q_{2}^{2} )
\left(
F_{1p} ( Q^{2} )
\langle  T'
{\mathcal{T}}_{z} \vert
P^{\mathrm{iso}}_{12} \frac{1+\tau^{3}_{2}}{2}
\vert  T
{\mathcal{T}}_{z} \rangle 
+
F_{1n} ( Q^{2} )
\langle  T'
{\mathcal{T}}_{z} \vert
P^{\mathrm{iso}}_{12}\frac{1-\tau^{3}_{2}}{2}
\vert  T
{\mathcal{T}}_{z} \rangle
\right) \, , \\
F^{(2)}_{T'T\mathcal{T}_{z}} &= \tilde{g} ( m^{2} , q_{2}^{2} )
\left(
F_{2p} ( Q^{2} )
\langle  T'
{\mathcal{T}}_{z} \vert
P^{\mathrm{iso}}_{12}\frac{1+\tau^{3}_{2}}{2}
\vert  T
{\mathcal{T}}_{z} \rangle 
+
F_{2n} ( Q^{2} )
\langle  T'
{\mathcal{T}}_{z} \vert
P^{\mathrm{iso}}_{12}\frac{1-\tau^{3}_{2}}{2}
\vert  T
{\mathcal{T}}_{z} \rangle
\right) \, , 
\end{align}
where $q_{2}$ is defined in \eqref{qwsa}, its component $\tilde{p}^{0}$ now being determined 
in  \eqref{mkjn}, and $\tilde{p}=r_{2}$. The matrix elements involving the isospin projections of nucleon 2 are
\begin{align}
\langle  0
{\mathcal{T}}_{z} \vert
P^{\mathrm{iso}}_{12}\frac{1\pm\tau^{3}_{2}}{2}
\vert  0
{\mathcal{T}}_{z} \rangle & =
\frac{1}{4} \pm \frac{1}{2} {\mathcal{T}}_{z} & 
\langle  1
{\mathcal{T}}_{z} \vert
P^{\mathrm{iso}}_{12}\frac{1\pm\tau^{3}_{2}}{2}
\vert  1
{\mathcal{T}}_{z} \rangle & =
-\frac{1}{4} \pm \frac{1}{6} {\mathcal{T}}_{z}  \nonumber\\
\langle  1
{\mathcal{T}}_{z} \vert
P^{\mathrm{iso}}_{12}\frac{1\pm\tau^{3}_{2}}{2}
\vert  0
{\mathcal{T}}_{z} \rangle & =
-\frac{\sqrt{3}}{4} \pm \frac{\sqrt{3}}{6} {\mathcal{T}}_{z} &
\langle  0
{\mathcal{T}}_{z} \vert
P^{\mathrm{iso}}_{12}\frac{1\pm\tau^{3}_{2}}{2}
\vert  1
{\mathcal{T}}_{z} \rangle & =
-\frac{\sqrt{3}}{4} \mp \frac{\sqrt{3}}{2} {\mathcal{T}}_{z} \, .
\end{align}
It is again useful to change integration variables in \eqref{cap3_lk_87_gt} from $\varPhi'$ and $\varPhi$ to $\phi$ and $\varphi$, according to \eqref{cap3_gh_39} and \eqref{cap3_gh_39p}.
The integration over $\varphi$ can then be carried out analytically, in analogy to the integration over $\varphi$
in diagram D.

As in the case of diagrams B and C, we encounter propagator singularities, whose location needs to be determined. For this purpose, the denominator of the factor 
$ ( \tilde{p}^{0}+\rho_{2}E(\tilde{p}) ) / ( E^{2}(\tilde{p})-(\tilde{p}^0)^2 ) $
which appears in \eqref{cap3_df345_1} and \eqref{cap3_df345_2}, is written as
\begin{align}
E^{2}(\tilde{p}) - (\tilde{p}^0)^2 
&=
Q^{2} - 2 Q ( \sinh ( \zeta + \eta ) E(q')  - \cosh ( \zeta + \eta ) q' \cos \varTheta' ) \, , \\
\intertext{with} 
\sinh \zeta &= -  \sqrt{\frac{K^2-Q^2}{Q^2}} \, .
\end{align}
We find that
$  E^2(\tilde{p})-(\tilde{p}^{0})^2 = 0  $
if and only if
$ a \le q' < q_{s} $ and $\varTheta'=\varTheta'_s$,
 where
\begin{align}
a &= \sqrt{ - m^{2} + 
\left [
\cosh ( \zeta + \eta )  E \left ( \frac{Q}{2} \right )       - \sinh ( \zeta + \eta ) \frac{Q}{2}
\right ]^{2} } \, , \\
B(q') &=
\frac{\sinh ( \zeta + \eta ) E(q')       -\frac{Q}{2}}{q'\cosh ( \zeta + \eta )} \, , 
\qquad
\varTheta'_s = \arccos B(q') \, .
\end{align}
With this information it is now possible to carry out the principal value integration over $\varTheta'$. The remaining integrations are straightforward.

Finally, diagram E is obtained from diagram F in complete analogy to the way diagram B is calculated from diagram C, namely
\begin{gather}
\begin{split}
\langle M'  \vert J_{E}^{0} \vert  M  \rangle 
&= 
( B ( \eta \hat{e}^{3} ) )^{0}_{\phantom{0}0} 
\langle M'  \vert J_{F}^{0} \vert  M  \rangle
-
( B ( \eta \hat{e}^{3} ) )^{0}_{\phantom{0}3}
\langle M'  \vert J_{F}^{3} \vert  M  \rangle
\, , \end{split} \\ 
\langle M'  \vert J_{E}^{1} \vert  M  \rangle
 = 
\langle M'  \vert J_{F}^{1} \vert  M  \rangle
\, , \\ 
\langle M'  \vert J_{E}^{2} \vert  M  \rangle
 =  
\langle M'  \vert J_{F}^{2} \vert  M  \rangle
\, , \\ 
\begin{split}
\langle M'  \vert J_{E}^{3} \vert  M  \rangle
& = 
- ( B ( \eta \hat{e}^{3} ) )^{3}_{\phantom{3}3} 
\langle M'  \vert J_{F}^{3} \vert  M  \rangle
+
( B ( \eta \hat{e}^{3} ) )^{3}_{\phantom{3}0}
\langle M'  \vert J_{F}^{0} \vert  M  \rangle
 \, . \end{split}
\end{gather}

\end{widetext}

\bibliographystyle{h-physrev3} 
\bibliography{FBpapers}

\end{document}